\documentclass[11pt,preprint]{aastex}
\usepackage{epsf,epsfig}
\begin{document}
\newcommand{\ts}{\thinspace}
\newcommand{\p}{$^{\prime}$\ts}
\newcommand{\pns}{$^{\prime}$\ts}
\newcommand{\nul}{$\nu${\it L}$_{\nu}$}
\newcommand{\degree}{\arcdeg}
\newcommand{\msun}{\hbox{M$_\odot$}\ }
\newcommand{\msunns}{\hbox{M$_\odot$}}
\newcommand{\micronns}{$\mu$m}
\newcommand{\lsun}{\hbox{{\it L}$_\odot$}\ }
\newcommand{\lsunns}{\hbox{{\it L}$_\odot$}}
\newcommand{\rqu}{r$^{-1/4}$\ }
\newcommand{\lstar}{L$^*$}
\newcommand{\lbol}{L$_{bol}$}
\newcommand{\arcsecns}     {\char'175}
\newcommand{\arcsecp}      {\char'175\llap{.}}

\title{SWIRE:  The SIRTF Wide-area InfraRed Extragalactic Survey}

\author{CAROL J. LONSDALE}
\affil{Infrared Processing \& Analysis Center\\
       California Institute of Technology 100-22, Pasadena, CA 91125, USA;\\
       {\it cjl@ipac.caltech.edu}}

\author{HARDING E. SMITH \altaffilmark{1}}
\affil{Center for Astrophysics \& Space Sciences and Department of Physics\\
       University of California, San Diego, La Jolla, CA 92093--0424, USA;\\
	{\it hsmith@ucsd.edu}}

\altaffiltext{1}{also, Infrared Processing and Analysis Center,
               California Institute of Technology}

\author{MICHAEL ROWAN-ROBINSON}
\affil{Astrophysics Group, Blackett Laboratory, Imperial College\\
       Prince Consort Road, London, SW7 2BW, UK}

\author{JASON SURACE, DAVID SHUPE and CONG XU}
\affil{Infrared Processing \& Analysis Center\\
       California Institute of Technology 100-22, Pasadena, CA 91125, USA}

\author{SEBASTIAN OLIVER}
\affil{Astronomy Centre, CPES, University of Sussex\\
       Falmer, Brighton BN1 9QJ, UK}

\author{DEBORAH PADGETT and FAN FANG}     
\affil{Infrared Processing \& Analysis Center\\
       California Institute of Technology 100-22, Pasadena, CA 91125, USA}

\author{TIM CONROW}          
\affil{Infrared Processing \& Analysis Center\\
       California Institute of Technology 100-22, Pasadena, CA 91125, USA}

\author{ALBERTO FRANCESCHINI}          
\affil{Dipartimento di Astronomia, Universita di Padova\\
       Vicolo Osservatorio 5, I-35122 Padua, Italy}

\author{ }

\author{NICK GAUTIER}          
\affil{Jet Propulsion Laboratory, 264-767, 4800 Oak Grove Drive\\
       Pasadena, CA 91109, USA}

\author{MATT GRIFFIN}          
\affil{Department of Physics and Astronomy, University of Wales Cardiff\\ 
       5 The Parade, Cardiff CF24 3YB, UK}

\author{PERRY HACKING}          
\affil{Department of Astronomy, El Camino College\\
	16007 Crenshaw Blvd., Torrance, CA 90506, USA}

\author{FRANK MASCI, GLENN MORRISON and JOANNE O'LINGER}          
\affil{Infrared Processing \& Analysis Center\\
       California Institute of Technology 100-22\\
       Pasadena, CA 91125}

\author{FRAZER OWEN}          
\affil{National Radio Astronomy Observatory\\
       P.O. Box O, Socorro, NM 87801, USA}

\author{ISMAEL P\'EREZ-FOURNON}          
\affil{Instituto de Astrofisica de Canarias\\
       38200 La Laguna, Tenerife, Spain}

\author{MARGUERITE PIERRE}          
\affil{CEA/DSM/DAPNIA, Service d'Astrophysique\\
       91191 Gif-sur-Yvette, France}

\author{RICK PUETTER  \altaffilmark{2}}          
\affil{Center for Astrophysics \& Space Sciences\\
       University of California at San Diego, La Jolla, CA 92093, USA}
\altaffiltext{2}{Pixon LLC, 9295 Farnham Street, San Diego, CA 92123, USA}
 
\author{GORDON STACEY}          
\affil{Department of Astronomy, Cornell University\\
       220 Space Science Building, Ithaca, NY 14853, USA}

\author{SANDRA CASTRO, MARIA DEL CARMEN POLLETTA, DUNCAN FARRAH, TOM JARRETT 
and DAVE FRAYER}
\affil{Infrared Processing \& Analysis Center\\
       California Institute of Technology 100-22, Pasadena, CA 91125}

\author{BRIAN SIANA}
\affil{Center for Astrophysics \& Space Sciences \\
       University of California, San Diego, La Jolla, CA 92093--0424, USA}

\author{TOM BABBEDGE, SIMON DYE and MATT FOX}
\affil{Astrophysics Group, Blackett Laboratory, Imperial College\\
       Prince Consort Road, London, SW7 2BW, UK}

\author{EDUARDO GONZALEZ-SOLARES and MALCOLM SALAMAN}
\affil{Astronomy Centre, CPES, University of Sussex\\
       Falmer, Brighton BN1 9QJ, UK}

\author{STEFANO BERTA}
\affil{Dipartimento di Astronomia, Universita di Padova\\
       Vicolo Osservatorio 5, I-35122 Padua, Italy}

\author{JIM J. CONDON}          
\affil{National Radio Astronomy Observatory\\
       520 Edgemont Road, Charlottesville, VA 22903, USA}

\author{HERV\' E DOLE}          
\affil{Steward Observatory, University of Arizona\\
       933 N Cherry Ave, Tucson, AZ 85721, USA}

\author{STEVE SERJEANT}          
\affil{Centre for Astronomy and Planetary Science, School of Physical 
       Sciences\\
       University of Kent at Canterbury, Canterbury, Kent CT2 7NZ}

\author{ }
\author{ }
\author{ }
\author{ }
        
        \begin{abstract}
The SIRTF Wide-area InfraRed Extragalactic survey (SWIRE), the largest
SIRTF Legacy program, is a wide-area, imaging survey to trace the evolution
of dusty, star-forming galaxies, evolved stellar populations, and AGN as a
function of environment, from redshifts z$\sim$3 to the current epoch.
SWIRE will survey 7 high-latitude fields, totaling 60 - 65 sq. deg. in all
7 SIRTF bands: IRAC 3.6, 4.5, 5.6, 8$\mu$m and MIPS 24, 70, 160$\mu$m.
Extensive modeling suggests that the Legacy Extragalactic Catalog may
contain in excess of 2 million IR-selected galaxies, dominated by (1)
$\sim$150,000 luminous infrared galaxies (LIRGs: L$_{FIR}>10^{11}$
L$_{\odot}$) detected by MIPS (and significantly more detected by IRAC),
$\sim$7,000 of these with z$>$2; (2) 1 million IRAC-detected early-type
galaxies ($\sim$ 2$\times$10$^5$ with $z > 1$ and $\sim$10,000 with $z >
2$); and (3) $\sim$ 20,000 classical AGN detected with MIPS, plus
significantly more dust-obscured QSO/AGN among the LIRGs. SWIRE will
provide an unprecedented view of the evolution of galaxies, structure, and
AGN.
  
The key scientific goals of SWIRE are: (1) to determine the evolution of
actively star-forming and passively evolving galaxies in order to
understand the history of galaxy formation in the context of cosmic
structure formation; (2) to determine the evolution of the spatial
distribution and clustering of evolved galaxies, starbursts and AGN in the
key redshift range, 0.5$<$z$<$3, over which much of cosmic evolution has
occurred; (3) to determine the evolutionary relationship between ``normal
galaxies'' and AGN, and the contribution of AGN accretion energy {\it vs.} 
stellar nucleosynthesis to the cosmic backgrounds.  The large area of SWIRE
is important to establish statistically significant population samples over
enough volume cells that we can resolve the star formation history as a
function of epoch and environment, {\it i.e.} in the context of structure
formation.  The large volume is also optimised for finding rare objects.
 
The SWIRE fields are likely to become the next generation of {\it large}
``Cosmic Windows'' into the extragalactic sky.  They have been uniquely
selected to minimize Galactic cirrus emission over large scales.  GALEX
will observe them as part of its deep 100 sq deg survey, as will Herschel.
SWIRE includes $\sim$9 sq deg of the unique large-area XMM-LSS hard X-ray
imaging survey, and is partly covered by the UKIDSS deep J \& K survey.  An
extensive optical/near-IR imaging program is underway from the ground.

The SWIRE data are non-proprietary; catalogs and images will be released
twice-yearly, beginning about 11 months after SIRTF launch.  Details of the
data products and release schedule are presented.
\end{abstract}


\section{Scientific Goals of SWIRE}

\subsection{Modes and Rates of Star Formation}

With SWIRE we seek to directly measure the light of both dusty galaxies and
evolved stellar populations at $z<3$ (Universe age $>$2 Gyr)
\footnote{H$_0$=75 km/s/Mpc, ${\Omega}_{\Lambda}=$0.7,
${\Omega}_m$=0.3, which we adopt throughout except for some length scales
and volumes quoted in H$_0=100 h$ km/s/Mpc units}, with enough galaxy and
volume cell statistics to fully sample the range of density environments
from dense clusters to voids.

From a theoretical viewpoint, the standard cosmological paradigm --- that we
live in a $\Lambda$CDM universe that began with the Big Bang followed by an
inflationary period --- has been exceedingly successfully in explaining the
major cosmological observations, including the cosmic microwave background
and its spatial structure, the power spectrum of the CMB, cosmic
nucleosynthesis, the large-scale filamentary structure of the cosmic web of
galaxies, and dark matter halos.

When it comes to galaxy formation and evolution, the
observational situation becomes much more complex because we most
easily study light, whereas theory only directly predicts the distribution
and merging history of the underlying mass distribution, and the relation
between the two depends on complex, usually non-linear, astrophysical
processes.  In spite of this, hierarchical theories have also been
remarkably successful at predicting the overall scheme of galaxy formation
and evolution: that it is dominated by merging of clumps within dark matter
halos and the steady accretion of gas into disks; that it was at peak
activity at z$\sim$1-3; that it seems to have occurred faster in higher
density environments; and that it has declined sharply since z$\sim$1.  The
morphology of galaxies seems to have evolved strongly, driven by both major
and minor mergers and by episodes of gas accretion, and the Hubble sequence
as we now know it seems to have come into being between redshifts 1 to 2.

Mid- and Far-infrared (IR) observations of evolving galaxies are essential 
because a large fraction of emitted starlight in galaxies is absorbed and 
re-emitted by dust in the thermal infrared.  The COBE detection of the 
Cosmic Infrared Background (CIB) 
has illustrated that $>$50\% of the total luminous energy density 
of the Universe emerges longward of $\sim$1$\mu$m \citep{hauser}.  An even 
higher fraction of a galaxy's light (90\% or more) 
can emerge in the far-infrared in starbursts, which may be
a significant mode of star formation over the history of the Universe.
The most extreme IR objects are the ULIRGs (ultraluminous IR galaxies), with
QSO-scale luminosities ($>10^{12}\lsunns$), which are very rare in the local
Universe.

The ISO satellite made several important surveys at 6.7, 15, 90 and
170\micronns, resulting in star formation histories that increase even more
steeply back in time than observed in the UV-optical \citep{mrr97,elbaz}.
Perhaps most surprisngly, a large population of faint submillimeter sources
has been discovered, using SCUBA on JCMT, which represent a population of
very luminous objects at redshifts $z=2$--4 with a much higher space density
than the local ULIRGs \citep{blainrev,chap03b}.  These populations, and the
integrated CIB, have been modelled successfully using phenomenological
models \citep{mrr01,charelb,xu01} some of which require high rates of
luminosity evolution, $L \propto (1+z)^{4}$. Moreover the star formation
efficiencies (SFEs) of the high z ULIRGs may have been much higher than
now; \citet{blain99} find the halo mass-to-infrared light ratio of a
typical merger at z=3 to be $\sim$200 times smaller than today.
   
These results challenge hierarchical models conceptually because it is
expected that galaxy masses grow slowly as smaller systems merge and gas
accretes.  Indeed, no CDM-based model, whether semi-analytic or N-body, has
successfully reproduced the observed numbers of high redshift ULIRGs from
the submm data \citep{guiderdoni,fardal,devriendt,somerville,somervillepc}.
Instead, very high star formation rates (SFRs) at early times are more
reminiscent of monolithic collapse theories than the standard hierarchical
theory. In an extreme case, too much star formation in bursts at high
redshift will conflict with the conversion rate of baryonic matter into
stars that is derived from deep K-band studies \citep{dickinson02}.

There are significant unknowns associated with these studies of 
the evolving IR galaxy populations.  A major one is the dust temperature: 
with only a single 850$\mu$m detection on the Rayleigh Jeans tail of 
the Planck spectrum, the temperature is unconstrained and the
luminosity and SFR can be uncertain by up to 2 orders of magnitude
\citep{kaviani}. \citet{mrr01} suggests that the high redshift IR
population might be dominated by cooler dust emission from larger disks
than the sub-kpc-scale warm dust sources that seem to characterize local
ULIRGs, and indeed \citet{chap03} find evidence for some cool ISO sources
at z$\sim$1.  Other unknowns include the timescale for starbursts at high
redshift, the form of the initial mass fucntion (IMF) in starbursts (an IMF
truncated at the low mass end would reduce conflicts between the observed
luminosities and the conversion rate of gas to stars \citep{frans01}), and
the contribution to the IR luminosities of dust-enshrouded AGN.

SWIRE is designed to provide unique and essential information about star
formation rates and modes between about 0.5$<$z$<$3.  The multiple bands
spanning the thermal infrared from 4 to 160$\mu$m provide unprecedented
coverage of the SED, which will allow an accurate estimate of the
luminosity of the warmer dust components out to redshifts $\sim$2, and
cooler components to lower redshifts.  Color-dependent luminosity functions
will elucidate starburst {\it vs.} quiescent star formation rates, and
starburst timescales {\it vs.} AGN processes.  SWIRE will address how star
formation in IR-luminous systems differed at early times from today.  The
majority of SWIRE dusty populations will be LIRGS ($L \sim 10^{11}\lsunns$)
at z$\sim$1, when bulges and disks were fully coming into being, so we can
study star formation rates associated with these processes directly.  The
z$>$2 SCUBA sources had much higher star formation efficiency than local
ULIRGs: did they simply have more prodigious nuclear starbursts (local
ULIRGs concentrate their starbursts within the central 1kpc, and frequently
the central 100pc)?  Or did they sport extensive disks with high star
formation efficiencies throughout, driven by high rates of gas accretion in
galaxy clusters?  Or are they enormous mergers with extensive regions of
distributed star formation?  The dust temperature associated with these
episodes will be a strong discriminator because higher density star forming
regions typically reach higher dust temperatures.

The large SWIRE fields will allow us to track these 
processes as a function of environment in hundreds of volume cells
from rich clusters to the ``field''.  For example, we can search for the
CDM-predicted trend for active star formation to be more closely confined to
the denser regions of the Universe at the higher redshifts, and as star 
formation rates decrease in overall strength with cosmic time, to move 
systematically to less and less dense environments.   

The large volume of SWIRE is also uniquely important for the discovery of
rare objects; a large ``shallow'' extragalactic survey can cover more
volume than the same amount of time spent on a smaller, deeper one
\citep{condon98} as demonstrated by the success of the Sloan Digital Sky
Survey at finding z$>$5 QSOs \citep{fan03}.  In particular, SWIRE has much 
more volume sensitivity to z$\sim$3  ULIRGs than current submm surveys, with
the potential to detect $\sim$100 z$>$3 ULIRGs per square degree by their 8
\&/or 24$\mu$m flux, according to the models of \citet{xu03}, which are
consistent with faint submm counts and redshift distributions, and the CIB.
Thus SWIRE will confirm the presence of a substantial population of high z
ULIRGs, improve estimates of their luminosities and SFRs, and trace their
clustering properties.  Intriguingly, there is tentative evidence to
suggest that high-$z$ sub-mm sources may be clustered
\citep{scott,almaini}, as might be expected if these events are tracing the
highest density environments at these redshifts.  According to semianalytic
star-formation scenarios (e.g, \citet{somerville}), ULIRGs at high-z may be
tracers of already formed massive halos.  Although at low redshift ULIRGs
are associated with violent mergers and are not found in massive halos
(rich clusters), at earlier epochs galaxy building occurred first in the
deepest potential wells, and distant ULIRGs may be the progenitors of
current cluster spheroids.  SWIRE may detect enough ULIRGs at z$>$2.5 with
halos of mass ${\ge}10^{13}$\msun to allow a statistical estimate of
${\sigma}_8$, the rms fluctuation of the mass distribution on 8/h Mpc
scales; the predicted number density of such halos differs by a factor of
$\sim$6 at z$\sim$2.5 for 0.7$<{\sigma}_8<$0.9
\citep{somervillepc}.

\subsection{Spheroids}

While dust emission in galaxies tracks the most recently formed stellar
populations, the SEDs of passively evolving older stellar populations peak
in the near-infrared, and the wavelength bands of SIRTF's IRAC camera were
selected (in part) to optimize study of them at high redshifts.  A
fundamental goal of SIRTF/IRAC is to establish the evolution of the mass
and luminosity functions of evolved stellar populations, and relate them to
the morphological and color evolution of galaxies and the establishment of
the Hubble sequence \citep{simpson}.  Is there a significant population of
high redshift evolved systems which formed at very high redshift in a
``monolithic collapse'' scenario, as perhaps indicated by the SCUBA
sources, or can hierarchical models fully explain the formation of massive
galaxies and spheroids at moderate redshifts from merging of pre-existing
galaxian units?  Are the stellar populations of distant spheroids being
formed in substantial amounts at 1$<$z$<$2, or are older stellar
populations being dynamically assembled into massive systems at these
moderate epochs?

Fossil evidence in the local Universe favors a uniform, high-redshift 
formation epoch for massive cluster ellipticals; in particular, the tightness 
of the fundamental plane and the enrichment of massive 
systems with $\alpha$ elements \citep{ellis97} point to rapid 
homogeneous star formation episodes at high redshift.  However,
recent HST imaging of distant spheroids has revealed substantially
more morphological than color evolution at z$<$2-3, revealing increasing
evidence for mergers and peculiarites with increasing redshift
\citep{conselice, phillips01}, and evidence for significant 
color inhomogeneities \citep{benson, menanteau, kajisawa}.   
Estimates for the stellar masses and the stellar mass 
build-up with redshift of a deep K-band sample in the HDF-N 
\citep{dickinson02}  indicate that only 3-14\% of the current day mass in 
stars had formed by z$\sim$2.7, and 50-75\% had formed by z$\sim$1. This is 
in agreement with some hierarchical models (between which there 
is significant dispersion in predictions), but apparently in disagreement
with models in which the bulk of stars in present day spheroids formed
at very early times (z$>>$2).  On the other hand, \citet{benson}
find that hierarchical models under-predict the proportion of high-redshift,
homogeneous, passively evolving objects, and \citet{labbe02} find a much 
larger population of red systems with z$_{phot}>$2 in the HDF-S than
\citet{dickinson02} do in HDF-N.  \citet{conselice} finds that
the rate of major mergers amongst the most massive systems increases
strongly with z, reaching $\sim$50\% at z$>$2.5, in qualitative agreement
with the hierarchical picture. On the other hand, massive systems with
rapid star formation must already be in place at z $\sim 3$, an observation
that is difficult to reconcile with slow stellar mass 
build-up predictions of some hierarchical models, such as 
\citet{cole00}.  Central to these discussions is the cosmological model
assumed; {\it eg.} \citet{phillips01} find morphological number counts in the 
HDFs inconsistent with the large ${\Omega}_{\Lambda}$ implied by current 
concordance cosmology.

SWIRE/IRAC will directly provide the accumulated mass in evolved stars 
measured from the r$'$-5$\mu$m SEDs. IRAC design was in part
optimized for measuring the stellar mass of distant systems, and the
GOODS Legacy program takes this science to the deepest possible SIRTF
limit in a very small region of sky.  Ground-based and HST NICMOS studies
are limited to ${\lambda}<2.5{\mu}$m, corresponding to z$<$0.6 for the 
${\sim}{\lambda}_{rest}=$1.6$\mu$m SED peak of evolved stellar systems.
IRAC therefore provides much more robust determination of stellar masses
to much higher redshifts \citep{sawicki02,dickinson02,frans03}.  
SWIRE/IRAC observations are very
well matched in depth to SWIRE dusty galaxy observations, sampling
the important z=1-2 redshift range over which
low matter-density cosmological models predict much evolution.

A key advantage of SWIRE's large area coverage is that these questions can
be addressed using large samples over many contiguous volume cells, thus 
minimizing cosmic variance problems, such as those implied between HDF-N and 
HDF-S, as mentioned above.
Moreover we naturally expect a strong dependence of cosmic evolution and 
the timescales of galaxy formation on the local matter density, so it
important to survey the full matter density range in a homogenous fashion.

\subsection{Active Galaxies}

The fundamental cosmogonic questions concerning active galactic nuclei (AGN)
include (1) the true distribution of physical processes underlying the zoo
of observational classes of AGN; (2) the connections between galaxy formation
and black hole growth and activity; (3) the importance of AGN contributions
to re-ionizing the Universe; and (4) the contribution of gravitational 
energy from AGN to the overall luminous energy density of the Universe as a 
function of redshift.

The most challenging aspect of AGN research over the years has been in
assembling complete samples, because AGN suffer very strong observational
selections effects of many kinds, which are very difficult to separate 
from intrinsic physical differences.  SWIRE will be one of the best ever 
surveys for AGN because many AGN classes emit strongly in the mid-IR where 
extinction is low and where SWIRE has excellent volume sensitivity.  
In particular,
many AGN have warm mid-to-far-infrared colors compared to star 
formation-dominated galaxies, and thus will be preferentially selected by 
the highly sensitive 8 and 24$\mu$m SIRTF bands.   

Moreover, evidence is now very strong for the existence of a substantial 
population of AGN which are heavily absorbed, and which are strong mid- to 
far-IR emitters.   Highly reddened and absorbed AGN are turning up in radio, 
hard X-ray and near infrared surveys \citep{maiolino98,gregg02,cutri}.  
Some of these populations are consistent with the classical
AGN unification picture featuring axisymmetrical distributions
of surrounding absorbing material, and others favor a young AGN turning 
on during a 
merger/starburst episode, deeply embedded in obscuring material from 
most or all vantage points.   Furthermore, the long-standing puzzle over
the mismatch between the spectral slope of the cosmic X-ray background and
that of its presumed consituent quasars may be solved now that the XRB is
close to being resolved by XMM and Chandra, and is found to be well
explained by a population of high column AGN which increases in number
density with redshift \citep{comastri,gilli,polletta03}.  

A major SWIRE goal is therefore to determine 
the evolving number density of AGN, particularly heavily obscured ones, and 
to quantify much better than any other survey their number density at z$>$1, 
and their contribution to the CIB; see \citet{polletta03} for a recent model.
A difficulty with this goal concerns
the most heavily obscured AGN, which can have columns over  
N${_H}{\sim}10^{25}cm^{-2}$ \citep{maiolino00}, 
and which can therefore be optically
thick even in the near- and mid-IR and thus not recognizable even in
SWIRE mid-IR colors.  Thus it is most important to undertake hard X-ray 
surveys in SWIRE fields.  The XMM-LSS field 
(Section \ref{xmmlss}) will be our prime one for obscured AGN studies.

The other key wavelength range for detection of heavily obscured AGN is the
radio, thus a systematic deep radio survey would be of great value in the
SWIRE fields.  The remarkably tight FIR/radio correlation for star forming
galaxies can be used as a tool for identifying radio-loud AGN.  Only with
radio and hard X-ray surveys can the population of dust-shrouded AGNs be
recognized and discriminated from starbursts.

SWIRE can be expected to make direct and unique contributions to the AGN
unification debate, due to the size and completeness of its AGN samples.
SWIRE will constrain the IR properties of AGN of all types over large
volumes and all environments: how many of the known AGN classes are IR-loud
and how does IR-loudness relate to other AGN properties and to the presence
of a starburst?  SWIRE will also track the IR evolution of all IR-loud AGN
classes over a substantial redshift range, addressing the global
evolutionary connections between them.

Another important SWIRE goal is to understand the ``starburst-AGN
connection''.  The locally observed correlation between the masses of black
holes and their surrounding stellar bulges \citep{magorrian} implies that
the formation of the two are linked, yet we do not understand how. Clues
must come from the relationship between star formation and black hole
accretion in active galaxies, since it is widely believed that
merger-driven starbursts are effective bulge-builders.  A close
phenomenological relationship between starbursts and AGNs has been recently
suggested by \citet{frans02}  based on deep combined IR and XMM-Newton 
observations. How are starbursts and AGN triggered in active galaxies? How
does AGN feedback affect starburst activity and vice versa?

In the local Universe, IR-luminous starburst and AGN activity is often
triggered by galaxy mergers, especially for the most luminous ULIRGs
($L_{1-1000}>10^{12}L_{\odot}$) \citep{surace00,farrah01}.  At high redshift, 
however, the trigger for starburst/AGN activity of the luminous submillimeter
sources is not known.   Possibilities include multiple mergers 
between many small galaxies, or even collapse of a large disk of gas to form 
a protogalaxy \citep{farrah02a}, this latter possibility is hinted at by 
models for high-$z$ sub-mm sources that suggest a cirrus origin for the 
sub-mm emission \citep{efstathiou}.  

Once triggered, the effects of starburst and AGN activity on each other are
currently the subject of debate. Theoretical arguments \citep{silk} suggest
that the onset of AGN activity will curtail a starburst, due to superwinds
from the AGN, and indeed modelling of starburst and AGN activity in
protogalaxies \citep{granato}, and of BH and spheroid formation
\citep{archibald} suggest that most of the star formation must have taken 
place before the onset of AGN activity. Conversely, there is substantial
observational evidence \citep{priddey, farrah02} that starburst and AGN
activity can be coeval in QSOs and Hyperluminous Infrared Galaxies
($L_{1-1000}>10^{13}L_{\odot}$); many of these systems contain a luminous
AGN together with star formation rates (inferred from sub-mm observations)
so high that it is unphysical to argue that they are not at or near their
peak.

\subsection{Large Scale Structure}

Theories of structure formation were strongly constrained by the
statistical measurements of clustering in some of the early galaxy
redshift surveys.  Surveys of infrared galaxies, in particular, were
able to rule out the then standard Cold Dark Matter Model \citep{e90,
s91}. Present day redshift surveys such as the 2dFGRS \citep{colless},
SDSS \citep{sdss} and, in the far-infrared, the Point Source Catalog
Redshift survey, PSC-z \citep{pscz} are now able to provide definitive
measurements of the galaxy clustering in the local Universe.

Despite this success, we have always known that galaxies are biased
tracers of the matter distribution.  The evidence for this is that
different types of galaxies cluster differently \citep{babul, lahav,
oliver96, norberg}.  If $\Omega_B \sim 0.3$ then some dynamical
comparisons of the galaxy velocity field and density fields are
consistant only if bias is included \citep{dekel94, strauss95, rr00}
while other comparisons do not require bias
\citep{1998ApJ...507...64W}, and recent studies of 2dFGRS
redshift-space distortions are also consistent with no bias
\citep{2002MNRAS.333..961L}.
Bias is expected because galaxies represent only a small fraction
of the total mass and even in very simplistic models where galaxies
form at the peak of the mass distribution this produces a bias
\citep{kaiser}.  The strong interactions seen in actively star-forming,
ULIRGs also provide circumstantial evidence
that galaxy formation depends on the environment in which galaxies
find themselves, which would also lead one to
expect a bias.  Until this question of bias is resolved the modern
galaxy surveys will be unable to probe the clustering properties of
the underlying mass field and thus test the cosmological models that
relate these to the CMB fluctuations.

Recently, models of structure formation have included physically
motivated mechanisms for galaxy formation to predict the spatial and
temporal evolution of galaxies, either using semi-analytic
prescriptions or hydrodynamic N-body simulations \citep{pearce,
somerville, benson, blanton2000}. By predicting the clustering
properties (and their evolution) of different types of galaxies such
models provide testable explanations of galaxy bias.  However, this
field is still in its infancy, partially due to a lack of
observational constraints over a significant redshift baseline.

SWIRE is able to make two key contributions to the understanding of
bias and constraining structure formation models of galaxy formation:
the survey samples very different populations of galaxies within the
same volume of the Universe and it provides a good sampling of large
volumes at high redshift.

\begin{figure}
\plotone{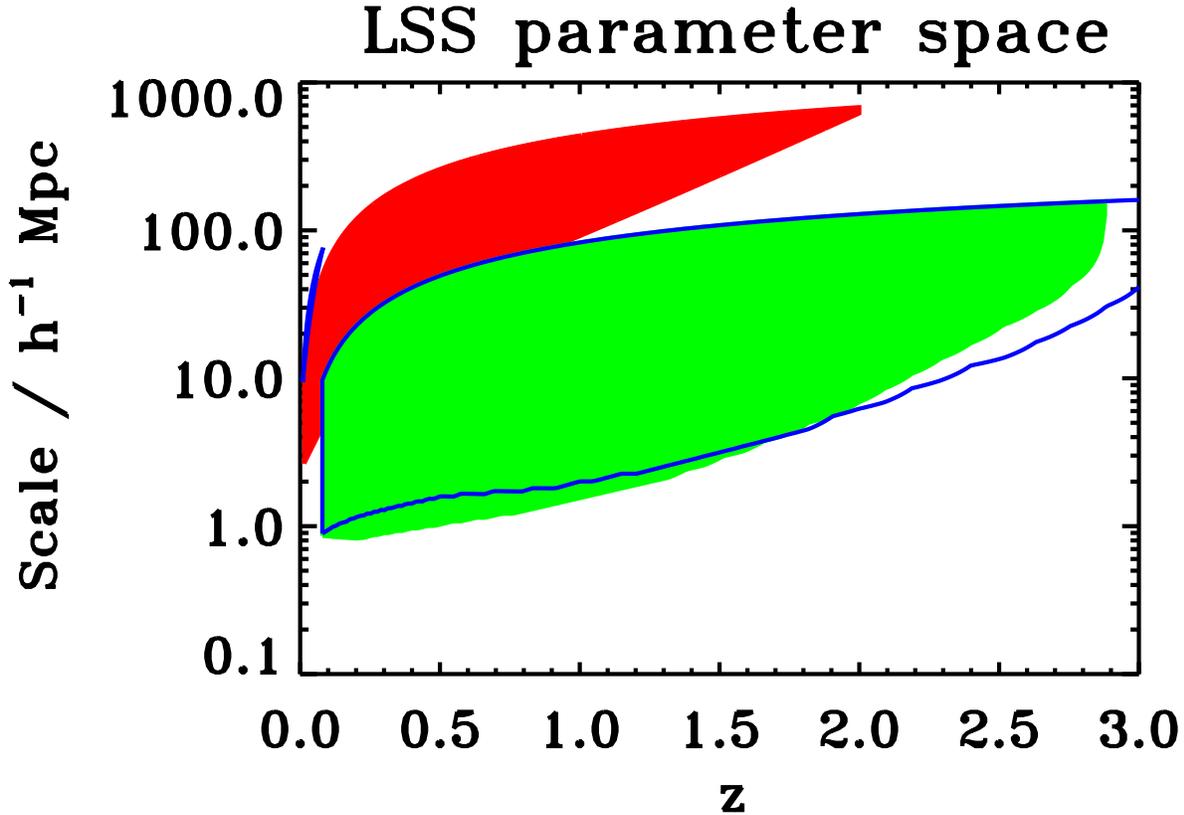}
\caption{Large-scale structure parameter space probed by various surveys.
We have assumed that to measure clustering statistics a survey needs at least 
100 cells above a threshold mean density of 1 galaxy per cell. 
Solid green and blue outline:  model spheroids at 3.6$\mu$m and model 
starbursts at 24$\mu$m from \citep{mrr01} respectively.  
For comparison we show the IRAS PSC-z sample (solid blue at low z) 
from \citet{pscz},
and the SDSS early data release spectroscopic sample (red) from 
\citet{sdss} to which we have fitted the following empirical fuction: 
$dn/dz = a z^2 e^{-(z/0.035)^{1.5}} + b z^2 e^{-(z/0.15)^{1.5}}$
with a=1.15$\times$10$^8$, b=3.1$\times$10$^5$.
The scale length (y-axis)
illustrates the size of cells that can be probed. The clustering that can be 
studied on a 
particular scale  (x-axis) is usually constrained by the number 
of cells criteria at low-$z$ and number density threshold at high-$z$.}
\label{fig:lss} 
\end{figure}

We illustrate the large-scale structure parameter space that SWIRE and
other surveys could probe in Figure \ref{fig:lss} and Table \ref{tbl:lss}.
The parameters of
interest are the spatial scale of clustering, $l$, and the redshift,
$z$.  To judge how a survey probes this parameter space we imagine the
survey out to $z$ to be divided into cells with volume $l^3$ and
assume that a cell has to have a mean density above 1 galaxy per cell to be
included, and that 100 cells are required for a useful clustering
analysis.  We also imposed a minimum redshift below which the mean
density of the survey is poorly determined due to sampling variance.

\subsection{Clusters and the XMM-LSS Field}\label{xmmlss}

As outlined above, measuring high redshift clustering is one of the 
challenges of modern observational cosmology.   Whereas galaxies are often 
considered as difficult objects to understand in terms of initial 
density fluctuations, clusters of galaxies, in a first approximation,  
are simpler and complementary. Indeed, an ab initio theory exists (and is 
well tested by N-body simulations) to describe how the cluster density 
variations relate to the dark matter density variation (biasing) 
while such a relation does not exist for galaxy formation. Moreover, 
clusters are the most massive relaxed entities of the Universe, located at 
the intersection of the cosmic filaments; cluster growth occurs by matter 
accretion flowing along the filaments. Studies of structure 
evolution and of cluster abundance can independently check cosmological 
parameter values determined from CMB and SN studies, as they do not rely on 
the same physical processes.  Lastly, from the purely physical point of 
view, clusters represent dense environments {\it i.e.} deep 
potential wells as well as high concentrations of galaxies, dark matter, and 
intra-cluster gas.   The entire cluster-group population thus provides an 
ideal range of conditions to study environmental effects on the formation of 
galaxies, active nuclei and the triggers of star forming activity.

The number of spectroscopically confirmed clusters beyond $z \sim 0.7$ 
is still quite small; the reason is simple: detection of high-redshift
clusters requires finely-tuned multi-color techniques in the optical 
wave-band because of the high surface density of faint background 
galaxies.  Proposed techniques,
{\it e.g.} adaptive matched-filters \citep{1999ApJ...517...78},
cluster elliptical red-sequence \citep{gladders}, photometric-redshift
classification \citep{1999MNRAS.302..152},
etc, are very model dependent owing to color evolution of the
cluster galaxies and, since they rely primarily on photometric redshifts,
they frequently provide large numbers of cluster candidates which are
simply portions of cosmic filaments viewed in projection. In parallel,
X-ray observations at high galactic latitude are an important tool for
detecting distant clusters because cluster emission, which can only be
produced by hot gas trapped in deep potential wells, is extended and easily
distinguishable from unresolved QSOs, greatly reducing confusion and
projection effects arising in the optical.

In this context, using the unrivalled sensitivity of the XMM-Newton
Observatory, the XMM Large Scale Structure Survey (XMM-LSS) has been
designed to investigate the large scale structure of the Universe out to a
redshift of $\sim$1 as traced by clusters. It will moreover probe the
presence of massive clusters out to z$\sim$2 and enable QSO studies out to
z$\sim$4 with a density of objects some 6 times larger than 2dF.  A
detailed description of the XMM-LSS can be found in \cite{pierre02}, and
preliminary results in \cite{pierre03}. The optical counterpart of the
X-ray survey is being provided by the CFHT Legacy
Survey\footnote{\it http://cdsweb.u-strasbg.fr:2001/Science/CFHLS/} and is
subject to a vigorous spectroscopic follow-up program 
\citep{valtchanov,willis}.

The XMM-LSS depth (z$_{max}\sim$1-2 for clusters, z$\sim$4 for QSOs)
matches that of SWIRE very well.  Consequently, the combination of the
XMM-LSS and SWIRE data set over an area of $\sim$9 sq. deg. will provide
the first coherent study of biasing mechanisms as a function of scale for
X-ray hot (XMM), dark (weak lensing), luminous galaxies (optical, SWIRE)
and obscured (SWIRE) material as well as unique new insights into the
physics of heavily obscured objects.

\subsection{Other Fields of Study with SWIRE}

Besides the extragalactic topics described above, which drove the 
design of the SWIRE survey, there are many other studies 
possible with SWIRE data.   These include nearby resolved galaxies, brown
dwarfs, evolved stars, circumstellar disks, cirrus emission, and asteroids.   
While we cannot address all of those topics here, we give a few highlights
with further details in Section \ref{sciresults}.

There are 4542 2MASS galaxies with diameter $>5\arcsec$ in the 2MASS Extended 
Source Catalog within the SWIRE fields, about 500 of which have
diameter $>20\arcsec$.   Many of these NIR-selected 
galaxies may be detected as extended by IRAC.   

SWIRE will be exceptionally powerful for brown dwarf detection due to its wide
area coverage and sensitivity, potentially able to detect a 5Gyr old, 275K 
brown dwarf at 10pc.   We give estimates of detection statistics in Section
\ref{browndwarfs}. 

SWIRE can also provide samples of new debris disks around young stars, 
unbiased by age or spectral type.   We may expect to detect up to 150 of
them with MIPS.

\section{The SWIRE Survey}

The SWIRE fields are listed in Table \ref{tbl:fields}.  The SWIRE survey
was designed to take maximum advantage of the unique capabilities of SIRTF
to further the study of cosmology and of galaxy formation and evolution; it
is the widest survey that can be made with the SIRTF time available that is
comensurate with robust data quality.  In addition, since it is a {\it
Legacy} survey, an over-riding principle in the design of SWIRE was to
ensure that the data products would be of long-lasting value for an
extremely broad range of scientific investigations, and not just for the
few that the SWIRE team itself would be able to undertake.

The depth, area-coverage and number of fields of the SWIRE survey
were the result of a trade-off analysis between redshift depth, maximum
volume cell size, number of volume cells, number of lines-of-sight required
to minimize cosmic variance, acceptable foreground cirrus noise levels and
total integration time.  An initial consideration was that SWIRE 
complement the already planned deep survey of the Guaranteed Time Observers
(GTO; \citet{faziogto,riekegto}), which covers 6$\times$0.4$\sq\degree$ to
5$\sigma$ photometric sensitivities,  $\sim$2$\mu$Jy at 3.6$\mu$m and
0.11mJy at 24$\mu$m.

A governing factor in the SWIRE design was the availability of SIRTF
astronomical observing templates (AOTs).  For MIPS a natural mode and
survey depth for a large area survey which nicely complements the GTO deep
survey is moderate scan speed with two passes of 4s integrations each.
(see Section \ref{mips} for details). This results in 80 seconds of
integration per point on the sky with a sensitivity at which many models
predict the extragalactic sky to be mildly confusion limited to the SIRTF
70$\mu$m beam (see Table \ref{tbl:sens}).  This then became the depth
yardstick for SWIRE.

The total area coverage was selected such that the survey would probe
several hundred volume cells of scale $\sim$100 Mpc; enough to sample many
different environments within the cosmic web.  The number of lines-of-sight
was a trade-off between cosmic variance considerations (more fields) and
maximizing the physical size-scale probed in each field (fewer, larger
fields), moderated by the availability of large sky areas with acceptable
cirrus noise levels.  The resulting 7 fields project between $\sim$130 and
250 100h$^{-1}$ co-moving Mpc at z=2, with about 50 100Mpc-scale co-moving
radial distance cells along each line of sight to that epoch.

The SWIRE fields and SIRTF prime observation windows are detailed in Table
\ref{tbl:fields}, with secondary windows listed for 3 fields.  The expected
SWIRE 5$\sigma$ photometric sensitivities compared to anticipated cirrus
noise (see below) and confusion limits are shown in Table \ref{tbl:sens}.
Regarding confusion noise, care must be taken in comparing different
predicitions, which depend both on the underlying source count model, the
method used to derive the confusion noise, and the accuracy with which the
PSF, the instrument and the data taking methods are modeled.  A detailed
treatment for MIPS is given by \citet{dole03}, and predictions based
specifically on the SWIRE models, data observation strategy and data
processing methods are presented by \citet{mrr03}.

\begin{deluxetable}{lccccllcc}
\tabletypesize \footnotesize
\tablewidth{0pt} 
\tablecaption{The SWIRE Fields\label{tbl:fields}}
\tablehead{
\colhead{Field} & \multicolumn{2}{c}{Center} & 
  \multicolumn{2}{c}{Area} & 
  \multicolumn{2}{c}{Primary$^1$} & \colhead{P.A.} & 
  \colhead{Background} \\ 
\omit & \multicolumn{2}{c}{(J2000)} & \multicolumn{2}{c}{(square degrees)} &
  \multicolumn{2}{c}{Window} & \omit & \colhead{I(100)$\mu$m} \\
\omit & \colhead{RA} & \colhead{DEC} & \colhead{MIPS} & \colhead{IRAC} & 
  \colhead{Start} & \colhead{End} & \colhead{({\degree}E of N)} & 
  \colhead{(MJy/sr)} \\
\omit & \colhead{h m s} & \colhead{d m s} & & & & & & \\}
\startdata
ELAIS-S1 & 00 38 30 & $-$44 00 00 & 14.32 & 14.26 & Oct 28 & Dec 25 & 293--249 & 0.42 \\
XMM-LSS  & 02 21 00 & $-$05 00 00 & 9.00 & 8.70 & Dec 25 & Feb 5 & 328--343 & 1.3 \\
CDF-S    & 03 32 00 & $-$28 16 00 & 7.14 & 6.58 & Dec 20 & Feb 19 & 303--351 & 0.46 \\
Lockman  & 10 45 00 &  +58 00 00  & 14.32 & 14.26 & Oct 26 & Dec 25 & 224--179 & 0.38 \\
Lonsdale & 14 41 00 & +59 25 00 & 6.70 & 6.69 & Nov 19 & Jul 15 & 259--28 & 0.47 \\
ELAIS-N1 & 16 11 00 & +55 00 00 & 9.00 & 8.70 & Dec 16 & Sep 3 & 255--1 & 0.44 \\
ELAIS-N2 & 12 36 48 & +41 01 45 & 4.45 & 4.01 & Jan 29 & Sep 16 & 216--357 & 0.42 \\
\enddata

{$^1$ Assuming 2003 August Launch. In the event that fields cannot be
scheduled in the primary window, secondary windows are: ELAIS-S1: May 27 --
Jul 26,  PA = 136--179; CDF-S: Jul 23 -- Sep 26,  PA=150--198; Lockman:
Mar 24 -- May 23,  PA = 70--24. }

\end{deluxetable}

\begin{deluxetable}{cccccccccc}
\tabletypesize \footnotesize
\tablewidth{0pt} 
\tablecaption{The Power of SWIRE to Probe LSS\label{tbl:lss}}
\tablehead{
\colhead{z} &
\multicolumn{5}{c}{Co-moving Length Scale} &
\colhead{Co-moving Volume} &
\multicolumn{3}{c}{Maximum number of cells} \\
\omit &
\multicolumn{5}{c}{$h^{-1}$Mpc} &
\colhead{/$10^9 h^{-3}$Mpc$^3$} &
\colhead{$10h^{-1}{\rm Mpc}^3$} &
\colhead{$50h^{-1}{\rm Mpc}^3$} &
\colhead{$100h^{-1}{\rm Mpc}^3$} \\
\omit &
\colhead{6\arcmin} &
\colhead{30\arcmin} &
\colhead{1\degree} &
\colhead{2\degree} &
\colhead{4\degree} &
\colhead{65 square degree} \\}
\startdata
0.1 & 0.51 & 2.6  & 5.1 & 10 &  21 &  0.16 & 168    & 1.34  & 0.16 \\
0.5 & 2.3  &  11  & 23  & 46 &  93 &  15   & 1.54e4 & 123   & 15   \\
1.0 & 4.0  &  20  & 40  & 81 & 162 &  82   & 8.21e4 & 657   & 82   \\
1.5 & 5.3  &  27  & 53  & 107 & 213 & 188   & 1.88e5 & 1.6e3 & 188  \\
2.0 & 6.3  &  32  & 63  & 127 & 253 & 315   & 3.15e5 & 2.5e3 & 315  \\
3.0 & 7.8  &  39  & 78  & 155 & 311 & 581   & 5.81e5 & 4.7e3 & 581  \\
\enddata

{This table
gives the co-moving length scales for given anglar separations at
different redshifts (columns 2-6).  The smallest SWIRE field has an
angular scale of $\sim$2\degree, the largest $\sim$4\degree.  The
total volume of the SWIRE survey at different $z$ is provided in
column 7 and the number cells of a given size that the survey can be
divided into is in columns 8-10.  We have assumed 
$H_0=100h^{-1}$kms$^{-1}$Mpc$^{-1}$, $\Omega_m$=0.3, $\Omega_{\Lambda}$=0.7.}
\end{deluxetable}

\begin{deluxetable}{llllllll}
\tabletypesize \scriptsize
\tablewidth{0pt} 
\tablecaption{Smaller Imaging Surveys Within the Large SWIRE Fields\label{tbl:prevsurv}}
\tablehead{
\colhead{SWIRE} & \colhead{Survey} & \colhead{Instrument} &
  \colhead{RA} & \colhead{Dec} & \colhead{Bands} & \colhead{Size} & \colhead{Depth}$^1$ \\
 & & & h m & d m & & $\sq\degree$ \\}
\startdata
ELAIS-S1 & ELAIS$^a$ & ISO & 0 35 & $-$43 28 & 7,15,90$\mu$m & 4 & 1mJy$^2$ \\
 & & BeppoSAX/MECS$^b$ & 0 35 & $-$4 28 & 2-10 kev & 1.7 & 36 ks \\ 
 & ES1 Radio Survey$^c$ & ATCA & 0 35 & -43 28 & 1.4GHz & 4 & 80$\mu$Jy \\
\tableline
XMM-LSS  & XMM/Moderate Survey$^d$ & XMM & 2 24 & $-$5 & 0.5-10 kev & 2 & 20 ks \\
 & Subaru/XMM-Newton & XMM & 2 18 & $-$5 00 & 0.5-10 kev & 1 & 50\&100ks \\
 & \ \ Deep Survey$^d$ & Subaru & & & R & 1.3 & 28$^3$ \\
 & & SCUBA/SHADES$^f$ & & & 450\&850$\mu$m & 0.25 & 60,3mJy \\
 & & VLA & & & 1.4GHz & $\sim$1 & 12$\mu$Jy \\
 & XMM/LSS VLA Survey$^g$ & VLA & 2 24 & $-$4 30 & 1.4GHz & 5.6 & 4mJy \\
\tableline
CDF-S & CDF-S$^h$ & Chandra & 3 32  & $-$27 48 & 0.5-8kev & 0.1 & 1Ms \\
 & & HST-WFPC$^i$ & & & VI & 17$\sq\arcmin$ & 28.2 \\
 & & HST-ACS$^j$ & & & BViz & 0.25 & V$\sim$28 \\
 & & SIRTF/GOODS$^k$ & & &  3.6-24$\mu$m & 300,50$\sq\arcmin$ & 4,0.02$\mu$Jy$^4$ \\
 & & SIRTF/GTO$^l$ & & & 3.6-160$\mu$m & 0.4 & 0.11mJy$^5$ \\
\tableline
Lockman  & LH-E & ROSAT-HRI$^m$ & 10 53 & 57 29 & 0.5-2 kev & 0.13 & 1.2 Ms \\
 & & ROSAT-PSPC$^m$ & 10 52 & 57 21 & 0.5-2 kev & 0.3 & 200 ks \\ 
 & & XMM$^n$ & 10 53 & 57 29 & 0.5-10 kev & 0.2 & 190 ks \\
 & & Chandra$^o$ & 10 53 & 57 29 & 0.5-8kev & 1.35 & 5 ks \\
 & & ISO$^{p,t}$ & 10 52 & 57 21 & 7,15,90,175$\mu$m & 0.1-0.5 & 3mJy$^2$ \\
 & & VLA$^q$ & 10 52 & 57 29 & 6,20cm & 0.35,0.09 & 11,30$\mu$Jy \\
 & & SIRTF/GTO$^k$ & & & 3.6-160$\mu$m & 0.4 & 0.11mJy$^5$ \\
 & & SCUBA 8mJy$^r$ & 10 52 & 57 22 & 450\&850$\mu$m & 130$\sq\arcmin$ & 8mJy \\
 & & VLA$^s$ & & & 20cm & 100$\sq\arcmin$ & 5$\mu$Jy \\  
 & & SCUBA/SHADES$^f$ & & & 450\&850$\mu$m & 0.25 & 60,3mJy \\
 & LH-W$^t$ & ISO & 10 34 & 58 & 90,175$\mu$m & 0.5 & 15mJy \\
 & & Chandra$^u$ & 10 34 & 57 40 & 0.5-8 kev & 0.4 & 40,70 ks \\
\tableline
ELAIS-N1 &  ELAIS$^a$ & ISO & 16 10 & 15 31 & 15,90,175$\mu$m & 2.6 & 1mJy$^2$ \\ 
 & FIRBACK$^v$ & ISOPHOT & 16 11 & 54 25 & 170$\mu$m & 1.98 & 50 mJy \\
 & ELAIS Deep Xray Survey$^w$ & Chandra & 16 10 & 54 33 & 0.5-8kev & 300$\sq\arcmin$ & 75ks \\
 & & XMM & & & 0.5-10 kev & 0.2 & 150 ks \\
\tableline
ELAIS-N2 &  ELAIS$^a$ & ISO & 16 37 & 41 16 & 7,15,90,175$\mu$m & 2.6 & 1mJy$^2$ \\ 
 & FIRBACK$^v$ & ISOPHOT & 16 36 & 41 05 & 170$\mu$m & 0.96 & 50 mJy \\
 & SCUBA 8mJy$^r$ & SCUBA & 16 37 & 41 02 & 450\&850$\mu$m & 130$\sq\arcmin$ & 8mJy \\
 & & VLA$^s$ & & & 20cm & 100$\sq\arcmin$ & 9$\mu$Jy \\  
 & ELAIS Deep Xray Survey$^w$ & Chandra & 16 36 & 41 01 & 0.5-8kev & 300$\sq\arcmin$ & 75ks \\
 & & XMM & & & 0.5-10 kev & 0.2 & 150ks \\
\enddata

{$^1$ ks for X-ray, limiting mag. for optical/NIR, $\sim$1$\sigma$ 
flux density for IR-radio;
$^2$ 15$\mu$m;
$^3$ R/r/r$'$ band; 
$^4$ 24$\mu$m;  
$^5$ 3.6$\mu$m;  
$^a$\citet{oliver00};
$^b$\citet{alexander01};
$^c$\citet{gruppioni99}; 
$^d${\it http://xmm.vilspa.esa.es/}
$^e$\citet{mizumoto};
$^f$\citet{dunlop}; 
$^g$\citet{cohen03}; 
$^h$\citet{giacconi01,rosati02};
$^i$\citet{schreier01};
$^j${\it http://www.stsci.edu/ftp/science/goods/abstract.html\#};
$^k$\citet{dickinsonpasp};
$^l$\citet{faziogto,riekegto};
$^m$\citet{hasinger98};
$^n$\citet{hasinger01};
$^o$\citet{kenter02};
$^p$\citet{fadda02,rodighiero02};
$^q$\citet{ciliegi03,deruiter97};
$^r$\citet{scott,fox};
$^s$\citet{ivison02};
$^t$\citet{kawara98};
$^u$\citet{yang03};
$^v$\citet{dole01};
$^w$\citet{manners03}}
 
\end{deluxetable}

\begin{deluxetable}{llllllll}
\tabletypesize \footnotesize
\tablewidth{0pt}
\tablecaption{Predicted SWIRE Photometric Sensitivity, Cirrus \& 
Confusion Noise Levels\label{tbl:sens}}
\tablehead{
\omit & \multicolumn{7}{c}{Photometric Sensitivity / Noise Levels} \\
\omit & \colhead{$\mu$Jy} & \colhead{$\mu$Jy} & \colhead{$\mu$Jy} & 
  \colhead{$\mu$Jy} & \colhead{$\mu$Jy} & \colhead{mJy} & \colhead{mJy} \\}
\startdata
IRAC Wavelength ($\mu$m) & 3.6 & 4.5 & 5.8 & 8.0 & & & \\
MIPS Wavelength ($\mu$m) & & & & & 24 & 70 & 160 \\ 
\tableline
Sensitivity 5$\sigma$ & 7.3 & 9.7 & 27.5 & 32.5 & 450 & 2.75 & 17.5 \\
\tableline
Cirrus noise 5$\sigma$ &&&&&& \\
$I_{\rm 100}$=0.5 MJy/sr (IRAS) & 2.8e-05 & 5.4e-05 & 1.3e-04 & 7.4e-04 & 3.1e-04 & 0.11 & 11 \\
$I_{\rm 100}$=1.0 MJy/sr  & 7.8e-05 & 1.5e-04 & 3.6e-04 & 2.1e-03 & 8.8e-04 & 0.32 & 30 \\
$I_{\rm 100}$=2.0 MJy/sr & 2.2e-04 & 4.3e-04 & 1.0e-03 & 5.9e-03 & 2.5e-03 & 0.90 & 84 \\
Assumed SED & 1.3 & 2.0 & 3.4 & 6.4 & 5.8e-02 & 0.49 & 2.6 \\
\tableline
Confusion limits$^1$ &&&&&& \\
Beam FWHM, arcseconds & 2.05 & 2.05 & 2.08 & 2.50 & 5.6 & 16.7 & 35.2 \\
${\Omega}_b$ (10$^{-6}$ sq. deg.) & 0.37 & 0.37 & 0.38 & 0.55 & 2.74 & 24.3 & 108 \\
\citet{xu03} model S3+E2 & 6.2 & 6.2 & 5.4 & 6.7 & 175 & 10.0 & 71 \\
\citet{mrr01} & 1.6  & 1.3 & 1.1 & 4.0 & 190 & 6.3 & 60 \\
\citet{rodighiero02, frans01} & 5.0 & 5.0 &   4.8 & 4.3 & 180 & 6.0 & 45 \\
\citet{lagache03, dole03} & \nodata & \nodata & \nodata & 4.0 & 125 & 6.5 & 56 \\
\enddata

{$^1$  Using a source density criterion of 40 beams per source.  The beam is
assumed to be gaussian with the quoted FWHM values, which were derived from
the pre-launch SIRTF beam profiles convolved with the band-dependant pixel
sizes (MIPS: \citet{dole03}; IRAC: this work).}

\end{deluxetable}

\subsection{Field Selection}\label{fieldselection}

We considered a number of factors in choosing our survey fields:
cirrus contamination; existing or future multi-wavelength coverage;
observability constraints; and avoidance constraints.  A detailed
discussion of these will appear in a future paper \citep{oliver03}. 
Here we discuss each constraint briefly.

Clouds within our own galaxy produces structured infrared emission,
``cirrus'', which can be confused with extra-galactic sources, and
which causes extinction in the soft X-ray and UV bands.   To
estimate the level of this cirrus noise we start from a scaling
relationship derived by \cite{helou}, derived in turn from the
power spectrum analysis of cirrus clouds performed by \cite{gautier}.

\begin{equation}\frac{\sigma_{\rm cir}}{\rm 1mJy}\approx
0.3\left(\frac{\lambda}{\rm 100\mu m}\right)^{2.5} \left(\frac{D}{\rm
m}\right)^{-2.5} \left(\frac{B(\lambda)}{\rm
1MJysr^{-1}}\right)^{1.5}.
\end{equation}    
(we modify this slightly by replacing $\lambda$ with the larger of
$\lambda$ and 5.8$\mu$m).  We have calculated the $5\sigma$ cirrus
noise for regions of 0.5, 1 and 2 MJy/sr and compare these to our
intended survey depth (Table \ref{tbl:sens}).  The cirrus noise is clearly not
important for $\lambda<24\mu$m, but if $I_{\rm 100}$ is above
$2MJy/sr$ the cirrus noise exceeds our depth at 160$\mu$m.

\begin{figure}
\epsfig{file=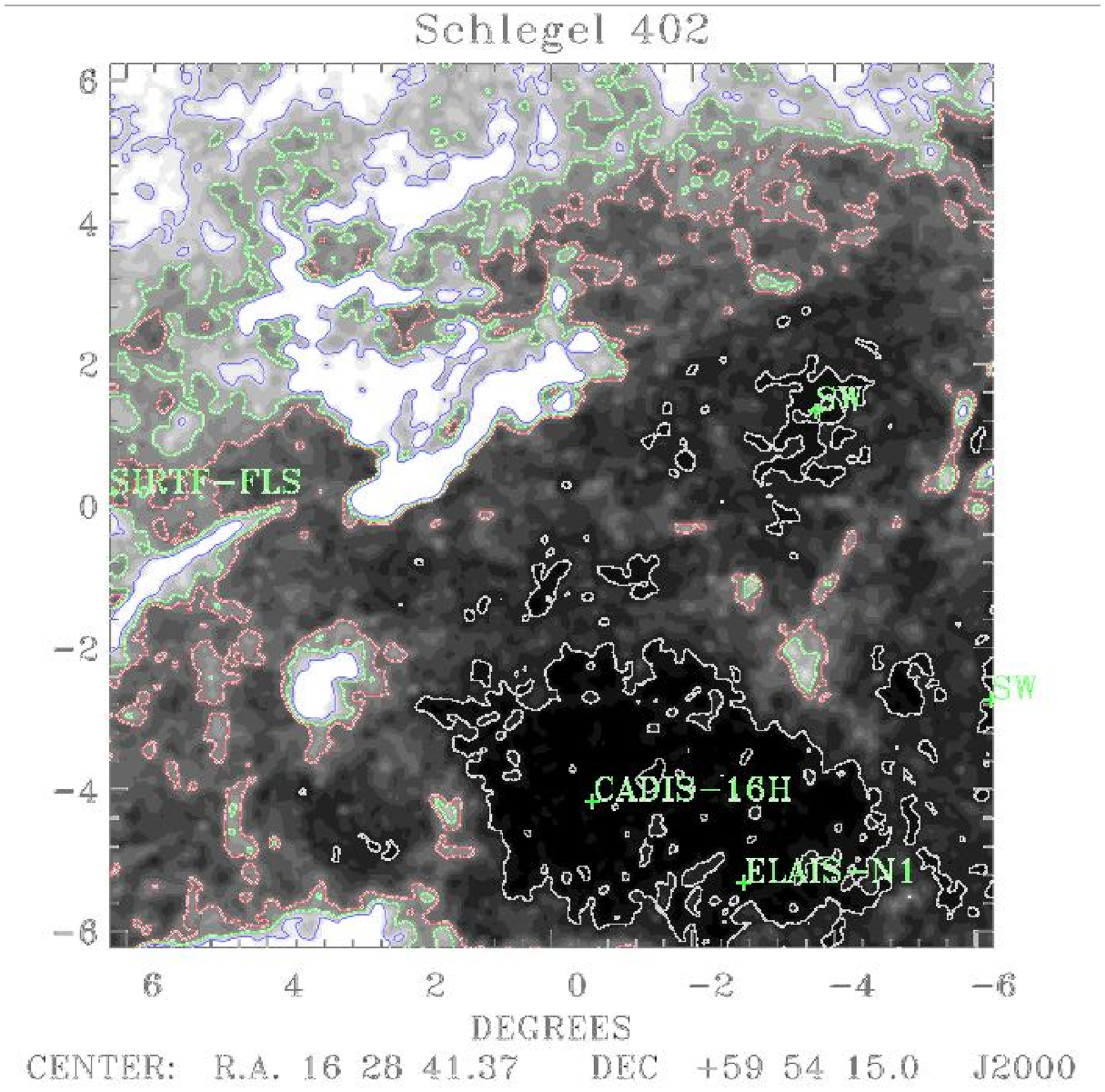, angle=90,width=9cm}
\epsfig{file=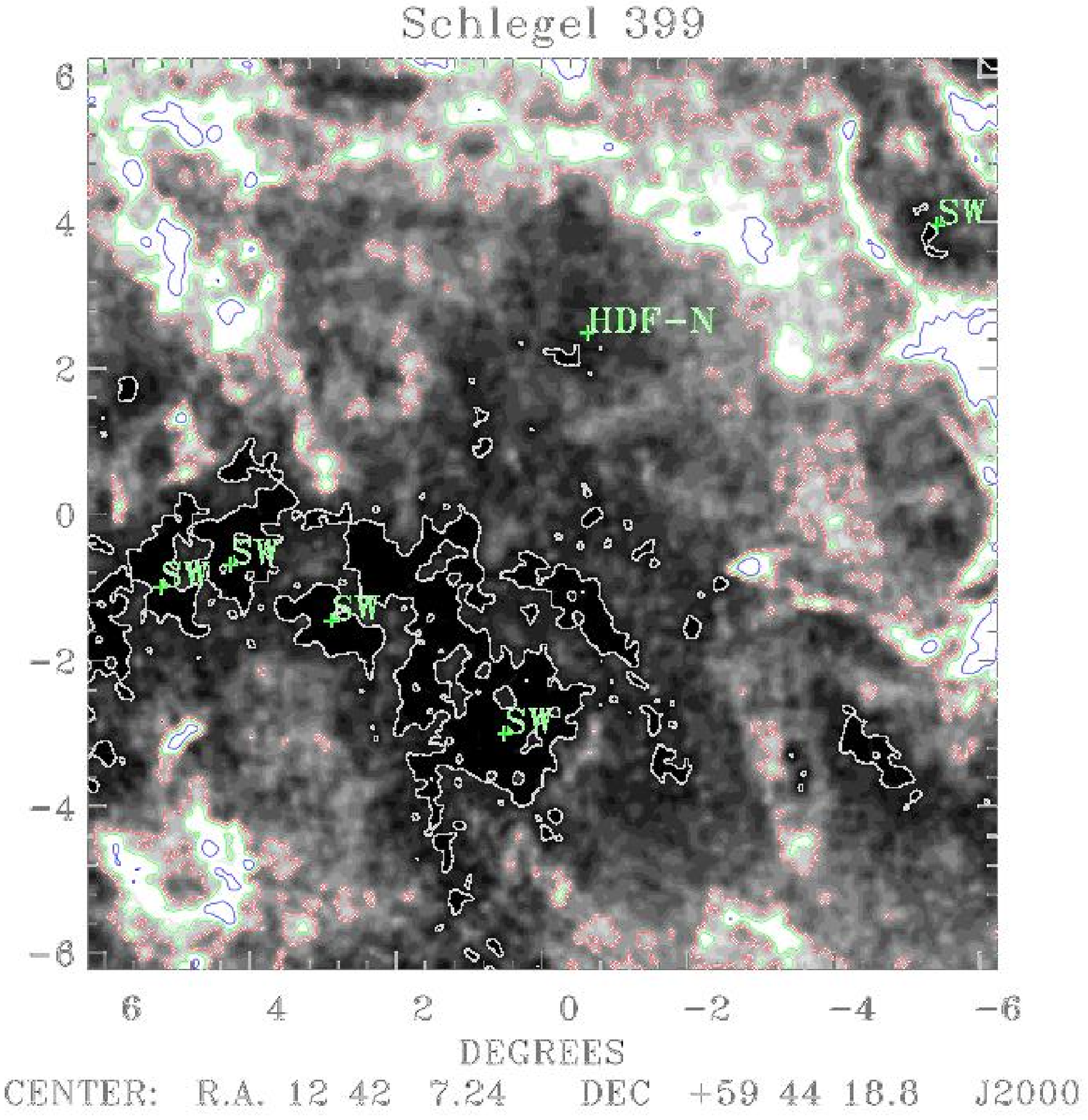, angle=90,width=9cm}
\caption{The emission from galactic cirrus.  The IRAS 100 $\mu$m
intensity map for our SWIRE field ELAIS N1 (left) and for a field we
rejected, the Hubble Deep Field, (right).  Previously identified
survey fields are marked, as are new field centres considered in our
selection process (marked ``SW''). Contours are at 0.5, 1.,1.2, \& 1.5
MJy/sr.  Note that the HDF itself is located in region of relatively
high I100 ($\sim$ 0.7 MJy/sr) and that ``holes'' ($<0.5$ MJy/sr) on
the same map are smaller than the ELAIS ``hole''.  The map boundaries are
ISSA plates 402 and 399 respectively, the map intensity comes from
IRAS data which has been normalised to COBE \protect\citep{schlegel}}
\label{fig:cir}
\end{figure}

When choosing our survey fields we thus did not consider any field
with a cirrus intensity much above 1MJy/sr.  Since the structure of the
cirrus is non-Gaussian and the source confusion limit is highly
uncertain we adopted a more conservative limit of 0.5MJy/sr for most
of our fields.  We performed an exhaustive search over the whole sky
considering all fields that met this latter criteria. The cirrus
map for one of the
fields that we finally selected is contrasted with one that we did not
in Figure \ref{fig:cir}.

As a legacy project we felt it was important to select fields that already
had a wealth of data.  We assembled an extensive list of fields which have
been extensively surveyed at a variety of wavelengths
\citep{hdf,oliver00}. We examined the cirrus in all of these fields
and rejected many where the cirrus was too high or where there were only
small patches of low cirrus.  One field, the XMM-LSS field, exceeded our
cirrus threshold, but the wealth of survey data over a large area (in
particular the XMM data) that this field provided was felt to outweight the
risk of compromising the longer wavelength data.  We only found it
necessary to include one field that did not have extensive multi-wavelength
data; this was a former WIRE field (Lonsdale).

We also considered the visibilty to SIRTF (and many other space
missions) and thus excluded any fields below an ecliptic latitude of
30$^{\circ}$, with the exception of XMM-LSS.

To aid ground based follow-up we included a similar number of fields
in the Northern and Southern Hemispheres and one equatorial field
(XMM-LSS).

Images of all our fields area available on the SWIRE web 
site\footnote{\it http://www.ipac.caltech.edu/SWIRE/}, and 
Table \ref{tbl:prevsurv} details previously observed smaller 
fields that lie within the large SWIRE areas (excluding 
extensive ground-based optical and NIR imaging, which is extensive
and complex).

\subsection{SIRTF Observations}

The SIRTF observations of the SWIRE fields are designed to return data with
high sensitivity and a reasonable number of samples, while still covering
large areas with both IRAC and MIPS.  Since the longest SIRTF Astronomical
Observing Requests (AORs) are limited to a few hours in duration,
our observing strategy requires stitching together dozens of AORs to
map each field.  Visualizations of IRAC and MIPS AORs for the SWIRE ELAIS
S1 field are shown in Figure 3.

\subsubsection{IRAC}\label{IRAC}

Our IRAC mapping strategy will result in four
30-sec exposures nominally covering each point.
We chose the 30 second frame time as the best 
trade between sensitivity at the shorter
wavelengths and obtaining enough overlapping images
for reliable data--particularly because shorter
IRAC exposures in the SWIRE regions are expected to
be limited by read noise.  The four exposures per point
are divided between two coverages separated in time
to allow discrimination against moving or transient
objects.  

Each coverage is made from several overlapping
AORs.  Each IRAC AOR will cover around a square degree.
Within each AOR, the map grid spacing is 280 
arcseconds, and two images are taken at
each grid point.  These exposures are offset slightly
using the small-scale cycling dither in the Astronomical Observing 
Template (AOT).
The AORs in each coverage overlap nominally by 120 arcsec
when rotation is neglected.  

The two coverages are spatially offset by 150 arcseconds
along both map grid axes, to place the center of the
grids of one coverage near the interstices of the
other coverage grid.

Table \ref{tbl:iracaor} lists the sizes of each AOR, both in terms
of grid cells and degrees, and the layout of these AORs
into the IRAC map.  The lower-coverage fringes have not
been included in the angular extents. 

\begin{deluxetable}{lcccc}
\tabletypesize \footnotesize
\tablewidth{0pt} 
\tablecaption{IRAC Field Sizes and AOR Coverages\label{tbl:iracaor}}
\tablehead{
\colhead{Field} & \multicolumn{2}{c}{Each AOR} & 
  \multicolumn{2}{c}{Field layout} \\
\omit & \multicolumn{2}{c}{(grid of images)}  & 
  \multicolumn{2}{c}{(grid of AORs)} \\
\omit & \colhead{cols$\times$rows} & \colhead{deg$\times$deg} & 
\colhead{cols$\times$rows} & \colhead{deg$\times$deg} \\}
\startdata
ELAIS S1  & 12$\times$13 & 0.94$\times$1.02 & 4$\times$4 & 3.62$\times$3.94 \\
XMM-LSS   & 13$\times$13 & 1.02$\times$1.02 & 3$\times$3 & 2.95$\times$2.95 \\
CDF-S     & 10$\times$13 & 0.78$\times$1.02 & 3$\times$3 & 2.23$\times$2.95 \\
Lockman   & 12$\times$13 & 0.94$\times$1.02 & 4$\times$4 & 3.62$\times$3.94 \\
Lonsdale  & 12$\times$11 & 0.94$\times$0.86 & 3$\times$3 & 2.71$\times$2.47 \\
ELAIS N1  & 13$\times$13 & 1.02$\times$1.02 & 3$\times$3 & 2.95$\times$2.95 \\
ELAIS N2  & 11$\times$16 & 0.86$\times$1.25 & 2$\times$2 & 1.65$\times$2.43 \\
\enddata
\end{deluxetable}

\subsubsection{MIPS}\label{mips}

Using the MIPS Scan Map AOT with the medium scanning
rate, each scan leg of each AOR at 70$\mu$m and 24$\mu$m will
yield ten overlapping 4-sec exposures at each point.  The
scan legs are spaced by 276 arcseconds for an overlap of
about 40 arcsec (four 70$\mu$m pixels) between scan legs.

Each sky position is covered by two AORs, separated by
hours as for IRAC to allow detection of moving or transient
sources.  The second coverage is offset in the cross-scan
direction by 150 arcseconds, to provide additional 160$\mu$m 
Ge:Ga array redundancy and counteract the memory effects 
present in the 70$\mu$m Ge:Ga array.

Several AORs are required in each of the two coverages.
The adjacent AORs in each coverage overlap by 70 arcseconds.
This will prevent gaps if adjacent AORs are rotated by less
than about 0.5 degrees relative to each other (corresponding
to about a day at our high ecliptic latitudes).

\begin{figure}
\plotone{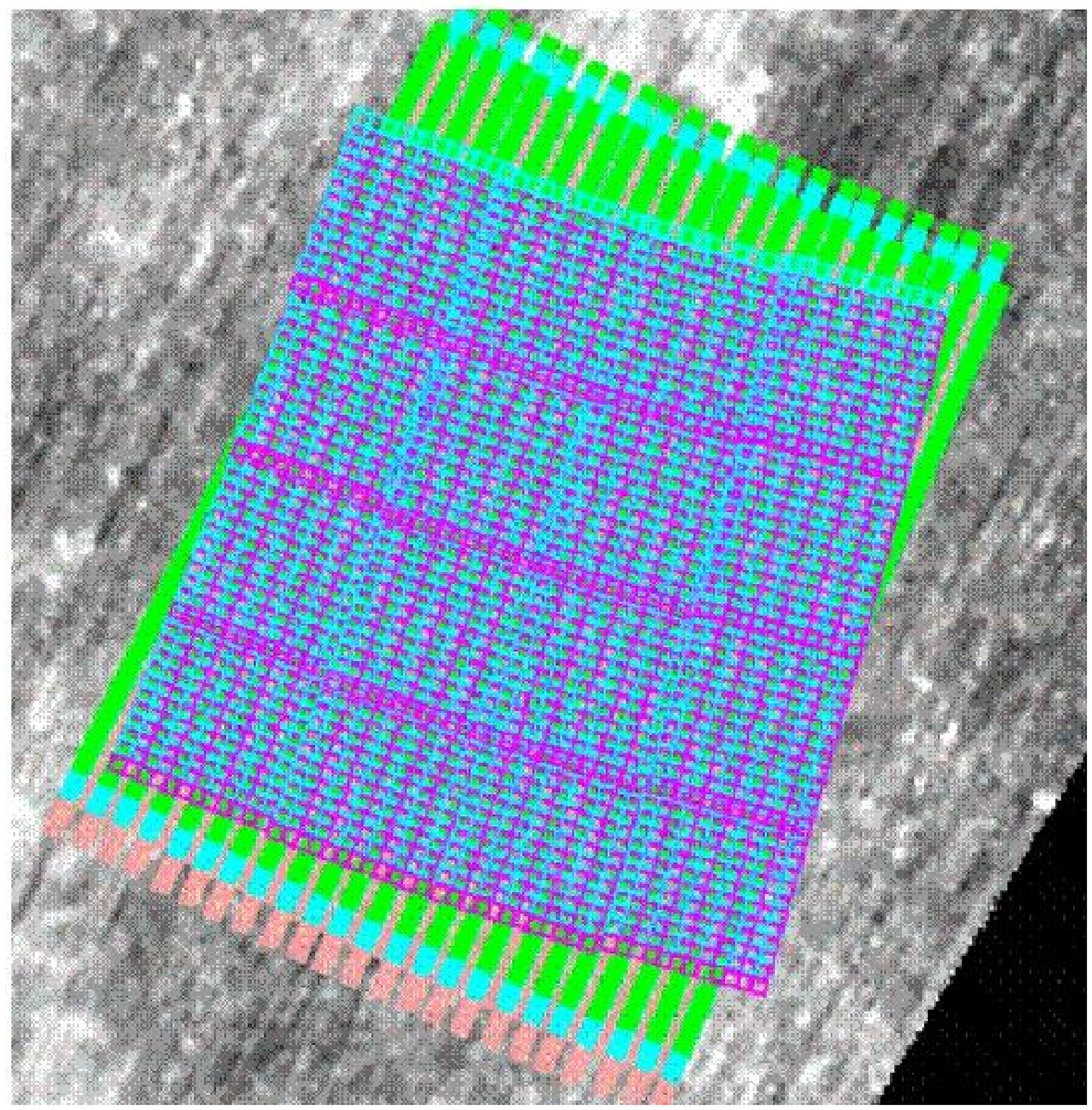}
\caption{Visualization of the MIPS (green)
and IRAC (blue) AORs for the ELAIS S1 field.  The AORs
have been rotated for MIPS observations occuring over
five days, and the IRAC observations spread over ten days.
The map in the background is the IRAS ISSA map at 100 microns.}
\end{figure}

Table \ref{tbl:mipsaor} gives details of the MIPS AORs.  

\begin{deluxetable}{lccccc}
\tabletypesize \footnotesize
\tablewidth{0pt} 
\tablecaption{MIPS Field Sizes and AOR Coverages\label{tbl:mipsaor}}
\tablehead{
\colhead{Field} & \colhead{\#Legs} & \colhead{Cross-scan} & \colhead{\#AORs}
     & \colhead{Cross-scan} & \colhead{Scan length} \\
\omit & \colhead{/AOR} & \colhead{size/AOR} & \colhead{/cvg} & \colhead{extent} & \omit \\
\omit & \omit & \colhead{deg.} & \omit & \colhead{deg.} & \colhead{deg.} \\}
\startdata
ELAIS S1  & 4 & 0.32 & 12 & 3.58 & 4.0 \\
XMM-LSS   & 5 & 0.39 &  8 & 3.00 & 3.0 \\
CDF-S     & 4 & 0.32 &  8 & 2.38 & 3.0 \\
Lockman   & 4 & 0.32 & 12 & 3.58 & 4.0 \\
Lonsdale  & 6 & 0.47 &  6 & 2.68 & 2.5 \\
ELAIS N1  & 5 & 0.39 &  8 & 3.00 & 3.0 \\
ELAIS N2  & 6 & 0.47 &  4 & 1.78 & 2.5 \\
\enddata
\end{deluxetable}

\subsubsection{Other considerations}

Since SIRTF observes with only one instrument at a time, our IRAC
and MIPS observations will be separated by several days, at least.
This will result in a relative rotation of a few degrees overall
between the MIPS and IRAC maps.

In the Lockman and CDF-S fields, the SWIRE observations overlap deeper 
GTO surveys.  Our AORs will be ``segmented'' into smaller 
regions in the areas of overlap, and the resulting small AORs that overlap the
GTO regions will be embargoed during the GTO proprietary period.
Details of the embargoed areas are not yet available because they depend on
the exactly when the SWIRE and GTO observations are made by
SIRTF.  When details
become available, they will be provided on the SWIRE public web pages:
{\it http://www.ipac.caltech.edu/SWIRE/}.

\subsection{Supporting Observations and Data Sets}\label{ancillary}

\subsubsection{Optical \& NIR Imaging}

Ground-based optical/near-infrared imaging data for the SWIRE fields
will be essential to:

\begin{enumerate}
\item  Obtain optical identifications for the roughly 2 million IR sources 
predicted to be detected by SWIRE; the present statistics on faint ISO 
sources in the HDFs indicates that of order of 90\%
of the SWIRE MIPS sources could be detected to $V,R \sim 25$; at the same
limits, and based on K-selected samples in the HDFN, 80\% of the IRAC sources 
may be detected.
                                                                   
\item Provide photometric redshifts for SWIRE sources.  Three-color optical
photometry will be esstential to supplement IRAC photometry for
high-quality photometric redshift estimation (see Section \ref{photoz}).
IRAC colors alone are powerful for stellar populations owing to the H$^-$
opacity feature at 1.6$\mu$m \citep{simpson}, however they suffer at
z$<$1.5 due to the degeneracy of the stellar population with age and
confusion by the 2$\mu$m CO bandheads \citep{sawicki02}.  For star-forming
galaxies MIPS colors are limited for photometric redshifts because there is
great variation in the ${\lambda}>10{\mu}$m SEDs of galaxies, where dust
emission dominates.  However three color optical, coupled with IRAC
(${\lambda}_{rest}<5{\mu}$m) photometry, yields good photometric redshift
discrimination.

\item Provide colors and rough morphologies for source classification and
the study of the effects of environment on morphology.

\item Optimize the discovery of rare sources  -- {\it e.g.} high z, high L, 
\& oddball SEDs -- which require good optical 
imaging for identification and follow-up; SWIRE has sensitivity to objects 
of as rare as 1 -in-10$^6$.
                                                                   
\item Produce independent optical samples for comparison with 
IR-selected samples.
\end{enumerate} 
                                                                
The original goal for optical/near-infrared ground-based imaging in the
SWIRE survey fields was to obtain multi-color optical imaging (SDSS $g'$,
$r'$, $i'$, or equivalent) down to the median optical magnitude ($r' \sim
25$) and galaxy redshift ($z \sim 1$) for the entire SWIRE survey area
with as much complementary near infrared ($J, K'$) data as could be
obtained.  Despite extensive observational facilities available to the
project at NOAO (KPNO/CTIO Mosaic Cameras, 2.1m FLAMINGOS), Palomar (LFC,
WIRC), ESO (2.2m WFI, VLT VIMOS), INT (WFS), etc. as well as existing data
for our ELAIS Survey fields, this goal has proven difficult to meet.
Moderate depth $R/r' \sim 24-25$ data will be available for most of SWIRE's
$\sim$65 $\sq\degree$, with a variety of additional imaging available with 
other
filters and to greater depths.  The available and projected ground-based
imaging data which will become accessible to the astronomical community is 
summarized by field in Table \ref{tbl:gb1}.

\subsubsubsection{NOAO: KPNO \& CTIO Mosaic Imaging}

Time has been granted under the NOAO-SIRTF Observing Program through the
original SWIRE Proposal for optical imaging in CDF-S, Lockman and Lonsdale
using the CTIO/KPNO Mosaic Cameras.  The Mosaic cameras image $0.6^\circ
\times 0.6^\circ$ in a single filter with scale 0.26--0.27\arcsec/px.  Imaging
has been obtained to a range of depths, as summarized in Table \ref{tbl:gb1}.

\subsubsubsection{Palomar LFC}

Further imaging in Lockman and Lonsdale is being obtained at Palomar
Observatory using the 5m Hale telescope and its Large Format Camera 
\citep{simcoe}.   The Palomar LFC has a field of 0.13
sq. deg. with 0.18 \arcsec/px which may be binned to 0.36\arcsec/px for
less-than-optimal seeing.

\subsubsubsection{INT}

A strong motivation for selecting fields from the ELAIS Survey regions was
the availability of ground-based imaging data as well as the 
original ISO Survey observations. In particular, the ELAIS N1 and N2
fields have been imaged through the Isaac Newton Telescope's Wide Field
Survey\footnote{\it http://www.ast.cam.ac.uk/\~wfcsur/index.php} covering
nearly  the entire SWIRE EN1 and EN2 fields.
Details of the overlap between the {\it INT WFS} and {\it SWIRE} will
depend upon SIRTF scheduling.  

Further INT Wide Field Camera observations in $r'$ have been undertaken by
S. Oliver, E. Gonzalez, and M. Salaman at U. Sussex (ISLES program).  The
goal of ISLES is to ensure complete $r' \sim 24$ coverage of the 
SWIRE northern fields, with deeper imaging over smaller
central fields. 

\subsubsubsection{ES1 - ESIS}

In the ELAIS S1 field the {\it ESIS}\footnote{ESO-SIRTF wide-area Imaging 
Survey:
({\it http://dipastro.pd.astro.it/esis/})} program is being carried out by 
A. Franceschini and colleagues at the University of Padova. {\it ESIS} is an 
optical
imaging survey over $\sim 6$ square degrees in 5 bands based on the ESO WFI
2.2m and VIMOS to $\sim 25-26$ mag.  The total amount of observing time
will be 27 nights with (2.2m WFI) and 8 nights with VIMOS.

\subsubsubsection{XMM-LSS}

The full $8^\circ \times 8^\circ$ XMM-Large Scale Structure Survey
area is being imaged by CFHT's MegaCam as part of the Canada-France-Hawaii
Legacy Survey.  Additionally, a large consortium of 
observatories and instruments are supporting further ground-based
observations, including NOAO, Suburu, and ESO.  These data will be available
to the community via CFH Legacy Survey and the XMM-LSS Consortium.  Their
ground-based program is summarized at the XMM-LSS 
website
\footnote{\it http://vela.astro.ulg.ac.be/themes/spatial/xmm/LSS/opt\_fu\_e.html}.

Also covering part of the SWIRE/XMM-LSS field is the NOAO Deep-Wide Survey 
in BRIJHK, reaching 26th magnitude in R and 21.4 in K \citep{jannuzi}.  
Depending on orientation of the SWIRE field, the overlap will be about 
$\sim$2$\sq\degree$.

\begin{deluxetable}{llccccccccc}
\tabletypesize \footnotesize
\tablewidth{0pt} 
\tablecaption{Ground-based Imaging\label{tbl:gb1}}
\tablehead{
\colhead{Field} & \colhead{Telescope/Instrument} &
	\multicolumn{8}{c}{Filters \& Magnitudes (Vega, 5$\sigma$, 3\arcsec)} 
  & \colhead{Area Covered} \\ 

\omit & \omit & \colhead{$u'$} & \colhead{$g'$} & \omit & \colhead{$r'$} 
   & \colhead{$i'$} & \colhead{$z'$} & \colhead{J} 
   & \colhead{K} & \colhead{[Projected]$^{1}$} \\
\omit & \omit & \colhead{U$^*$} & \colhead{B$^*$} & \colhead{V$^*$} & 
   \colhead{R$^*$} & \colhead{I$^*$} & \colhead{Z$^*$} & H$^*$ & 
   \colhead{K$_s$$^*$} & \colhead{(sq deg)} \\}
\startdata           
ELAIS S1 & ESO 2.2m/WFI      & \nodata & 26$^*$ & 25.5$^*$ & 25.5$^*$ &
 \nodata & \nodata & \nodata & \nodata & [6.25] \\
	 & ESO VLT/VIMOS     & \nodata & \nodata & \nodata & \nodata & 
  25$^*$ & 24$^*$ & \nodata  & \nodata  & [6.25] \\
XMM-LSS  & CFHT/Megacam$^2$ & 25.5 & 26.5 & \nodata & 25.7 & 25.5 & 24.0 & 
  \nodata & \nodata & [64] \\
         & Palomar 5m/LFC & \nodata & \nodata & 25 & \nodata &
       \nodata & \nodata & \nodata & \nodata & 0.5 \\
	 & UKIRT/WFCAM$^3$ & \nodata & \nodata & \nodata & \nodata & 
  \nodata & \nodata & 22.5 & 21 & [8.75]  \\
CDF-S    & CTIO 4m/Mosaic II & 24 & 25.7 & \nodata & 25 & 24 & 23.5 & 
  \nodata & \nodata & 1.6 \\
         & CTIO 4m/Mosaic II & 27$^*$ & 27 & \nodata& 26.5 & 25.8 & \nodata & 

  \nodata & \nodata & 0.36 \\
         & Las Campanas/WIRC & \nodata & \nodata & \nodata & \nodata &  
  \nodata & \nodata & \nodata & K$'$=20.5 & 0.6 [1.0] \\
Lockman  & KPNO 4m/Mosaic I  & \nodata & 25.7 & \nodata & 25 & 24 &
  \nodata & \nodata & \nodata  & 2.0  \\
         & KPNO 4m/Mosaic I  & \nodata & 26.7 & \nodata & 26 & 25 & 
  \nodata & \nodata & \nodata  & 0.36 \\
	 & KPNO 2.5m/Flamingos   & \nodata & \nodata & \nodata & \nodata &    
  \nodata & \nodata & \nodata & K$'$=19.5 & 0.09 \\
	 & Palomar 5m/LFC    & \nodata & 25.7 & \nodata & 25 & 24 & 
  \nodata & \nodata & \nodata  & 1.5   \\
	 & INT/WFC (ISLES)   & \nodata & \nodata &\nodata & 23.8 & 
  \nodata & \nodata & \nodata & \nodata & 9.2   \\
	 & Palomar 5m/WIRC   & \nodata & \nodata & \nodata & \nodata &    
  \nodata & \nodata & 21 & 20$^*$ & [1.0] \\
	 & UKIRT/WFCAM$^3$ & \nodata & \nodata & \nodata & \nodata & 
  \nodata & \nodata & 22.5 & 21 & [8.75]  \\
Lonsdale & KPNO 4m/Mosaic I  & \nodata & 25.7 & \nodata & 25 & 24 & 
  \nodata & \nodata & \nodata & 0.7   \\
	 & INT/WFC (ISLES)   & \nodata & \nodata & \nodata & 23.8 & 
  \nodata & \nodata & \nodata & \nodata & 6.5   \\
ELAIS N1 & INT WFS           & 23.3$^*$ & 24.7 & \nodata & 23.8 & 23.0 
  & 21.7$^*$ & \nodata &  \nodata  & 9.0$^4$ \\
	 & UKIRT/WFCAM$^3$ & \nodata & \nodata & \nodata & \nodata & 
  \nodata & \nodata & 22.5 & 21 & [8.75]  \\
ELAIS N2 & INT WFS           & 23.3$^*$ & 24.7 & \nodata & 23.8 & 23.0 & 
  21.7$^*$ & \nodata & \nodata & 4.5$^4$ \\
\enddata

{$^1$ Based on time allocated or programs already approved \\
$^2$ The CFH Legacy Survey, {\it http://www.cfht.hawaii.edu/Science/CFHLS/}
(AB magnitudes).\\
$^3$ UKIDSS \\
$^4$ Overlap between SWIRE area and INT fields depends on field orientation.}
\end{deluxetable}
  
\subsubsection{Near-Infrared Imaging}

FLAMINGOS imaging of the SWIRE Lockman Field in $K'$ using the KPNO 2.1m
telescope was obtained in 2001 Dec and 2002 February.  Poor weather
restricted observations to a total of 8 pointings of 0.09 sq deg
(0.6\arcsec pixels) to a limiting magnitude $K' \sim 19.5$.  Deeper
near-infrared observations in the Lockman Field were carried out in February 
2003 using the Cornell Wide-field InfraRed Camera on the 5m telescope at 
Palomar, by G. Stacey {\it et al.}.  

A survey is being conducted by A. Cimatti in ELAIS-S1 to J=22 and K=20 
over 1$\sq\degree$ using the ESO 3.5m NTT/SOFI.   Approximately  
half of the observations have been carried out with the remainder to be 
completed by the end of 2003.

The 2MASS survey covers the entire SWIRE survey area.  In the SWIRE Lockman
Hole field a unique 2MASS deep survey exists 
\citep{beichman}.  $\sim$ 1 magnitude deeper than the main 2MASS survey, 
the 24$\sq\degree$ area overlaps most of the planned SWIRE Lockman Hole 
field (dependent on the final orientation of the SWIRE/SIRTF observation), 
and includes 69,115 sources to 90-95\% completeness levels of 
17.8, 16.5 \& 16.0 at J, H \& K$_s$.  
  
A UK Consortium --- The UKIRT Infrared Deep Sky Survey\footnote{UKIDSS:
{\it http://www.ukidss.org/}} is planning to observe 8.75 sq deg in each of
three SWIRE Fields as part of their Deep Extragalactic Survey.  Using
UKIRT's Wide Field Infrared Camera (WFCAM; 0.19 sq deg FOV, 0.4\arcsec/px),
UKIDSS will begin by covering approximately 3 sq deg in each of Lockman,
XMM-LSS and EN1 to J=22.5 and K=21 in the first 2 years.  Subsequent
observations will complete the J and K imaging over 8.75 sq deg and will
image a smaller area to H$\sim$22.  WFCAM is due to be commissioned in
late 2003 and the UKIDSS Deep Extragalactic Survey will require 118 nights
over 7 years.

\subsubsection{Optical \& NIR Data Processing}

The bulk of the new imaging data reduction has been or is being carried out at
the Cambridge (UK) Astronomical Survey Unit via their image reduction
pipeline, led by M. Irwin.  The ELAIS N1 and N2 INT data have been fully 
processed at CASU.  KPNO, CTIO and Palomar observations of SWIRE's CDF-S, 
Lockman and Lonsdale Fields are also being carried out through the CASU 
pipeline under supervision of M. Rowan-Robinson and colleagues at Imperial 
College.

ESO observations by the Padova group and INT observations by the Sussex
group are being reduced by the groups themselves.

Source extraction for all optical and \& NIR imaging is being undertaken in 
a uniform manner at IPAC, using SExtractor \citep{bertin}, and following the procedures 
described for SWIRE/IRAC data processing and validation.  Details will be 
provided in a forthcoming publication.

\subsubsection{ISO Mid-IR Surveys in the ELAIS Fields}

Roughly 4 square degrees in each of the ELAIS N1, N2 and S1 fields have been
observed with the Infrared Space Observatory at 15 and 90$\mu$m 
\citep{oliver00}.  \citet{lari01,vaccari03}
report $\sim$2000 catalogued 15$\mu$m sources in these areas.
These samples appear to be complete for fluxes brighter than
$S_{15\mu}\simeq 1.5$mJy and include sources down to $\simeq$0.8mJy. 
Though much shallower
than the planned SWIRE observations, these data may prove useful to complement
the SWIRE photometry between 8 and 24$\mu$m.

\subsubsection{A Deep Radio Survey in the Lockman Hole}\label{deepvla}

We have conducted an ultradeep A/C/D-array VLA imaging survey at 20cm, 
centered at 10h46m, +59d01m, and reaching $\sim$ 3$\mu$Jy 
{\it rms} \citep{owen03}, which matches the deepest ever VLA 
imaging.  This field was picked to be the best place in the entire 
$\sim$65 sq degree SIRTF SWIRE Legacy survey for such a deep survey, with 
respect to elevation, \& brighter radio sources in the primary beam and 
sidelobes.   The goal is to determine if and how the radio-FIR 
relation evolves with redshift (Compton losses might be expected to increase
with z, for example), whether AGNs and star formation are more 
closely connected at higher z, and to identify populations of heavily 
obscured AGN.  The radio image is extremely well matched in depth to SWIRE;
we expect to detect $>$90\% of the SWIRE population and to measure evolution 
of the star formation rate out to z=1-3 with a sensitivity equivalent to 
$\sim$10\msunns/yr at z$\sim$1 and $\sim$200\msunns/yr at z$\sim$3 
(if the local
radio/IR correlation holds).   This radio survey is not in the same location 
as previous ROSAT, ISO, SCUBA, Chandra and XMM surveys in the Lockman Hole area 
because those fields are not as optimal for deep radio imaging (see Table 
\ref{tbl:prevsurv}).

\subsubsection{Broader Radio Imaging}

It would be highly desirable to image the SWIRE area deeply in the radio
but impractical with current VLA capabilities to match SWIRE depth over an
appreciable fraction of the VLA-accessible SWIRE areas, a total of about
43$\sq\degree$.  However a VLA survey with $5''$ FWHM resolution and $40\,
\mu$Jy rms noise at 1.4~GHz could yield a complete catalog of 
$\sim 2.5 \times 10^4$ radio sources stronger than $200\,\mu$Jy.  This is
substantially deeper than the FIRST survey and would detect of order 13\%
of the SWIRE MIPS sources, about 14,600 in total, according to the models
of \citet{xu03}.  Most of these sources would be star-forming galaxies
obeying the remarkably tight FIR/radio correlation, and the sample would
also include many radio loud AGN.  The FIRST survey itself overlaps
$\sim$34$\sq\degree$ of the SWIRE area (Lockman, Lonsdale, ELAIS-N1 and
ELAIS N2), and might be expected to detect $\sim$850 SWIRE sources.  For
comparison \citet{ivezic} estimate about 1350 SDSS-FIRST sources within the
boundaries of the SWIRE fields, about 200 of which are radio-loud quasars.

\citet{gruppioni99} have observed 4$\sq\degree$ of the ELAIS S1 field with
the Australia Telescope Compact Array at 1.4GHz, to a sensitivity of 
$\sim$80$\mu$Jy and with a resolution of 8$\times$15 arcsec, detecting about
600 sources.

\citet{cohen03} present results of a VLA A-array 74MHz survey of the
entire XMM/LSS field, including the subset to be observed by SWIRE,
with a flux limits of 275mJy/beam, resolution of 30$''$, and a source 
density of 2/$\sq\degree$.  They also present 325MHz VLA A-array imaging of 
5.6$\sq\degree$ to 4mJy/beam with a resolution of 6.3$\arcsec$, and 
a source density of 46/$\sq\degree$ (see Table \ref{tbl:prevsurv}).   

\subsubsection{X-ray Imaging}

The XMM-LSS survey will have a sensitivity of $\sim$3${\times}10^{-13}  erg/s/cm^2$ 
for point sources, and ${\sim}10^{-14} erg/s/cm^2$ for extended sources,
in the [0.5-2] keV band.  The SWIRE area will be covered by $\sim$9$\times$9 
10ks XMM/EPIC pointings separated by 20$\arcmin$.  Also within the
SWIRE/XMM-LSS field are (a) the Subaru/XMM-Newton Deep Survey 
(see Table \ref{tbl:prevsurv}), which covers 1$\sq\deg$ with 7$\times$50ks 
pointings
and one 100ks pointing \citep{mizumoto}, and (b) 
most of two XMM medium depth surveys (P.I.s K. Mason and M. Watson), which 
cover a total 
of $\sim$2$\sq\degree$ with a mean exposure time of 20ks.

A mosaic of 4 deep (100 ksec) integrations with XMM-Newton (P.I. Fabrizio
Fiore) has been approved in AO2 to cover 1$\sq\degree$ in ELAIS-S1.
With this survey, flux limits of 2$\times$10$^{-15}$, 3$\times$10$^{-15}$, 
and 3$\times$10$^{-16}$ $erg/cm^2/s$ will
be reached in the 2-10, 5-10, and 0.5-2 keV bands, respectively.

\section{Sky Simulations}

Simulations are used extensively by SWIRE, both for predicting the
SWIRE source populations and thereby constraining the cosmological 
models once the data are available, and to validate the SWIRE pipeline 
data reduction and source extraction process, which is
complex and involves numerous non-linear components. Understanding the
behavior of this pipeline can only be accomplished by generating simulated
datasets with known inputs, and examining the resulting output. Metrics
examined include the completeness and reliability of extracted sources,
accuracy of bandmerging of multiple wavelengths, effective beamshape, and
accuracy of derived positions.

The sky simulations have three ingredients:  model source populations, 
truth images reflecting the expected band-dependent instrumental PSFs,
and full image reconstructions taking into account the observing
parameters of the SWIRE survey and the expected instrumental effects.
We describe each step in turn below.

The SWIRE sky simulator can generate images with a range of tunable 
parameters.   Here we describe a simulation which is 
tuned to a specific $\sim$0.5$\times$0.5$\degree$ SWIRE field in the 
Lockman Hole.  It has three model components, galaxies, stars and cirrus, 
which are described in turn below  The model star and galaxy source 
populations were derived
from the actual positions of real stars and galaxies detected within this
field in an $r'_{lim}{\sim}$26.7 image.  Source extractions were done using
SExtractor, and stars were separated from galaxies using a stellarity
index (a measure of difference in FWHM from the PSF) of $>$0.7.   A model 
star or galaxy was matched to each real
source by $r'$ magnitude, and model sources predicted to have an r$'$ 
magnitude too faint to be visible on the image
were assigned a random position within the image.

\subsection{Source Populations}

Both dusty galaxies and E/S0s, the latter to be detected presumably only in
the IRAC bands, have been simulated using the models of
\citet{xu03}. Models S1, S2 \& S3 for dusty galaxies exploit a large
library of Spectral Energy Distributions (SEDs) of 837 local IR galaxies
(IRAS 25$\mu m$ selected) from the UV (700{\AA}) to the radio (20cm),
including ISO-measured 3--13$\mu m$ unidentified broad-band features
(UIBs).  The basic assumption is that the local correlation between SEDs
and Mid-Infrared (MIR) luminosities can be applied to earlier epochs of the
Universe. By attaching an SED appropriately drawn from the SED library to
every source predicted according to the evolved luminosity functions, the
algorithm enables the comparisons with surveys in a wide range of wavebands
simultaneously.  Therefore the model parameters are tightly constrained by
available surveys in the literature. Three populations of dusty galaxies
are considered in the model: (1) normal late-type galaxies, (2) starburst
galaxies, and (3) AGNs, each with a different evolutionary function.  In
models S1 \& S3 most of the increase in IR luminosity density to early
times is attributed to starbursts, while in S2 a large fraction is
attributed instead to quiescent star formation in galaxy disks, or
``cirrus'' emission.  The predictions of these models for the global star
formation rate as a function of redshift are compared with observations in
Figure 9 of \citet{xu03}.  All models adequately fit all available optical,
near- mid- \& far-IR, submm and radio number counts, redshift distributions
and the CIB, though each has its shortcomings \citep{xu03}.  We have chosen
model S3 as the best overall representation of the IR Universe for most of
our predictions discussed in Section \ref{sciresults}, but we used model
S1, which has the highest FIR count rates of the three models, for
constructing the simulated images in order to provide a conservative upper
limit to the confusion noise we will likely encounter on the sky.

Model E2, for E/S0 galaxies, follows a simple passive evolution approach.
The basic assumption is that there has been no star formation in an E/S0
galaxy since its initial formation.  Consequently, its radiation in
different bands (i.e.  the SED and the L/M ratio) evolves passively with
the ever-aging stellar population.  Instead of assuming that all E/S0's
formed at once together (as in the classical monolithic galaxy formation
scenario), the E/S0 galaxies are assumed to form in a broad redshift
range. The dependence of the formation rate of E/S0 galaxies to the cosmic
time is assumed to be a truncated Gaussian function, specified by a peak
formation redshift $z_{peak}=2$, an e-folding formation time scale
$\omega=2$ Gyr, and a starting redshift $z_0=7$.  SEDs of E/S0s of
different ages are calculated using the code GRASIL \citep{grasil}.  Note
that this model is intended to roughly approximate the hierarchical merging
of systems in a CDM prescription but should not be considered a full CDM
treatment; an independent modeling effort addresses SWIRE predictions in a
CDM-based scenario \citep{fang03}.

In Figure \ref{fig:zdist} predicted redshift distributions in 6 SWIRE/SIRTF
bands models are presented.  Color-color distributions are shown in Figure
\ref{fig:mipscol}.  For these figures we use the SWIRE 5$\sigma$
photometric sensitivity limits for 3.6-24$\mu$m, and the confusion limits
of the \citet{mrr01} model at 70 and 160$\mu$m (see Table \ref{tbl:sens}).

The stars in the SWIRE simulation are based on point sources
(stellarity$>$0.7) identified in the Lockman $r'$-band image. The
distribution of spectral types and luminosities for the simulated stars was
generated using a 2MASS stellar population model developed by T. Jarrett
\citep{jarrett94,cambresy02}. Once assignments of spectral class were made,
the K-band brightness of the star was calculated using the $r'$-K {\it vs.} 
spectral and luminosity class from the model.  Finally, stellar fluxes in
the IRAC and MIPS filters were predicted using a blackbody extrapolation
from K-band to the mid- and-far-IR bands.

To properly assess the impact of cirrus on our observations we have
simulated the cirrus in each of the SWIRE fields.  To do this we add
synthetic cirrus at a higher resolution to the real low-resolution
cirrus maps that we have from IRAS. The synthetic cirrus is cloned
from real low-resolution maps of a larger area of sky and ``shrunk''
to our field size and resolution requirement.  The cloning,
shrinking and adding the synthetic data to the real data 
is done to produce a map whose power spectrum matches an extrapolation
of the real power spectrum.  This method is described in more detail
by \cite{oliver03}.

\subsection{Truth \& Simulated Images}

The truth-image generator uses the instrument point spread function (PSF),
provided by the SIRTF Science Center, to place sources from a model, such
as that described above, onto a ``truth'' image.  The PSFs are theoretical
PSFs based on the measured, as-built SIRTF optical train, including all
expected optical aberrations, and are expected to be very similar to the
in-flight performance. They are quantized with pixels typically 4-8x
smaller than the actual instrument pixels and extend to many 10s of
beamwidths. There is no time-dependence, but unlike HST (which has
sun-induced "breathing" modes) there is no a priori expected time variance
to the SIRTF PSF.  The truth-image generator can create random positions
when they are missing in the source list, as is the case for model sources
faiter than the r$'$ limit of the Lockman field image.  For galaxies
correlated positions can also be made based on a predicted 2-point angular
correlation function.

\begin{figure}
\plotone{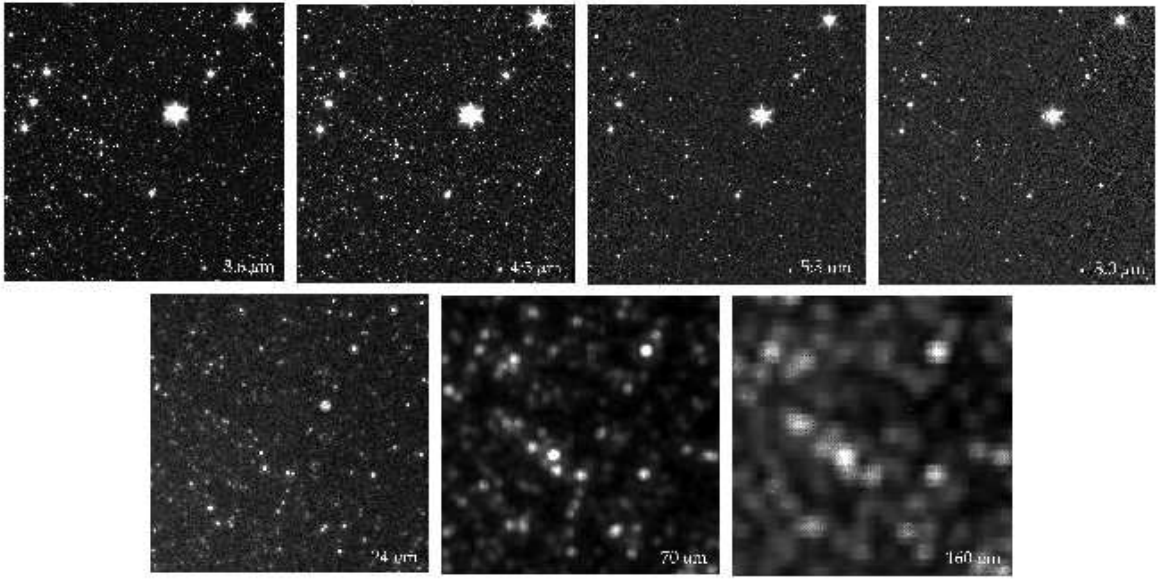}
\caption{Simulated full depth SWIRE images, 10\arcmin\ on a side, in each of 
the 7 SIRTF bands.}
\label{fig:7band}
\end{figure}

\begin{figure}
\plotone{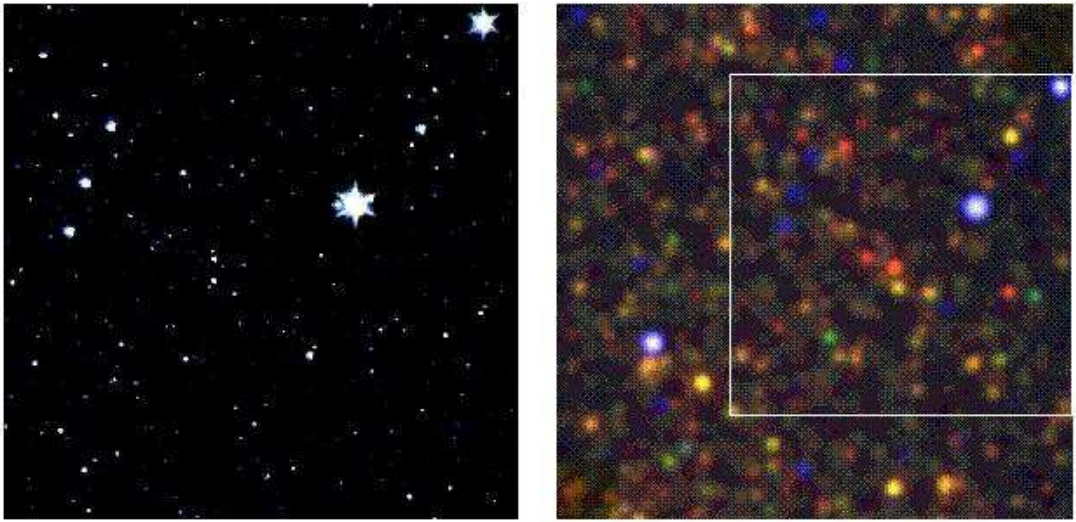}
\caption{Pseudo-truecolor images constructed from simulated IRAC data at
3.6, 4.5, and 5.8\micronns, 10\arcmin\ on a side (left) and IRAC/MIPS data at 
8, 24, 70\micronns, 15\arcmin\ on a side, with the coverage of the IRAC 
3-color image outlined (right).  The
images are made from simulated detector data
using the actual SWIRE observation parameters, and then mosaicked with
the SWIRE pipeline using Montage (see text).  The data have been smoothed to 
the spatial resolution
of the longest wavelength for the purpose of coaddition.} 
\label{fig:simimages}  
\end{figure}

In the simulation of the Lockman field illustrated in Figures 
\ref{fig:7band} and \ref{fig:simimages}, the optical detections of stars 
and galaxies in a 0.25 square deg area were matched in r-band flux with
model sources to yield 36,271 galaxies and 1356 stars with positions and
IRAC and MIPS fluxes.  The truth-image generator also generated 292,746
random positions for fainter ``un-matched'' galaxies from the model.  The
truth image pixels are upsampled by a factor of 4 in each dimension, giving
pixel sizes of 0.30$\arcsec$ for IRAC, and 0.64$\arcsec$, 2.46$\arcsec$,
and 4.00$\arcsec$ for MIPS 24, 70, and 160$\mu$m truth images,
respectively.  The instrument PSF was further upsampled so that a source is
placeable to 1/8 of the truth image pixel, or 1/32 of the instrument pixel
in both dimensions.  For computing efficiency the central 1/8 of the PSF
image is used for sources fainter than 1mJy.  The separate cirrus frames
are then combined to yield the final truth images.  The total number of
model sources in the simulation that we would expect to detect in one or
more SWIRE bands is 24,500 (5$\sigma$ photometric sensitivity at
3.6-24$\mu$m; confusion limits at 70 and 160$\mu$m).

The IRAC simulated images, calibrated in units of surface brightness, were
generated from the truth images using the {\it IRAC Science Data Simulator}
\citep{ashby}.  The command files produced by SPOT, the SIRTF observation
planning tool, were used to produce a simulation of the IRAC data-taking
process, including simulated slews using the commanded pointing and all
detector effects such as the flat-field response, non-linearity, etc.
Generally, most simulations are generated to mimic the form of the reduced,
calibrated data that will be provided by the SIRTF Science Center.  In this
form the simulations are individual IRAC images with the expected
sensitivity and noise properties of IRAC.

The MIPS simulations were made with a modified version of the
WIRE science image simulator \citep{shupe96}.
Owing to the complex behavior of the Ge detectors and the onboard
data-taking procedures, our simulation efforts have focused
on producing the basic calibrated images, in units of surface
brightness.  Read noise and photon noise are simulated, and the
simulated pixels are scanned across the truth images in the
manner planned for MIPS scan maps.

\section{Data Processing and Products}

\subsection{Data Processing}

SWIRE data processing consists primarily of five steps: organizing and
tracking the data, mosaicing/coaddition, source extraction, bandmerging,
and catalog
building.  The SWIRE data system is designed under a Sun UNIX
environment.  As much as possible, existing off-the-shelf
software has been used to reduce development time and costs
and take advantage of well-understood software properties 
and community support.  Architecturally, the
software consists of individual modules executable from a UNIX command
line and connected via PERL wrapper scripts.  Additional PERL modules
perform various housekeeping functions such as file transfer and
reformatting.  Commercial database software administrated by the
Infrared Science Archive (IRSA) at IPAC is used for tracking each
individual observation.
Furthermore, the generated data products such as source catalogs are
loaded into a database which is accessible from the IRSA data system,
allowing use of their advanced search and data mining tools.
A web-based interface is used for most functions that require
interaction with the SWIRE science team.

After receipt from the Deep Space Network, the SIRTF Science Center
(SSC) performs an initial processing of the data.  Automated pipelines
remove nearly all known instrumental signatures from the data.  These
include but are not limited to: dark current subtraction,
flat-fielding, bias removal, cosmic ray detection, and image latent
tracking.  Additionally, the SSC performs basic calibration tasks,
notably flux calibration (in units of surface brightness) and pointing
reconstruction.  The result of this processing is known as ``Basic
Calibrated Data'', or BCD. It is with this data that the SWIRE
pipelines normally begin their processing.

Data (both raw instrumental data and calibrated data) are initially
received in the SWIRE data system in a tar bundle sent by the SIRTF Science
Center.  A daemon process running within the data system automatically
unpacks this data upon its arrival and stores it in a terabyte disk array.
The data is registered by the daemon into a science operations database
(SODB), which stores its location within the system, its header contents,
its current processing status, and various statistics about the pixel
contents.  If needed, frame-level reprocessing of the raw data into
calibrated data, using the SSC pipeline within the SWIRE system, is
performed.  This reprocessing is likely to be needed during the early
mission when the instrument calibration is changing rapidly.  Later in the
mission when the calibration and data reduction have stabilized this will
become unnecessary.  Web pages are automatically generated which present
thumbnail images and statistics for every image received, along with direct
access to the original images and on-line analysis tools.  Team members are
then contacted via an automated system to perform quality analysis checking
from their remote institutions.  The team members connect to the SWIRE
server, examine the data, and submit an evaluation of the data quality.
Data quality is assessed by both qualitative and quantitative guidelines. A
set of basic quantitative statistics are used to provide a lower level of
basic data rejection. In addition to these quantitative pass/fail criteria,
all the data is examined by eye for unknown or unanticipated defects (for
example, unexpected time-variability). When such a defect is found it is
handled on a case-by-case basis.  When sufficient data graded of high
quality has accumulated for a given SWIRE field, SWIRE pipeline processing
is initiated.

The individual frames are then processed in groups, with corrections
made to individual frames based on the behavior of the data group.
This allows for correction of time-dependent effects, such as drifts in
the background bias level, which is a known effect in IRAC data.
This is also expected to be a major step in the processing of MIPS
data, as the MIPS germanium detectors exhibit numerous transient
effects. In essence, this step will result in destriping of the data
and elimination of large-scale detector-based variability.

At this point the data are still in the form of individual images, one per
telescope pointing.  The data will be reprojected onto a common TAN-TAN
spatial projection (one per each of the seven SWIRE fields).  At the same
time detector distortion is corrected.  Reprojection is done via the
Montage software\footnote{Montage ({\it http://montage.ipac.caltech.edu})
is an image mosaic service under development by the Center for Advanced
Computing Research, California Institute of Technology, the Infrared
Processing and Analysis Center, California Institute of Technology and the
Jet Propulsion Laboratory.  The images presented here were derived with an
evaluation version of Montage.}.

Pre-defined tiles one-half degree on a side are then generated by coadding
all of the reprojected data lying within the specified tile.  Outlier
detection (e.g. cosmic rays) is performed by examining the contribution
from each input pixel to a given output pixel and identifying high sigma
outliers.  The data are then coadded using Montage onto a reprojected grid
finer than the original data scale to minimize aliasing with the large
detector pixels.  Coverage maps for 1 sq. deg. IRAC and MIPS mosaics are
shown in Figure 6.

After coaddition, sources are detected in the coadded and mosaicked
tiles.  Source densities are expected to be of the order 100
sources per 5 arcminutes frame in our densest IRAC filter (3.6
\micronns ).  This source density is similar to or less than the
confusion limit.  For this data we are using the SExtractor software
written by Emmanuel Bertin.  This reads the coadded data and weight
images generated during the mosaicing process, detects sources, and
performs aperture and isophotal photometry.  For objects identified as
point sources, extracted circular aperture photometry is used, to
which is applied an aperture correction based on the accurately known
point response function for the instrument.  For sources identified as
extended, isophotal magnitudes are used.  IRAC data is expected to have a 
spatial resolution of
1--2\arcsecns , sufficient that many SWIRE galaxies will be slightly
extended.  In the case of MIPS, the
large SIRTF beam ($\approx$45\arcsec at 160\micronns ) ensures that
all extragalactic targets will be point sources.  Additionally,
because of the large beam and SIRTF's great sensitivity, nearly all of
the MIPS data will suffer significantly from source confusion.  In
this case point source fitting is a more optimal extraction approach
than aperture photometry, and so either DAOPHOT and the SSC's APEX
extraction software are under investigation.

Moving targets will be identified by multi-epoch
iteration of the SWIRE coaddition and source extraction pipeline.  The
data itself are taken in multiple passes separated by a time period
optimized for asteroid detection.  A post-processor will sort out all
non-repeating targets by location, and then locate probable matches
based on magnitudes and assumptions about target velocities. Lists
will be compiled both of non-repeating (transient) targets as well as
moving targets.

The resulting lists of source fluxes and positions are then bandmerged
into catalogs which consist of a single source position and fluxes (or
upper limits) in each of the SIRTF bands, based on source
positions and their associated uncertainties, using the SSC pipeline 
bandmerge module, which is based on that by the 2MASS project.  For all 
sources in a
given wavelength source list a most likely counterpart is identified in
adjacent wavelengths.  Chains of associations are then generated and 
resolved.  Bandmerging of
IRAC data is expected to be relatively straightforward due to the accurate 
SIRTF
positions and the relatively small IRAC beam.  Bandmerging of MIPS data
will be more problematic because this data suffers from a large
beamsize, resulting in significant source confusion, and also spans a
much larger range in wavelength and therefore a wider range in possible
spectral energy distributions.  The process is being optimized using extensive
simulations with a wide range of known SED shapes.  Initially, only the 
data taken with a
given instrument will be bandmerged, with bandmerging between the
instruments to follow later.   We are also investigating more sophisticated 
bandmerging techniques based on the coaddition of multi-wavelength images 
({\it eg.} \citet{szalay99}).

Details of this data processing, along with expected performance
including an analysis of the completeness and reliability of extracted
sources, will follow in an additional paper.

\subsection{Data Products and Delivery Schedule}\label{delivery}

The SWIRE SIRTF data will be archived and served to the community by
the SIRTF Science Center, and the SWIRE ancillary data will be archived and
served to the community by the InfraRed Science Archive, IRSA, at IPAC.
Access to the two different data sets via these different archives
will be seamless from the perspective of the user.   In addition to the 
ancillary data described in Section \ref{ancillary}, the SWIRE/IRSA data 
archive  will provide cross-links to data in the major large area catalogs 
and data holdings, including 2MASS, ROSAT, SDSS, FIRST, NVSS, NED \& 
SIMBAD.  

The SIRTF data products will include both processed images, subdivided into
0.5$\sq\degree$ tiles, and source catalogs.  Anticipated image data volumes
range from 3.3 Gbytes for ELAIS N2 to 10.6 Gbytes for Lockman and ELAIS S1,
including 4 images per tile: flux image, noise image, artifact map and
coverage map.  The predicted catalog sizes range from about 130,000 sources
in ELAIS N2, to as many 450,000 sources in Lockman \& ELAIS S1.  A
prototype catalog data record is shown in Table \ref{tbl:catalog} (subject
to change).

\begin{deluxetable}{lll}
\tabletypesize \footnotesize
\tablewidth{0pt} 
\tablecaption{Prototype SIRTF Catalog Source Entry, Minimum Column 
Set\label{tbl:catalog}}
\tablehead{
\colhead{Parameter} & \colhead{\# Columns} & \colhead{Description} \\}
\startdata
RA, Dec \& uncs. & 4 & final bandmerged position \\
flux (PSF-fit) \& uncs, or limit & 14 & one per band \\
flux (aperture) \& uncs, or limit & 14 & one per band \\
SNR & 7 & one per band \\
source reliability & 7 & one per band \\
bandmerge flags & 6 & one per neighboring band pair \\
confusion flags & 7 & one per band \\
cross-id links & $\sim$10 & one per catalog \\
extent flag & 7 & one per band \\
major axis & 1 & from multi-band coadd  \\
minor axis & 1 & from multi-band coadd  \\

position angle & 1 & from multi-band coadd \\
SED fit; best model SED & 1 & one per source \\
photometric redshift \& unc. & 2 & one per source \\
coverage depth & 7 & one per band \\
Total & 96 & \\
\enddata
\end{deluxetable}

Documentation describing the SWIRE data products will include
an Ap. J Supplement-style paper describing the methods and results in full,
metadata released with the data products, and archive 
help files for each catalog and each parameter.

\begin{deluxetable}{ll}
\tabletypesize \footnotesize
\tablewidth{0pt} 
\tablecaption{Data Release Schedule$^1$ \label{tbl:datarelease}}
\tablehead{
\colhead{Product} & \colhead{Date} \\ 
}
\startdata
V1.0 Sky simulations & April 2003 \\
\tableline
V1.0 Products: SIRTF image tiles \& catalog; ancillary \& pre-exisiting 
images \& catalogs; XIDs $^2$ & March 2004 \\
V 2.0 Sky simulations & \\
\tableline
V1.0 Data analysis tools & Sept 2004 \\
V1.0 Products: SIRTF image tiles \& catalog; ancillary \& pre-exisiting 
images \& catalogs; XIDs $^3$ & \\
V2.0 Products: SIRTF image tiles \& catalog; ancillary \& pre-exisiting 
images \& catalogs; XIDs $^2$ & \\
\tableline
V3.0 Products: SIRTF image tiles \& catalog; ancillary 
images \& catalogs; XIDs & March 2005 \\

V1.0 Bandmerged catalog, IRAC-only and MIPS-only & \\ 

\tableline
V4.0 Products: SIRTF image tiles \& catalog; ancillary 
images \& catalogs; XIDs & Sept 2005 \\
V2.0 Bandmerged catalog, including IRAC-MIPS & \\
V1.0 Lower SNR source release with validation analysis \& flags & \\
V1.0 Moving object catalog, pending resources & \\
V1.0 Selection function (coverage maps convolved with final noise maps) & \\
V3.0 Sky simulations & \\
\tableline
V5.0 Data products: final astrometry, bandmerging, confusion processing
 using cross-band 
information & Sept 2006 \\
V5.0 XIDs between all V5.0 data products & \\
V4.0 Science simulations for fully developed science models & \\
\enddata

{$^1$ Subject to revision due to launch date or unanticipated on-orbit 
variances.
$^2$ SWIRE fields delivered to team by $\sim$Oct 2003.   Best efforts only 
on MIPS Ge.  
$^3$ Remaining SWIRE fields.}

\end{deluxetable}

The SWIRE data product delivery schedule is detailed in Table 
\ref{tbl:datarelease}.  The 
releases occur at bi-yearly intervals, beginning about 11 months after SIRTF
launch.  Successive deliveries will steadily expand the products with respect
to  increased area coverage, refinement of data processing techniques and 
improvement of the ensuing data products, decreasing SNR levels while 
maintaining high levels of completeness and reliability of the source 
extractions (C\&R), validation to increasing levels of accuracy, and
cross-matching to increasing numbers of bands and other catalogs.

\begin{figure}
\plottwo{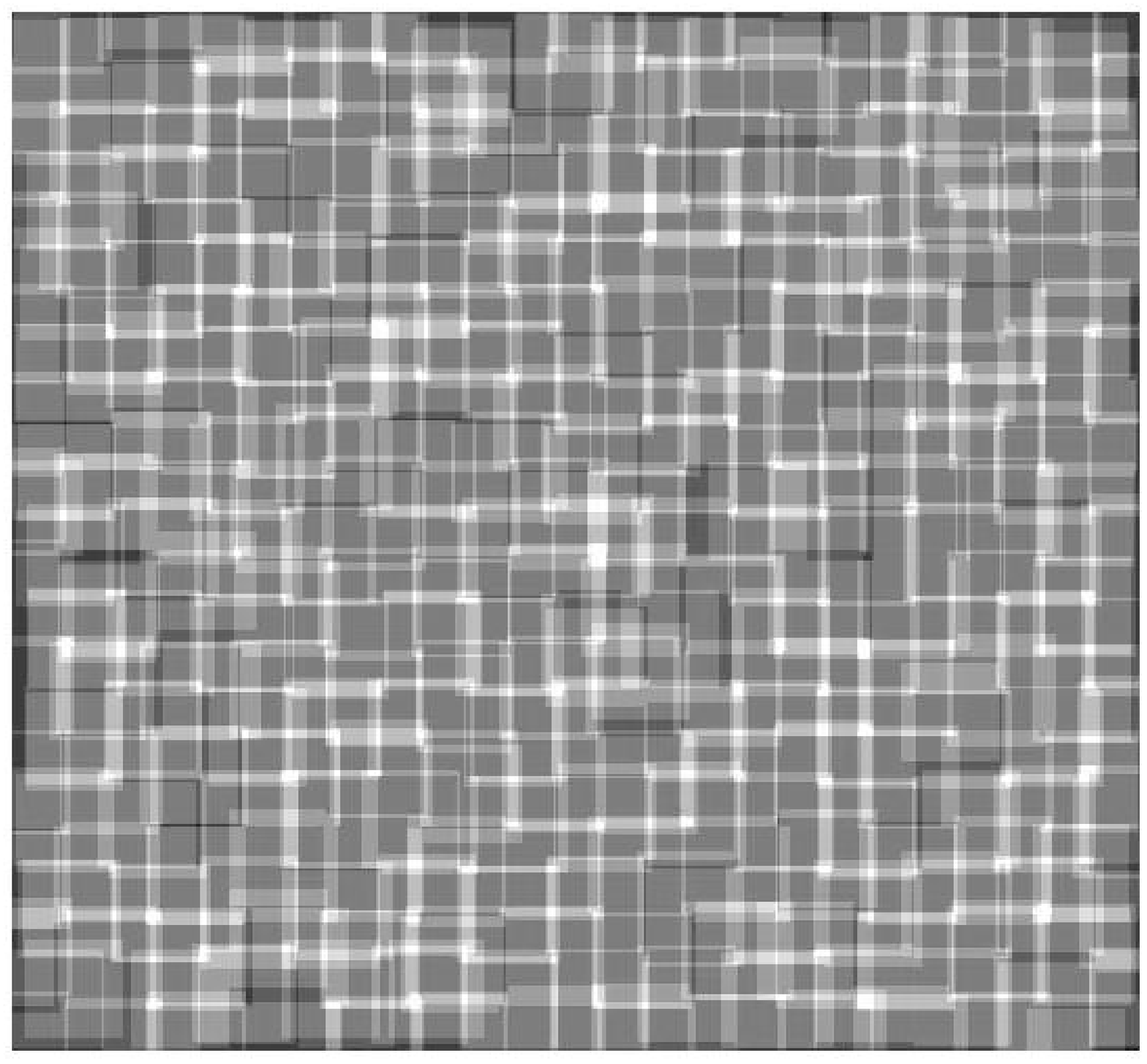}{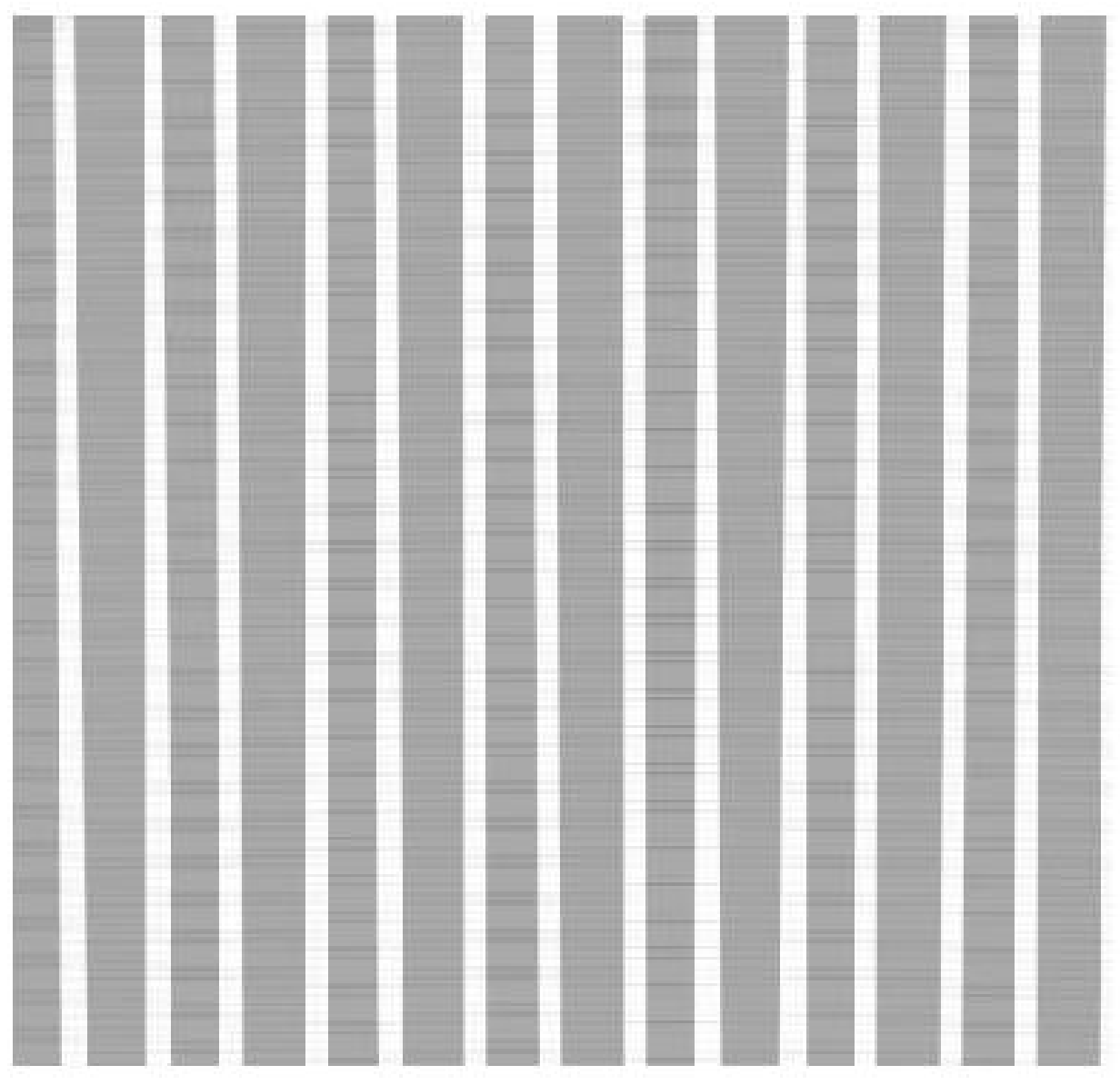}
\caption{Simulated full depth coverage map of a 1.0$\times$1.0 $\sq\degree$
IRAC 3.6$\mu$m image (left) and a 0.6$\times$0.6 $\sq\degree$
MIPS 70$\mu$m image (right).  Lighter areas indicate deeper coverage.}
\end{figure}

\section{Photometric Redshifts and Source Characterization}

SWIRE science goals depend strongly on accurate redshift estimates
and source classification from our Optical-IRAC-MIPS photometry.  To 
complement SIRTF observations in the IR (3.6--160$\mu$m), it is our goal to 
have optical ($R/r^{\prime}$-band) observations for the whole sample; 
additional 
optical, near-IR, radio and X-ray observations will be accessible for a 
sub-sample to allow detailed source characterization for representative 
populations.

\subsection{Photometric Redshifts}\label{photoz}

By combining optical ground-based data with SIRTF data we will be able
to determine photometric redshifts for large samples of SWIRE sources, 
as well as to determine key parameters characterizing their infrared
emission.  The approach is based on the photometric codes of Rowan-Robinson 
(2002), which fit a set of 6 optical sed templates for normal galaxies, with 
the option to determine the extinction $A_V$.  This code has been tested on
available photometric catalogues in HDF-N and HDF-S and gives robust
estimates of redshift accurate to 10$\%$ in (1+z).  The incidence of
aliasing has been shown to be smaller than most other photometric
redshift estimation codes.  For application to SIRTF data
additional cirrus, starburst, and AGN dust torus components have been
incorporated.  An optical AGN accretion disk SED is also an option
and this is being tested against SLOAN quasar catalogues.

The code has been run on pre-launch optical data prepared for SWIRE.  Where
4 or more optical bands are available, redshifts are demonstrated to be
reliable at the 98$\%$ level ({\it i.e.} no significant alias).  Where only
3 optical bands are available the reliability drops to 70$\%$.  Where UV
data are available, an important bi-product is the star-formation rate in
the galaxy.  An optical SED-type $n_{typ}$ is also determined.  Provided at
least 4 bands are available an estimate of $A_V$ is made.  [An
intercomparison of photometric redshift codes is being carried out by
members of the SWIRE team and will be reported in a future publication.

The majority of SWIRE galaxies are expected to be detected in the IRAC
bands, especially the 3.6 and 4.5 $\mu$m bands.  Since little contribution
from dust emission is expected at the shorter IRAC wavelengths, IRAC
photometry can be used to improve photometric redshift estimates for
systems with multi-band optical data.  For systems with only $R/r^\prime$
imaging, approximate photometric redshifts may be obtained from R-IRAC
photometry alone.  A bonus is that an estimate of the total stellar mass in
the galaxy can then be made for galaxies with IRAC detection, at least at
low redshifts. Where galaxies are detected at 4.5, 5.8 and 8 $\mu$m an
estimate can also be made of the luminosity in the cirrus component, after
subtraction of the predicted stellar contribution.  The ratio of $L_{cirr}$
to, say, $L_B$ should be related to the $A_V$ value.

If there are also detections at 24$\mu$m and beyond, estimates can be made
of the luminosity in the starburst and AGN dust torus components.  We describe
the procedures we are further developing to address these questions in more 
detail in the next section.  There
are insufficient SIRTF bands for accurate photometric estimates of 
redshifts to be made from SIRTF data alone, given the presence of several
components with distinct spectral signatures in the mid- and fir-IR,
and the large observed dispersion of dust temperatures found in ULIRGs 
\citep{blain03,farrah03}.

The above procedure has been prototyped against (1) IRAS galaxy samples, 
and (2)
simulated SWIRE data generated by C.K.Xu.  The IRAS data is at 
different wavelengths from SIRTF and the galaxies tend to be at low redshift
($<$ 0.3), but the code does succesfully recover both the redshifts and
the different SED components.  The simulated data is, of course, highly
dependent on the model assumptions, which are slightly different from those
used in the deconvolution.  The results of the simulations are again
satisfactory.

In conclusion, depending on the number of detected bands, we expect to be
able to output z, $n_{typ}$, $\dot{\phi_*}$, $A_V$, $M_*$, $L_{cirr}$, 
$L_{sb}$, $L_{tor}$, for large samples of SWIRE galaxies, which will be 
powerful diagnostics of galaxy evolution.

\subsection{SEDs and Source Characterization}\label{sourcechar}

In addition to photometric redshift determination we are developing methods
to classify SWIRE sources based on the optical-radio spectral energy
distributions (SED) in order to distinguish the main emitting components
(starburst, AGN) and the contribution of the stellar population.  Our
method is similar to the template fitting technique applied to estimate
photometric redshifts, \citep{hyperz}. This methods relies on the
identification of spectral features strong enough to be preserved after the
integration of the spectrum through the filter transmission
curves. Emission lines can be useful if they are strong enough relative to
the continuum flux to be detectable within the broad-band photometry. For
objects with featureless spectra, the success of template fitting depends
on the extent to which there exist unique continuum spectral shapes for
source populations.

The far- (60-1000 $\mu$m), mid- (6-60 $\mu$m) and near- (1-6 $\mu$m) IR
spectra of galaxies are characterized by different complexity and show
distinct emitting components. The far-IR spectrum is featureless thermal
radiation emitted by dust associated to an AGN or a starburst or the galaxy
disc with temperatures ranging from a few tens to a few hundreds degrees
Kelvin. The mid-IR spectrum is significantly more complex because of the
variety of the emitting components: photo-dissociation regions (PDRs), HII
regions~\citep*{tran98} and AGN-heated dust.  Star-forming systems are
dominated by the unidentified infrared bands (UIB), which are produced in
the PDRs at the interface between HII regions and molecular clouds. Very
small grains associated to HII regions produce a more intense and steeper
continuum at $\lambda>$12$\mu$m~\citep*{cesarsky96,verstraete96,roelfsema96}.  
The prominence of this component depends largely on the relative intensity
of the UV radiation field of the galaxy. The spectrum of an AGN may show
UIB features from star-forming regions or from the host galaxy
disc~\citep*{mirabel99,clavel00,moorwood99,alexander99}, but they are
usually weaker than in starburst galaxies because of dilution from the
strong continuum emitted by AGN-heated dust. The near-IR spectra of
galaxies are dominated by stellar photospheric radiation. The stellar
contribution can vary in galaxies with different star formation rates and
stellar populations. The warm dust continuum is very faint at
7$\mu$m~\citep*{cesarsky96,verstraete96,tran98}, and is not generally
detected in starbursts where the PDR emission (i.e., the UIB features)
dominates \citep*{sturm00}. Emission from AGN-heated dust can also
contribute to flattening of the spectrum. If an AGN component or an
extremely young ($\simeq$1 Myr) stellar population does not dominate in the
near-IR, a prominent feature at 1.6$\mu$m appears in the near-IR spectrum
of galaxies, caused by the minimum in the opacity of the H$^{-}$ ion
present in the atmosphere of cool stars~\citep{john88}.

In general, a starburst can be identified by strong PAHs, faint emission in
the NIR and cool MIR colors (F(60$\mu$m)/F(25$\mu$m)$>$5) and an AGN by
weak PAHs, strong NIR continuum and warm MIR colors
(F(60$\mu$m)/F(25$\mu$m)$\leq$5).  However, these spectral differences
between an AGN and a starburst are not always so distinctive.  A
significant continuum at short wavelengths (5-8 $\mu$m) is not necessarily
a tracer of an AGN, but could reveal a higher emission from \ion{H}{2}
regions in the ISM with respect to PDRs (see e.g. some dwarf
metal-deficient starburst galaxies like NGC 5253 and IIZw40
\citep{rigopoulou99}).  Moreover, strong PAHs can also be observed in
type-2 AGNs where the hot dust continuum is absorbed by material along the
line of sight \citep{clavel00}.

This variety of IR spectra makes any source characterization based on colors
alone problematic. Our method is based on matching all observed
data of SWIRE sources with templates representative of the known galaxy
population. Existing template libraries derived from observed
SED \citep{cww} or from models \citep{gissel, pegase, starburst, grasil,
stardust} either do not contain templates for AGN, do not include dust
emission, or do not extend behind NIR wavelengths.  Regrouping the available
models of a few galaxies obtained using different methods  
\citep{papovich02} produces a heterogeneous sample which will not be 
representative of the full observed palette of IR spectra in SWIRE. 

\begin{figure}
  \centerline{\psfig{figure=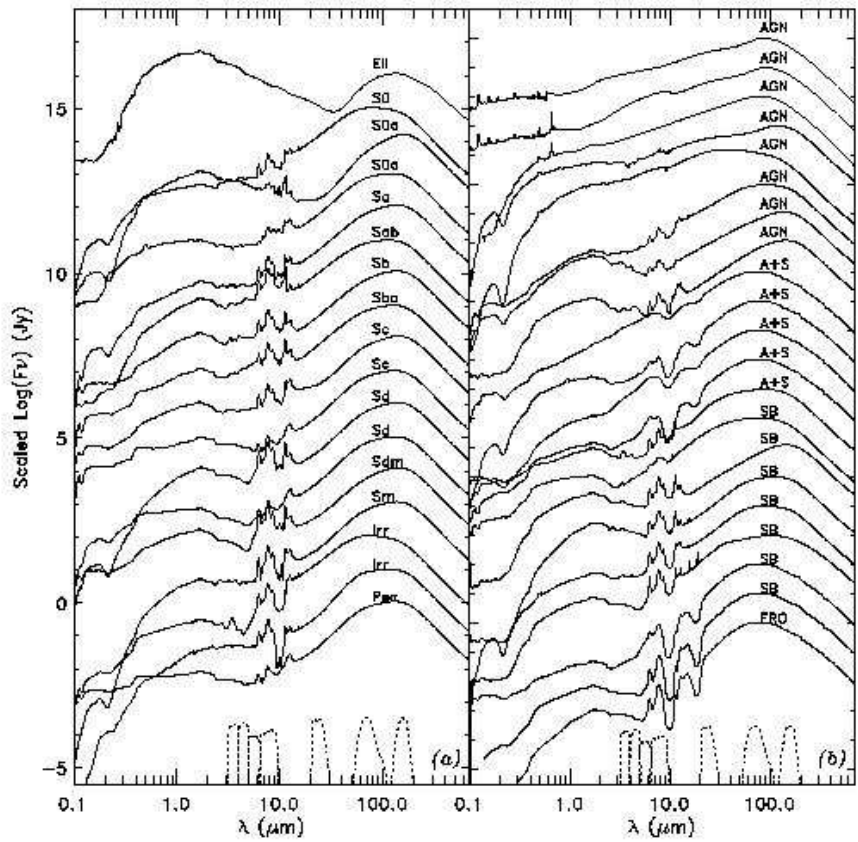,width=14cm}}
  \caption{Galaxy templates for normal galaxies $(a)$ and starburst, AGN 
	   and ULIRGs $(b)$. The IRAC and MIPS filter transmission curves 
           are shown as dotted curves.}
           \label{templates}
\end{figure}

In order to create a complete, representative library of galaxy
templates, we characterized the dispersion of galaxy IR spectra and
then selected objects with well sampled SED that could describe the observed
variety of IR spectra. The optical-radio SEDs of 837 objects from the IRAS
25$\mu$m sample \citep{xu01} and the ISO PHT-S and
-L \citep{kessler96} NIR SEDs of a sample of 175 normal and active galaxies
(N Lu, private communication) were analyzed in order to select SEDs which 
span the full range of galaxy characteristics. The final catalogue contains 
37 SEDs with redshift from 0.0007 to 0.08 and
two objects at redshift 1.06 and 1.44. In the
optical and NIR, each SED was modeled using a spectrophotometric
synthesis code \citep{berta03a,berta03b,fritz03} derived from the spectral
synthesis program by \citet{poggianti01}. In the MIR we adopted the observed
spectrum between 2.4 and 11 $\mu$m observed with ISO PHT-S and -L and the
SEDs as modeled by \citet{xu01} at longer wavelengths. Templates for quasars
were derived by combining composite optical spectra from the FIRST Bright
Quasar Survey\footnote{\it http://sundog.stsci.edu/first/QSO/Composites/}
\citep{fbqs} and the observed spectrum of a red quasar \citep{gregg02}
with ISO PHT-S and PHT-L NIR SEDs and photometric data in the IR of quasars
with optical SEDs well-fitted by the composite spectra.
The library contains
templates for ellipticals, spirals, irregulars, starburst, ULIRGs and active
galaxies covering the wavelength range between 300\AA\ and 20$cm$.    
All the current templates are shown in Figure~\ref{templates}.

The completeness of the current library is being tested using multicolor
photometric catalogues of galaxies at low and high redshifts with different
sets of filters and selection effects, like the catalogues of the
HDFN \citep{fsoto01}
and of CDF-S \citep{barger02}, the quasars observed with
ISO \citep{polletta00,haas00,andreani02, kuras03}, the CfA Seyfert
sample \citep{rodriguez01}, ULIRG samples \citep{klaas01,farrah03}, and the 
IRAS 25$\mu$m sample \citep{shupe98, xu01}.
New templates will be added to represent observed SEDs that
are not fitted by the current templates. Our library contains
mostly SED of local galaxies and only two high-$z$ objects. Distant
galaxies likely differ from local galaxies in their IR
SED \citep{xu01,dole03} and therefore new templates may be added to the
existing library to represent the high redshift galaxy population when more
knowledge about this population is available.

The bulk of SWIRE sources will be detected only at shorter wavelengths
(IRAC). Thus, full coverage of galaxies IR SEDs with accurate photometry may
only be possible for relatively nearby or bright objects. Therefore, the
limitations  of our method are being investigated on simulated catalogues
with solely IRAC data or IRAC plus one optical band. In particular the
degeneracy between redshift, dust temperature, age, and extinction are under
study.

Our galaxy template library provides a tool to investigate which colors can
be used to distinguish different types of objects. \citet{sawicki02} claims
that SIRTF NIR data alone should be sufficient to identify most galaxies at
$z\geq$1.5 that are dominated by stellar populations older than $\geq$20 Myr
through the 1.6$\mu$m bump. Galaxies at lower redshifts or those dominated 
by very
young stellar populations suffer from severe degeneracies and their SEDs can
be fitted if either IR observations at shorter wavelengths or optical data
are included. The color analysis cannot provide an accurate source
classification for the entire SWIRE sample because of the degeneracies among
different objects. However, some objects have unique colors and can be
easily identified: (1) ULIRGs can often be identified through 
the 70/8$\mu$m ratio; (2) type-1 AGN with 0.1$\leq z \leq$1.6 using the
IRAC 5.8$\mu$m/IRAC 3.8$\mu$m ratio; and (3) and objects like Mrk231, a 
type-1 AGN and ULIRG using the blueness of the IRAC or optical-IRAC colors. 
The FIR
color F(70$\mu$m)/F(160$\mu$m) can provide estimates of the FIR luminosity
(L$_FIR$), depending on how well the correlation observed for local galaxies 
holds for galaxies up to redshift 3 \citep{chap03, blain03}.  The optical/IR
flux ratio may also work well as a luminoisty estimator; see Figure
\ref{fig:radir}b.

Additional tools to characterize SWIRE sources will be provided by the
morphological analysis in optical images and by X-ray data. Optical imaging
will provide morphological information about the objects that can be used to
distinguish spheroids from disk galaxies and identify mergers. X-ray
emission is a good tracer of AGN even if starforming galaxies can also emit
X-rays through X-ray binaries, supernov\ae\ and hot diffuse gas. However,
the X-ray emission of starburst galaxies is fainter and softer than the AGN
X-ray emission.

The final library of galaxy templates with data and spectral models, the
method for SWIRE source classification and the tests on existing
photometric catalogue will be published in a separate paper.

\section{Early Science with SWIRE}\label{sciresults}

Since SWIRE is a homogenous, multi-wavelength survey covering a 
large area of sky and producing several million sources, the science 
possibilities with it are enormous.  In this section we very briefly
outline the main science topics that the SWIRE team plans to address 
initially.

\subsection{Modes and Rates of Star Formation}

\begin{figure}
\plotone{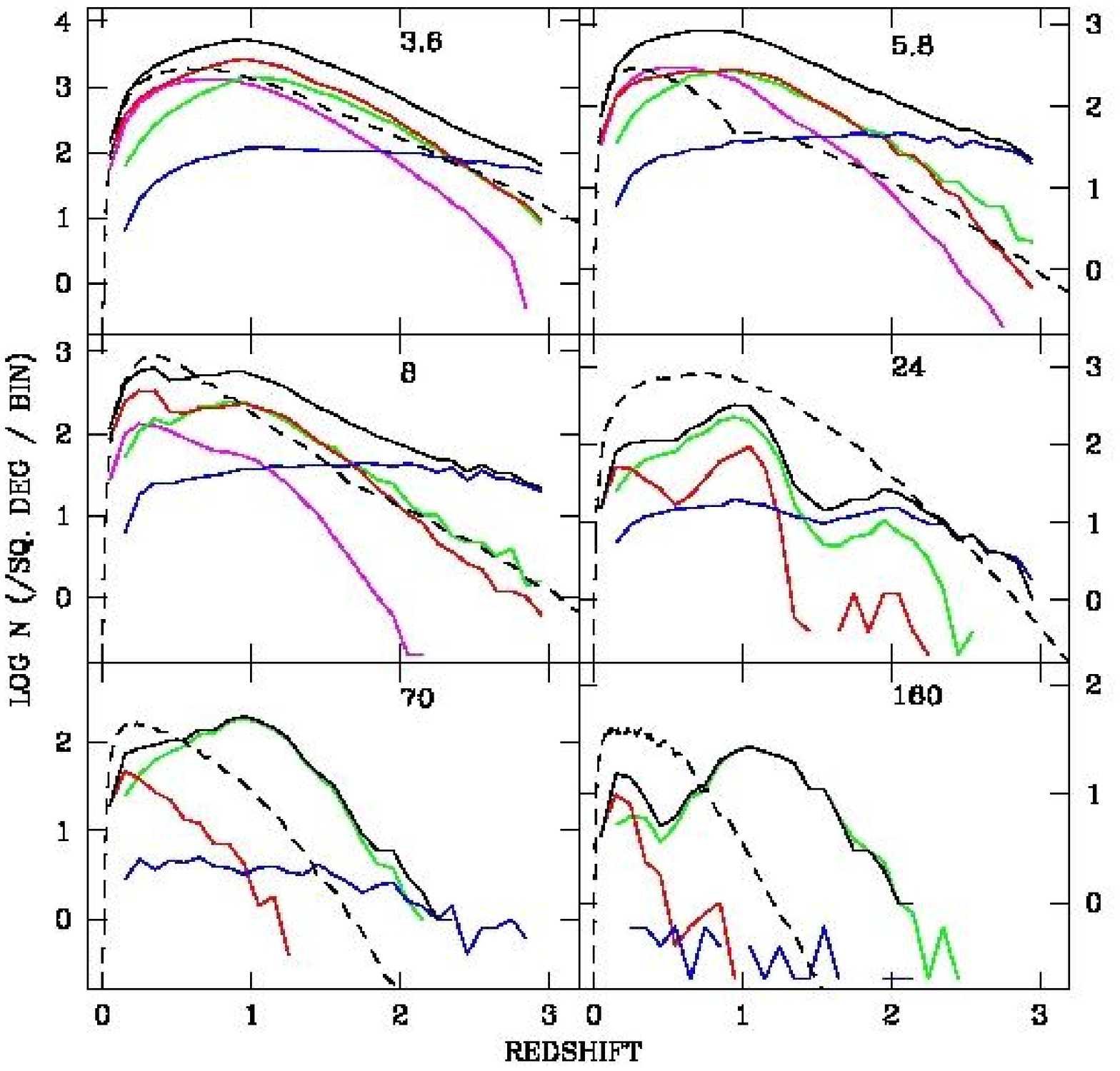}
\caption{Redshift distributions per 0.1 redshift bin per square degree in 
6 SWIRE SIRTF bands.  The models depicted are S3\&E2 
of \citet{xu03} (solid lines), and \citet{mrr01} (dashed line).  The 
flux limits are the 5$\sigma$ sensivity limits from
Table \ref{tbl:sens} except where confusion noise is expected to dominate,
namely S$_{lim}(70){\sim}$6 mJy, S$_{lim}(160){\sim}$60 mJy.  
For the \citet{xu03} models the
contribution of each galaxy type is indicated by color:  spheroids (magenta);
disks (red); starbursts (green); AGN (blue); total (black).}
\label{fig:zdist}
\end{figure}

As described in Section 1, a fundamental question for galaxy evolution is
the relative importance of quiescent star formation {\it vs.} starbursts 
as a function of epoch and matter density/environment.
SWIRE will directly measure the mid- to far-IR SEDs of $\sim$100,000 
IR-luminous galaxies.   The expected SWIRE source detection statistics are 
summarised in Figure \ref{fig:zdist}, which shows
results for two contrasting phenomenological models from \citet{xu03} and 
\citet{mrr01}, (see Section 4.1.1).  Model S3 of \citet{xu03} is dominated by 
starbursts at high redshift, while \citet{mrr01} attributes more sources to 
cooler, disk-dominated galaxies.  

\begin{figure}
\plottwo{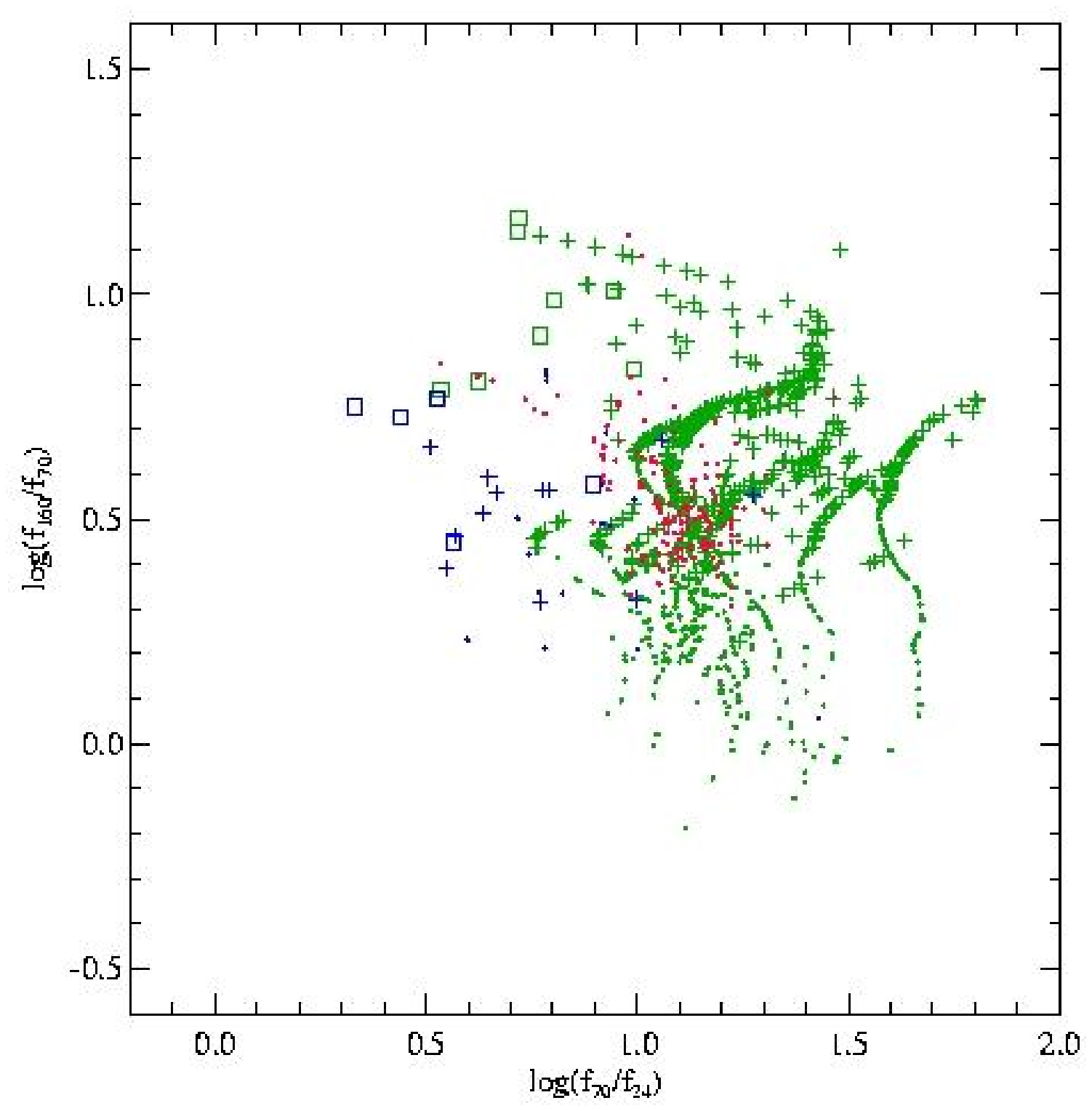}{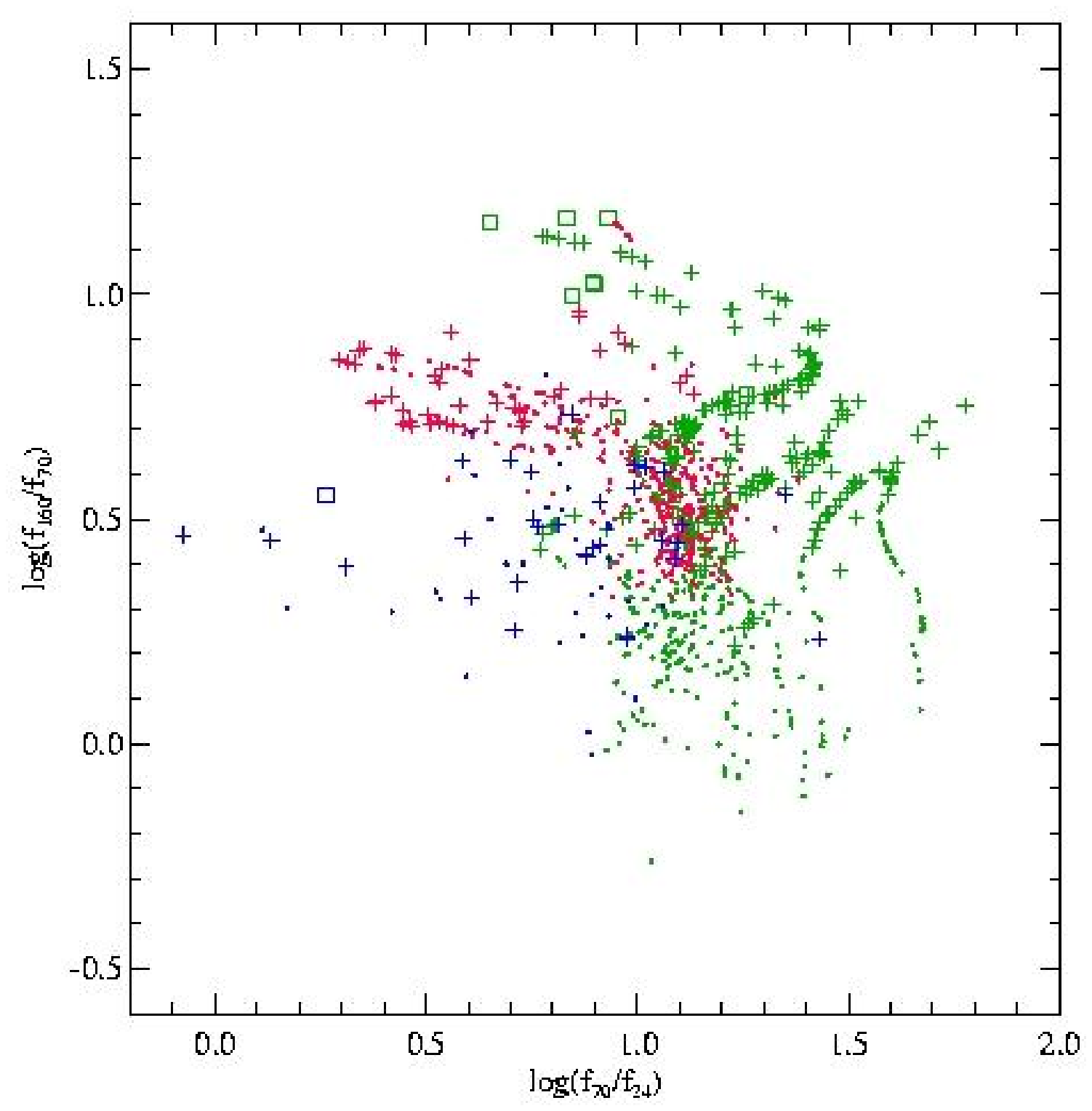}
\caption{SWIRE dusty populations within 1 square degree for model S3 (left)
and mode1 S2 (right).  Disks are red, starbursts green and AGN blue.  
Symbols denote
redshift range: z$<$1 (points), 1$<$z$<$2 (pluses), z$>$2 (squares).} 
\label{fig:mipscol}
\end{figure}

\begin{figure}
\plottwo{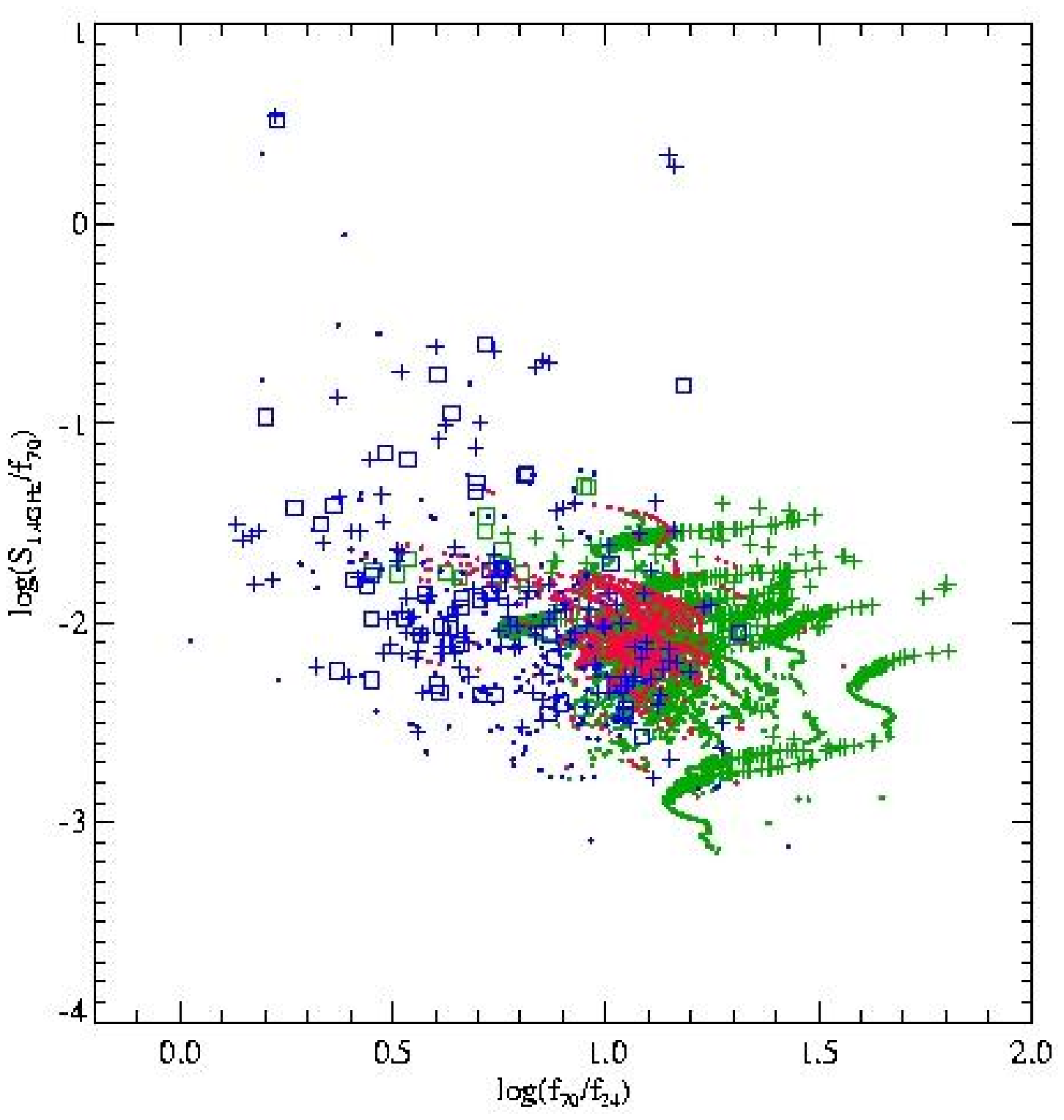}{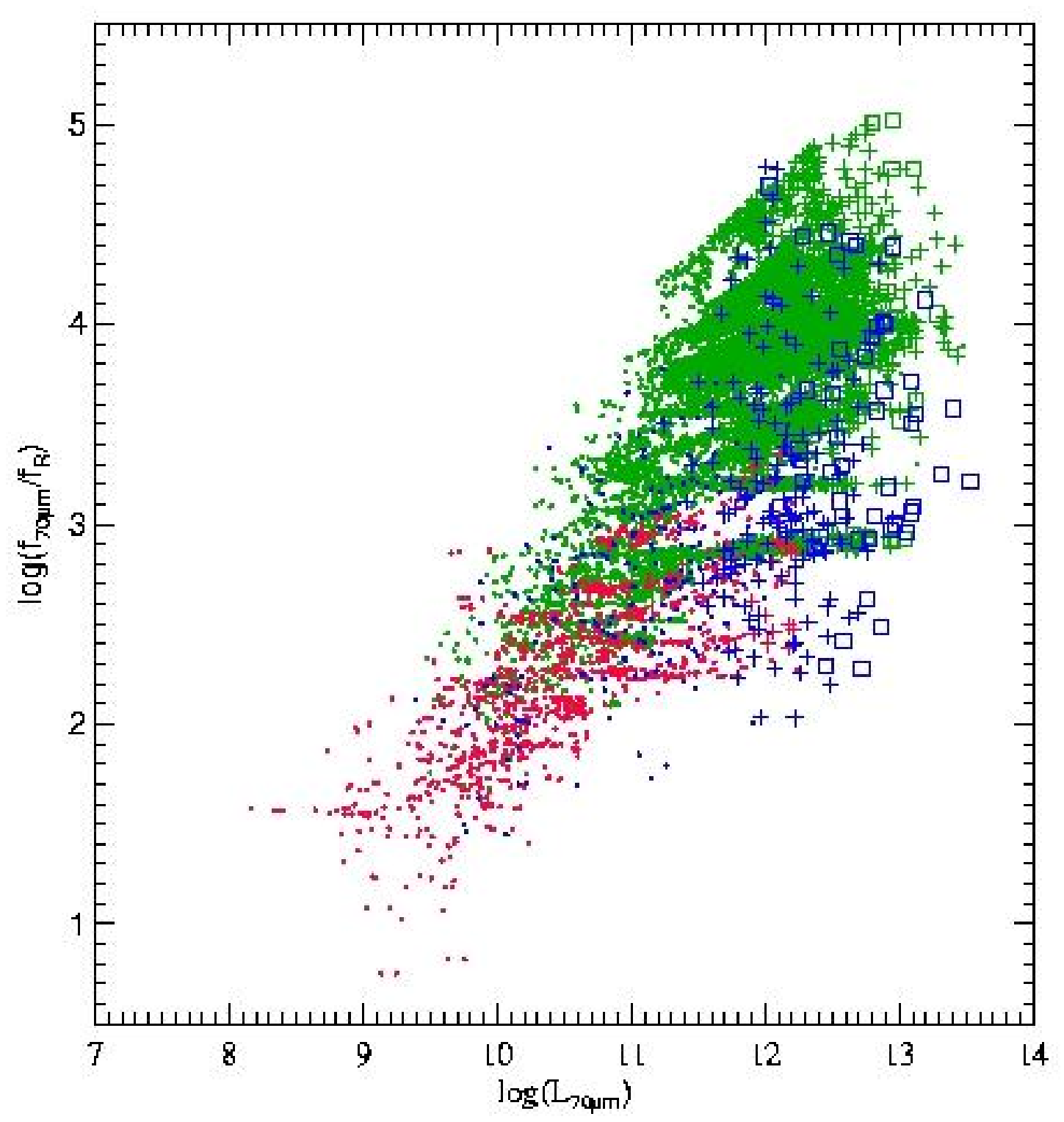}
\caption{Left:  MIPS/20cm color-color plot for the same population described 
in Figure \ref{fig:mipscol}.  The radio flux density limit is 16$\mu$Jy, 
corresponding to $\sim$5$\sigma$ in our ultradeep VLA image.  Right: 
70$\mu$m/R color vs L$_{70{\mu}m}$}
\label{fig:radir}
\end{figure}

The 70$\mu$m band will sample the broad peak of typical dust temperature 
components ($\sim$35K) out to redshifts of about 0.5, and 
warmer ones ($\sim$70K) to redshifts about 1, while the 
less sensitive 160$\mu$m band can sample the most luminous sources in
these temperature ranges to redshifts above 1 and 3 respectively.
Over these redshift and temperature ranges the SWIRE multiwavelength
SEDs will provide an accurate measurement of the bolometric luminosity of the
galaxy and the dust temperature range present.  Using photometric
redshifts (see Section \ref{photoz}) derived from the optical/IRAC data, we 
will construct luminosity functions in many separate volume cells
and for different dust temperature ranges.  Prior to the
availability of photometric redshifts, the same questions can be addressed
to lesser accuracy using number/color/flux data, {\it eg.} 
Figure \ref{fig:mipscol} to \ref{fig:radir}b.

Many systems will be detectable in the 8 and 24$\mu$m bands to redshifts,
z$\sim$3, or higher, although SWIRE will not sample their 
IR SEDs at peak, and so will not directly determine their IR luminosities.

A primary goal of SWIRE is to determine the cause for the strong increase
in the global star formation rate between z$\sim 0$ to z$\sim 1$ as
measured by the IR energy density. SWIRE will focus on the nature of the
most luminous sources, which seem to be much more numerous at high
redshift, and which provide the most critical challenges to hierarchical
models in terms of the extreme star-formation rates required at high
redshift.  We will address this by finding these systems (with MIPS),
studying their old stellar populations (with IRAC), observing their
morphologies (with the optical imaging) and determining their environments
(all data).  We can investigate the relative importance of quiescent star
formation {\it vs} starbursts from 160/70$\mu$m colors (Figure
\ref{fig:mipscol}) and morphology, looking, for example, for large disks
rather than the major mergers which dominate local ULIRGs.  Quiescent,
distributed star formation ocurring over galaxy disks should be cooler than
nuclear starbursts, thus we can directly track these modes, and test models
which attribute the high fluxes of high redshift ISO 170$\mu$m and
submillimeter sources to distributed ``cirrus'' emission \citep{mrr01,
efstathiou, xu03, kaviani} {\it eg.} compare the models of \citet{xu03} and
\citet{mrr01} in Figure \ref{fig:zdist}, and \citet{xu03} model S2 in
Figure \ref{fig:mipscol}.

In Figure~\ref{fig:radir}a 
we show the anticipated color distribution in the
70/24 {\it vs.} 1.4GHz/70$\mu$m color-color plane for model S3 for our deep
VLA field (Section \ref{deepvla}).
In this model, $\sim$90\% of the SWIRE MIPS population is detected at 
20cm, which will allow an interesting
investigation of the radio/IR correlation at high redshifts (the model assumes
no ratio evolution).  Note that the 
very large dispersion in observed 1.4GHz/70$\mu$m color is dominated by
the dispersion in redshift, but also incorporates the intrinsic
dispersion in the template SEDs (see \citet{xu03} for full details).  It
remains to be seen whether the local radio/IR correlation holds up
at high redshifts.

Figure \ref{fig:radir}b shows the well known infrared/optical
vs L$_{ir}$ relation from the local Universe \citep{soifer89}.  The apparent 
horizontal tracks in this figure are due to individual template SEDs,
which are fixed in luminosity, sampled by the model galaxy population over a 
range of redshifts.   Whilst this ``k-correction'' dispersion degrades the
use of this relation as a luminosity indicator at the higher luminosities,
we can use it to estimate minimum luminosities, and hence a lower limit
to luminosity distance using the measured flux.  This technique of course 
assumes that the high redshift SEDs have been well represented by the model,
which may not turn out to be the case.

\subsection{The Mass Function of Spheroids}

SWIRE will detect $\sim$1 million spheroids by the emission of evolved
stars in the shorter IRAC bands, reaching z$\sim$2.5 (Figure
\ref{fig:zdist}).  The stellar mass of these systems can be
estimated by SED fitting using a library of stellar population synthesis
models ({\it eg.} \citet{dickinson02}).  This method has been prototyped
for SWIRE using the models of \citet{poggianti01} by \citet{frans03}, who
found fairly large uncertainties in estimated stellar masses for
ISO-detected HDF galaxies when using only optical/NIR data, especially if a
there exists a substantial stellar population with significant extinction. 
The addition of photometry in the IRAC bands will
provide much greater accuracy because older populations dominate these
longer wavelength bands, rendering the SEDs very uniform \citep{simpson,
sawicki02}.  The longer wavelength IRAC bands also enable robust mass
estimation over a much larger redshift window.

Using the SWIRE $r'-5{\mu}$m SEDs we will assess the build-up rate of stellar
mass in massive stellar systems since z$\sim$2.5 and compare it directly
to the global star formation rates in IR-luminous systems, for both 
quiescent and starburst modes, over the same redshift range.  Again, the
dependence of the results on the matter density field and the nearby galaxy 
environment will be of primary importance to constraining galaxy formation
models and the cosmological model.
  
\subsection{Active Galactic Nuclei}

As described in Section 1, a major SWIRE goal is to determine the evolving 
number density of 
AGN, particularly heavily obscured ones.  The difficult aspect of this goal 
lies in identifying the AGN, because
the most heavily obscured ones can be very optically thick even in the 
near- and mid-IR where we expect the unique AGN identification signatures,
specifically warm dust associated with the torus in the 
3$<{\lambda}<40{\mu}$m range, and very red colors due to extinction in the 
near-IR, such as being found for moderately obscured 2MASS 
AGN \citep{lacy, cutri}.    Our initial studies
of SWIRE AGN populations will be based on identifying candidates with
indicators of heavy extinction from red optical-IRAC colors, and 
warm dust from 3-70$\mu$m colors.  

The blue curves in Figure \ref{fig:zdist} illustrate the detectabilty of
{\it optically-selected} AGN by SWIRE.  If obscured populations such as
those required to explain the slope of the XRB, increase with redshifts and
are actually more numerous than low obscuration AGN by factors of up to 4,
observed in the Deep Chandra Surveys \citep{gilli02}, then a substantial
fraction of the SWIRE ``starburst'' sources (green curves) will harbor such
AGN.  Moreover these AGN may possibly warm the MIPS colors above those
assumed in our models, which would boost the relative 24 and 8$\mu$m counts
still further.

Statistical studies of the starburst/AGN connection as a function of
luminosity will provide major insights into how starbursts and AGN coexist
and to possible evolutionary relationships.  Furthermore, we can
examine starburst-AGN connections as a function of redshift (thus 
increasing gas mass, decreasing dust content and
decreasing metallicity), environment and Hubble type.

\subsection{Large Scale Structure}

Our first large-scale structure goal will be to provide an estimate of the
clustering of both active and passive galaxies with mean redshifts higher
than any substantial surveys to date, using 2-D Auto-correlation
functions. Such measurements are expected to provide powerful constraints
on models of galaxy formation.

We will measure the angular-autocorrelation functions of the SWIRE
galaxies selected in each of the seven SIRTF bands, comparing each of
these seven auto-correlation functions with predictions from existing
numerical and semi-analytic models of galaxy formation and with
extrapolations from existing surveys at other wavelengths and lower
redshifts.  Inversion techniques will provide an estimate of
three dimensional clustering, against which future models can be
directly compared.  This will require model redshift distributions or
real redshift distributions as they become available.

Other two-dimensional two-point statistics including the
power spectrum and variance of counts-in-cells may also be
investigated.

2-D cross-correlations can provide a better handle on the issue of galaxy
bias, by allowing us to investigate the relative clustering of different
sub-samples of the SWIRE populations.  We will measure the angular
cross-correlation between galaxy catalogues selected at different SIRTF
wavelengths.  Initially we will focus on comparing the IRAC and MIPS
catalogues as these probe passive and active star-formation systems
respectively. As our classification of galaxies on the basis of their
multi-wavelength colors improves we will compare distinct populations {\it
e.g.} star-burst galaxies, spheroids and AGN.  Measurements of
cross-clustering statistics will be compared with predictions from existing
models and inversion techniques will provide estimates of three dimensional
cross-clustering.

Once we have established a reasonable photometric redshift system we
will be able to provide an estimation of the evolution of the
clustering, independently of other low-redshift surveys, by repeating the 
above analyses in photometric redshift slices.  We can
also use these photometric redshifts to provide better estimates
of three dimensional clustering without resorting to inversion
techniques.

As a more subtle test of galaxy formation models we also intend to use
counts-in-cells techniques to investigate the higher order statistics
of galaxy clustering. We will perform two-dimensional counts-in-cells
analyses on individual populations and on joint populations. We will
also use photometric redshifts to take these analyses into three
dimensions.

Ultimately, in collaboration with the wider astronomical community, we
hope to obtain sparse redshift surveys to provide a better precision
and check on our photometric redshift based studies.

\subsection{Clusters in the XMM-LSS Field}

A major goal for the XMM-LSS field is comparative cluster detection in
terms of galaxy and cluster properties.  In parallel to optical multi-color
and X-ray cluster detection techniques, we shall investigate clustering
methods in the SIRTF wavebands; this is a totally new approach with the
potential to yield distant clusters of AGN as well.  Clusters will principally
be detected by late-type stellar emission of individual cluster members in the
shortest IRAC bands, but at the higher redshifts star formation and
AGN activity is 
expected to increase, detectable by dust emission in the longer SIRTF bands. 
We will compare
cluster existence/richness/morphology derived in the different
wavebands. Although these techniques provide comparable results in the
local Universe, there are strong hints that the overlap between optically
and X-ray selected clusters beyond z$>$0.5 is only $\sim60\%$ (Donahue et
al. 2001, 2002). It is most interesting to understand this discrepancy. One
possibility is that optical cluster selection is very sensitive to the
star-formation rate owing to the effect on galaxy color by active star
formation.

A further goal is to determine the location of the SWIRE sources within the
cosmic network.  Environmental studies of the XMM-LSS data (X-ray and
optical) will provide a unique view of the LLS out to z$\sim$1.  Redshift
measurements of the SWIRE sources will allow us to subsequently locate them
within the cosmic network (field, filaments, groups, clusters). This will
provide decisive clues as to the effect of environment on star, galaxy and
AGN formation. Ultimately, one can relate environmental conditions to the
initial density fluctuations from which the clusters originated.
Conversely, we shall be in a position to investigate how galaxy activity
affects the global properties of clusters like metallicity, temperature
etc.  and thus quantitatively address such questions as the effect of
preheating by star formation and AGN on the intra-cluster medium. Another
point will be to investigate whether star forming and nuclear activity are,
as expected, most efficient in medium dense environments (i.e. groups),
since in these regions, galaxy velocities are moderate, thus optimal for
efficient interactions.

A third study will use clusters as gravitational telescopes to flag and
study very distant SWIRE sources.  Weak lensing studies using the
Canada-France-Hawaii Legacy Survey data on the most massive X-ray detected
clusters will offer a truly new window to the distant Universe and we
expect as well several SWIRE sources per cluster to be the magnified images
of very distant (obscured?) objects. We shall thus gather a sample of
several tens of pencil beam surveys, providing a statistical complement to
the study undertaken by GOODS and, in addition, less affected by possible
instrumental confusion effects.

\subsection{Mergers, Environment \& Morphological Evolution}

SWIRE will provide a unique window into the development of the Hubble 
sequence by providing true AGN and starburst luminosities  
associated with merger-driven events and other environmental factors such 
as cluster-cluster mergers, over large co-moving size scales at z$\sim$1-2.  
Moreover, the depth and angular resolution (particularly at IRAC wavelengths) 
of the SIRTF observations will be very high for
local objects.  For these we can address the detailed structural
relationships between the optical/near-IR/submm structure and the
mid-IR structure in a wide variety of galaxies.

A few of the planned follow-up projects for the SWIRE data include pointed
high resolution optical imaging of ULIRGs at medium and high redshift,
adaptive optics surveys of galaxies lying in the SWIRE fields, a study of
optically identified interacting galaxies using the SWIRE optical and
infrared data, a high resolution submm study of SWIRE interacting, infrared
luminous galaxies, optical asymmetries in UV vs. IR-selected galaxies, the
star formation rate as a function of environmental richness, and an
unbiased high spatial resolution optical survey of a large fraction of a
SWIRE field.

\subsection{Rare Objects}

A major strength of SWIRE is the large cosmological volume available to it,
therefore the sample will contain unusual objects of up to 1-in-10$^6$
rarity, to redshifts $>$3.  These will include both object types we already
know about such as Extremely Red Objects (EROs), Lyman Break Galaxies
(LBGs), distant ULIRGs such as discovered by SCUBA, powerful QSOs \& radio
galaxies, and brown dwarfs.  We may also anticipate previously unknown
categories of objects that SWIRE may select in favor of, such as galaxies
with unusually warm mid-IR colors.  The individual objects will be
extremely interesting in their own right, and also as elements in the major
science themes discussed in previous sections.  For example, very distant
and luminous ULIRGs and AGN detected at 8 \&/or 24$\mu$m will be used as
potential tracers of massive halos.  A major goal is to measure the mid-
and far-IR luminosity distributions of LBG candidates \citep{siana}.

Rare object candidates will be selected by their flux-color signatures in
the multi-wavelength SWIRE databases and followed-up extensively, both on
the ground and with SIRTF deeper imaging and IRS spectroscopy (beginning
with a small GTO IRS program to follow up $\sim$6 SWIRE sources in CDF-S;
M. Werner, PI).

\subsection{Nearby Galaxies}

The large area of the SWIRE survey, together with SIRTF higher sensitivity
and better angular resolution than previous IR space telescopes allow us
and the community to undertake a number of investigations of galaxies in
the local Universe.  Galaxies with diameters larger than $\sim$10\arcsec\
will be spatially resolved by SIRTF in the IRAC bands, and those larger
than $\sim$15\arcsec\ also in the 24$\mu$m MIPS band. The largest of the
nearby galaxies present in the survey area could also be resolved at 70 and
160$\mu$m.  The SIRTF Legacy project SINGS, (SIRTF Nearby Galaxy Survey,
\citet{kennicutt}) is designed to study the physics of the star-forming ISM
and galaxy evolution by a comprehensive imaging and spectroscopic study of
75 nearby galaxies (D$<$30 Mpc).  SWIRE will extend some of these
investigations to a larger sample of galaxies at larger redshifts and in
very different galaxy environments, including isolated galaxies,
interacting galaxies, groups and clusters.

\begin{figure}
\plotone{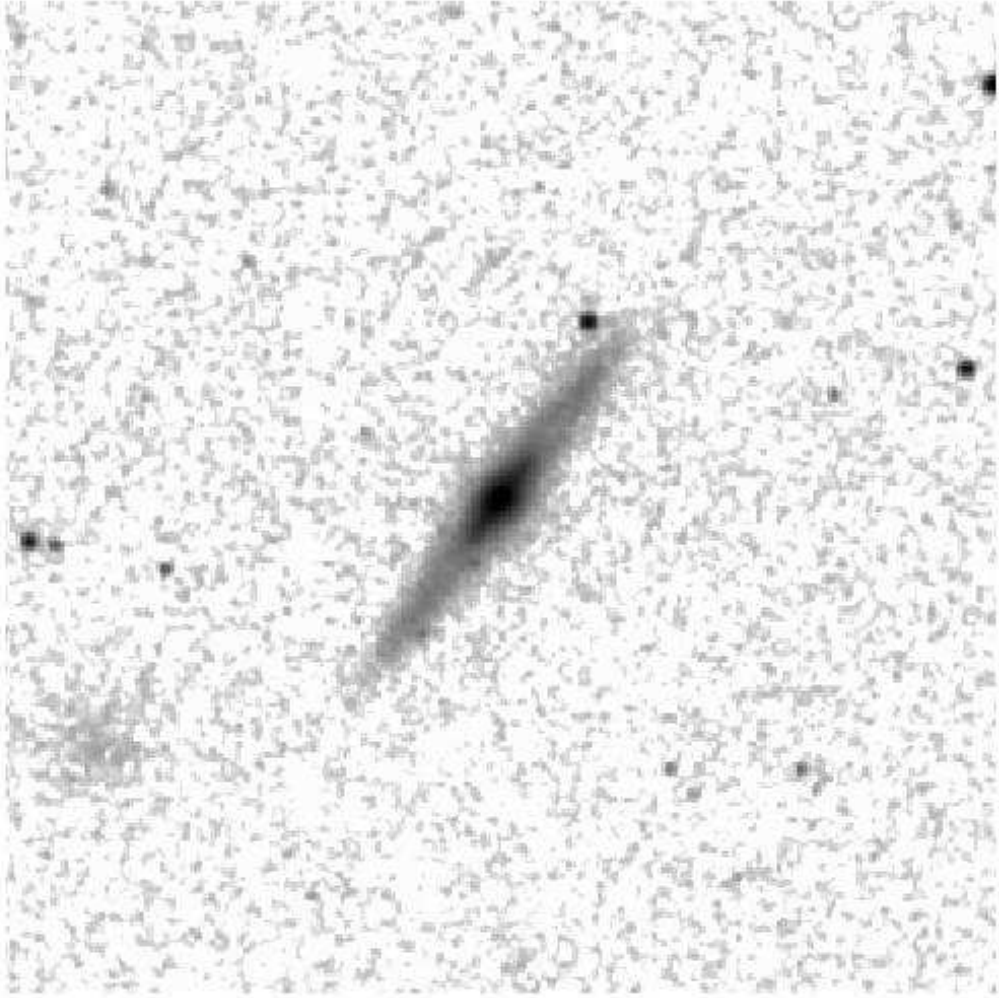}
\caption{NGC 5777 as seen in the combined 2MASS near-infrared bands.
The 1$\sigma$ surface brightness is 21.8 mag/arcsec$^2$ and the resolution
is $\sim$3\arcsec.  To the lower left (south-east) of the image center
is the dwarf companion UGC 09570.  The image size is 5.7$\arcmin$ on a side.}
\label{fig:n5777}
\end{figure}

We have defined a protoype SWIRE nearby galaxy sample from the 2MASS
final-reduced database, searching in the SWIRE areas for objects with
isophotal diameter at 2.2$\mu$m (the 2MASS band closest in wavelength to
the IRAC spectral range) greater than 10\arcsec. The sample consists of
several hundred galaxies with angular diameters up to $\sim$100\arcsec.  A
small number of galaxies are contained in the 2MASS Large Galaxy Atlas
\citep{jarrett}.  One of the largest is NGC 5777 in the Lonsdale field,
which is illustrated in Figure \ref{fig:n5777}.  Most of the galaxies with
a published redshift are at z$<$0.1.

The SIRTF broad-band images, combined with data at other wavelengths ({\it
eg.} UV imaging from GALEX) will allow us to study in detail the physics of
star formation as a function of location on the galaxy, ISM properties,
galaxy morphological and spectral type and galaxy environment, particularly
for objects of relatively low IR luminosities, which will not be detectable
at higher redshift. The sample will also be used to construct composite
SEDs for different galaxy types which will be useful to interpret the SEDs
of more distant objects.

\subsection{Brown Dwarfs}\label{browndwarfs}

During the past few years, a large number of very low mass stars and
substellar objects ("L'' dwarfs) have been discovered, primarily in the
near-infrared 2MASS and DENIS surveys \citep{oppenheimer} and more recently
in the SDSS \citep{geballe02}. These objects lie below the M dwarfs in
effective temperature and are primarily distinguished by their very red JHK
colors. However, even cooler objects, designated "methane or T dwarfs"
\citep{burgasser}, are more difficult to find in these near-IR surveys,
since their methane bands give them a blue color shared by many ordinary
disk stars. The IRAC 3.6/4.5 micron bands are ideal for identification of T
dwarfs. The 3.6 micron channel encompasses a major methane band, while the
4.5 micron channel samples the SED peak for these cool sources. Thus, the
F$_{4.5}$-F$_{3.6}$ color should easily pick out nearby T dwarf candidates
in our source list.  SWIRE will provide the best SIRTF Legacy database for
serendipitous brown dwarf discoveries due to its wide area coverage and
high sensitivity. SWIRE will have sufficient 4.5 \micron\ sensitivity and
areal coverage to potentially detect very low mass old brown dwarfs in the
field.  Comparing the SWIRE sensitivities (Table \ref{tbl:sens}) to the
predicted fluxes for old BD from \citet{burrows}, we find that SWIRE could
detect a 5 Gyr old, 275 K brown dwarf at a distance of 10 pc at 4.5
\micron.

We simulated the detectability of field brown dwarfs in the 
SWIRE legacy survey using a Monte Carlo model similar to \citet{martin},
 who estimated brown dwarf number counts for
a range of IRAC survey sizes and sensitivities.  Fluxes in 
the four IRAC bandpasses were tabulated 
vs.  effective temperatures using the online models of Marley.  
T$_{eff}$ vs. mass and age was taken from evolutionary models by 
A. Burrows.  Brown dwarf initial mass functions 
with various power law indices $\alpha$ were simulated, with total counts 
normalized to the adopted stellar space density of 0.057 pc$^{-3}$ 
\citep{reid}.  A star formation rate proportional to
[1-exp(-t/$\tau_1$)$\times$exp(-t/$\tau_2$)] was used, with $\tau_1= 0.2$
and $\tau_2= 8.0$ Gyrs.  Random masses, ages, and X, Y, Z coordinates
within a sun-centered cubical volume were then used as inputs to the
evolution and mid-IR flux models, which were interpolated to give the
observables. For $\alpha=$ 1.5, 1.0, and 0.5, the model predicts 
SWIRE detection of 23, 8, and 1 objects respectively in IRAC channel 
2 (4.5 $\mu$m). 

\subsection{Circumstellar Debris Disks}

Among the most important questions in astronomy today are the frequency of
extrasolar planetary systems and their similarity to our own solar system.
SIRTF's extremely high sensitivity at long wavelengths enables
investigation of circumstellar disks which are plausible signposts of
debris from planet formation. Disks similar to the one observed around HR
4796A \citep{koerner} show excess emission from dust at wavelengths beyond
8$\mu$m with a steep rise between 24 and 70$\mu$m, ideal for study with
MIPS. IRAS detected debris disks around massive stars at distances of tens
of parsecs. SIRTF will be able to detect much more tenuous disks around
solar-type stars out to 100 pc and analogs of HR4796A (A0 star) at a
kiloparsec. Even considering the high galactic latitudes of our
survey fields, we expect to detect the photospheres of at least 1000 A - K
spectral type stars at 24$\mu$m. The frequency of dust excess among stars
in the solar neighborhood is at least 15\% (\citet{lagrange}, PPIV), so we
would plausibly expect to discover of order 150 new debris disk systems
during our MIPS surveys. Unlike the other SIRTF legacy debris disk
searches, SWIRE will have no age or spectral type bias in its survey of
debris disks.

\subsection{Small Asteroids}

Current knowledge of the asteroid size-frequency distribution is limited to
sizes larger than about 10 km. The population of small asteroids is almost
completely uncharacterized, but is critical for understanding the parent
population of the Earth-crossing asteroids \citep{evans}. SIRTF will be
sensitive to thermal emission at 8 and 24$\mu$m from main belt asteroids as
small as 1 km. Estimates of the number of asteroids in SIRTF 5' x
5' images at ecliptic latitude 0$^\circ$ range over at least an order of
magnitude. Although most of our survey fields are located at 
high ecliptic latitude, the XMM-LSS (10 sq. deg.) field is at 17$^\circ$
latitude.  We will repeat the MIPS scans across this field at an interval
of several hours, allowing us to identify asteroids and to estimate
their mean motion.

\begin{acknowledgements}

SWIRE is supported by NASA through the SIRTF Legacy Program under contract
number 1407 with the Jet Propulsion Laboratory.  The research described in
this paper was carried out, in part, by the Jet Propulsion Laboratory,
California Institute of Technology, and was sponsored by the National
Aeronautics and Space Administration.  H. E. Smith wishes to express
gratitude to IPAC for providing continued support as a visitor.  Seb
Oliver's travel was supported by the Nuffield Foundation grant NAL/00240/G.
Eduardo Gonzalez-Solares was supported by PPARC grant PPA/G/S/2000/00508.
Malcolm Salaman is supported as a PPARC postgraduate student number
PPA/S/S/2000/03082A.  We are deeply indebted to Mike Irwin and his
colleagues at CASU for allowing SWIRE use of their facilities and for
substantial help and support.  SWIRE acknowledges generous ground-based
observing support through the NOAO-SIRTF Legacy agreement; NOAO is a
facility of the US NSF, operated under cooperative agreement by AURA.  We
acknowledge generous allocation of observing time by the European Southern
Observatory for southern fields.  Montage is funded by the NASA Earth
Sciences Technology Office Computational Technologies project under
Cooperative Agreement Notice NCC5-626. Finally, we thank referee Michael
Strauss for cogent comments which improved this manuscript.

\end{acknowledgements}


\begin{thebibliography}{}
  

\bibitem[Alexander et~al.(1999)]{alexander99} 
Alexander, D. M., et~al. 1999, \mnras, 310, 78

\bibitem[Alexander et~al.(2001)]{alexander01} 
Alexander, D. M.,  et~al. 2001,  \apj, 554, 18
 
\bibitem[Almaini et~al. (2003)]{almaini} 
Almaini, O., et~al. 2003, \mnras, 338, 303

\bibitem[Andreani et~al.(2003)]{andreani02} 
Andreani, P., Cristiani, S., Grazian, A., LaFranca, F., \& Goldschmidt, P.
	2003, \aj, 125, 444

\bibitem[Archibald et~al.(2002)]{archibald} 
Archibald, E., Dunlop, J., Jiminez, R., Friaca, A., McClure, R., Hughes, D. 
	2002, \mnras, 336, 353

\bibitem[Ashby et~al. (2003)]{ashby} 
Ashby, M., Surace, J., \& Hora, J. 2003, in 
	{\it Infrared Telescopes and Instruments}, SPIE, in press

\bibitem[Babul \& Postman (1990)]{babul} 
Babul, A. \& Postman, M. 1990, \apj, 359, 280

\bibitem[Barger et~al.(2002)]{barger02} 
Barger, A. J.,  et~al. 2002, \aj, 124, 1839

\bibitem[Beichman et~al. (2003)]{beichman} 
Beichman, C.A., Cutri, R., Jarrett, T.J., Steining, R. \& Skrutskie, M. 2003, 
	\apjl, in press

\bibitem[Benson et~al. (2001)]{benson} 
Benson, A. J., Frenk C. S., Baugh C. M., Cole S., \& Lacey C. G., 2001, 
	\mnras,  327, 1041

\bibitem[Berta et~al.(2003a)]{berta03a} 
Berta, S., et~al. 2003a, \aap, submitted

\bibitem[Berta et~al.(2003b)]{berta03b} 
Berta, S., et~al. 2003b, in preparation

\bibitem[Bertin \& Arnouts (1996)]{bertin} 
Bertin, E. \& Arnouts, S. 1996, \aap\ Suppl., 117, 393

\bibitem[Blain et~al. (1999)]{blain99} 
Blain, A., et~al. 1999, \mnras, 309, 715

\bibitem[Blain et~al. (2002)]{blainrev} 
Blain, A. W., Smail, I., Ivison, R.I., Kneib, J.-P., \& Frayer, D.T. 2002, 
	Physics Reports, 369, 111

\bibitem[Blain et~al. (2003)]{blain03} 
Blain, A.W., Barnard, V.E. \& Chapman, S. C. 2003, \mnras, 338, 733

\bibitem[Blanton et~al.(2000)]{blanton2000} Blanton, M., Cen, R., 
Ostriker, J.~P., Strauss, M.~A., \& Tegmark, M.\ 2000, \apj, 531, 1 

\bibitem[Bolzonella, Miralles \& Pell\'o (2000)]{hyperz} 
Bolzonella, M., Miralles J.-M. \& Pell\'o R. 2000, \aap, 363, 476

\bibitem[Brotherton et~al. (2001)]{fbqs} 
Brotherton, M., Tran, H., Gregg, M., Becker, R., Laurent-Meuhleisen, S. \& 
     White, R. 2001, \apj, 546, 775

\bibitem[Bruzual \& Charlot(1993)]{gissel} 
Bruzual, G. A. \& Charlot S. 1993, \apj, 405, 538

\bibitem[Burgasser et~al. (2002)]{burgasser} 
Burgasser, A., et~al. 2002, \apj, 564, 421

\bibitem[Burrows et~al. (1997)]{burrows} 
Burrows, A., et~al. 1997, \apj, 491, 856

\bibitem[Cambresy et~al. (2002)]{cambresy02} 
Cambresy, L., Beichman, C., Jarrett, T.,  \& Cutri, R. M. 2002, \aj, 123, 2559

\bibitem[Cesarsky et~al.(1996)]{cesarsky96} 
Cesarsky, D., et~al. 1996, \aap, 315, L309

\bibitem[Chapman et~al.(2003a)]{chap03} 
Chapman, S., Helou, G., Lewis, G., \& Dale, D. 2003a, \apj, in press 
	(astro-ph/0301233)

\bibitem[Chapman et~al. (2003b)]{chap03b} 
Chapman, S., Blain, A., Ivison, R., \& Smail, I.  2003b, \nat, 422, 1540

\bibitem[Chary \& Elbaz (2001)]{charelb} 
Chary, R. \& Elbaz D. 2001, \apj, 556, 562

\bibitem[Ciliegi et~al. (2003)]{ciliegi03} 
Ciliegi, P., Zamorani, G., Hasinger, G., Lehmann, I, Szokoly. G \& Wilson, G. 
	2003, \aap, 398, 901

\bibitem[Clavel et~al.(2000)]{clavel00} 
Clavel, J. , et~al. 2000, \aap, 357, 839

\bibitem[Cohen et~al. (2003)]{cohen03} 
Cohen, A. S., et~al. 2003, \apj, in press (astro-ph/0303419)

\bibitem[Cole et~al. (2000)]{cole00} 
Cole, S., Lacey, C. Baugh, C., \& Frenk, C. 2000, \mnras, 319, 168

\bibitem[Coleman, Wu \& Weedman(1980)]{cww} 
Coleman, G. D., Wu C.-C. \& Weedman D. W. 1980, \apjs, 43, 393

\bibitem[Colless et~al. (2001)]{colless} 
Colless, M., et~al. 2001, \mnras,  328 1039

\bibitem[Comastri et~al. (2001)]{comastri} 
Comastri, A., Fiore, F., Vignali, C., Matt, G., Perola, G., \& 
	La Franca, F. 2001, \mnras, 327, 781
 
\bibitem[Condon et~al. (1998)]{condon98} 
Condon, J. J., et~al. 1998, in 
	{\it Observational Cosmology with the New Radio Surveys},
	ed. M. N. Bremer et~al. (Kluwer: Dordrecht), p37

\bibitem[Conselice (2003)]{conselice} 
Conselice, C. 2003, in {\it Galaxy Dynamics}, ed. C. Boily, P. Patsis, 
C. Theis, S. PortegiesZwart, R. Spurzem, (EDP Sciences), in press 
(astro-ph/0212468)

\bibitem[Cutri  (2001)]{cutri} 
Cutri, R. 2001, in {\it AGN Surveys}, ed. R. Green, E Kachikian, \& D. Sanders,
	(ASP:San Francisco), p5.

\bibitem[Dekel (1994)]{dekel94} 
Dekel, A. 1994, \araa, 32, 371

\bibitem[de Ruiter et~al. (1997)]{deruiter97} 
de Ruiter, H. R., et~al. 1997, \aap, 319. 7

\bibitem[Devriendt, Guiderdoni \& Sadat (1999)]{stardust} 
Devriendt, J.E.G., Guiderdoni, B., \& Sadat R. 1999, \aap, 350, 381

\bibitem[Devriendt \& Guiderdoni(2000)]{devriendt} 
Devriendt, J.E.G. \& Guiderdoni, B. 2000, \aap, 363, 851

\bibitem[Dickinson et~al. (2003)]{dickinsonpasp} 
Dickinson, M., et~al., 2003, \pasp, this volume

\bibitem[Dickinson et~al. (2003)]{dickinson02} 
Dickinson, M., Papovich, C., Ferguson, H., Budavari, T. 2002, 
	\apj, 587, 25

\bibitem[Dole et~al. (2003)]{dole03} 
Dole, H., Lagache, G., \& Puget, J.-L. 2003, \apj, 585, 617

\bibitem[Dole et~al. (2001)]{dole01} 
Dole, H., et~al. 2001, \aap, 372, 364

\bibitem[Donahue et~al. (2001)]{donahuea} 
Donahue, M., et~al. 2001, \apj, 552, 93

\bibitem[Donahue et~al. (2002)]{donahueb} 
Donahue, M., et~al. 2002, \apj, 569, 689

\bibitem[Dunlop et~al. (2003)]{dunlop} Dunlop, J., et~al 2003,
 {\it http://www.roe.ac.uk/shades/index.html}

\bibitem[Elbaz et~al. (1999)]{elbaz}
Elbaz, D., et~al. 1999, \aap, 351, L37

\bibitem[Ellis et~al. (1997)]{ellis97} 
Ellis, R., et~al. 1997, \apj, 483, 582

\bibitem[Efstathiou \& Rowan-Robinson (2003)]{efstathiou} 
Efstathiou, A. \& Rowan-Robinson, M. 2003, \mnras, in press
(astro-ph/0304555)

\bibitem[Efstathiou et~al. (1990)]{e90} 
Efstathiou, A.  et~al. 1990, \mnras, 247, 10

\bibitem[Evans et~al. (1998)]{evans} 
Evans, R. W., et~al. 1998, Icarus, 131, 261

\bibitem[Fadda et~al. (2002)]{fadda02} 
Fadda, D., et~al. 2002, \aap, 383, 838

\bibitem[Fan et~al.(2003)]{fan03} 
Fan, X., et~al. 2003, \aj, 125, 1649

\bibitem[Fang et~al.(2003)]{fang03} 
Fang, F., et~al. 2003, in preparation

\bibitem[Fardal et~al. (2001)]{fardal} 
Fardal, M. A., Katz, N., Gardner, J., Hernquist, L., Weinberg, D., 
	\& Dave, R. 2001, \apj, 562, 605

\bibitem[Farrah et~al.(2001)]{farrah01} 
Farrah, D., et~al. 2001,  \mnras, 326, 1333

\bibitem[Farrah et~al.(2002a)]{farrah02a}
Farrah, D., Verma, A., Oliver, S., Rowan-Robinson, M., \& McMahon, R. 
	2002a, \mnras, 329, 605

\bibitem[Farrah et~al.(2002b)]{farrah02} 
Farrah, D., Serjeant, S., Efstathiou, A., Rowan-Robinson, M., \& Verma, A.
	2002b, 	\mnras, 335, 1163

\bibitem[Farrah et~al. (2003)]{farrah03} 
Farrah, D., Afonso, J., Efstathiou A., Rowan-Robinson, M., Fox, M. \& 
	Clements, D. 2003, \mnras, in press (astro-ph/0304154)

\bibitem[Fazio et~al. (2003)]{faziogto} 
Fazio, G., et~al. 2003, SIRTF Guaranteed Time Observation Program,\\ 
{\it http://sirtf.caltech.edu/SSC/geninfo/gto/absabs/pid8}

\bibitem[Fern\'andez-Soto et~al. (2001)]{fsoto01} 
Fern\'andez-Soto A., Lanzetta K., Chen H.-W., Pascarelle S., \&
	Yahata N. 2001, \apjs, 135, 41

\bibitem[Fioc \& Rocca-Volmerange (1997)]{pegase} Fioc, M. \&
Rocca-Volmerange, B. 1997, \aap, 326, 950

\bibitem[Fox et~al. (2002)]{fox}  
Fox, M. J., et~al. 2002, \mnras, 331, 839

\bibitem[Franceschini et~al. (2003)]{frans03} 
Franceschini, A., Lonsdale, C., \& the SWIRE Co-Investigator Team, 2003,
	in {\it The Mass of Galaxies at Low and High Redshift}, p. 338

\bibitem[Franceschini et~al. (2003)]{frans02} 
Franceschini, A., Braito, V., \& Fadda, D. 2002, \mnras, 335, L51

\bibitem[Franceschini et~al. (2001)]{frans01} 
Franceschini, A.,  et~al. 2001, \aap, 378, 1

\bibitem[Fritz, J., et~al. (2003)]{fritz03} 
Fritz, J., et~al. 2003, in preparation

\bibitem[Gautier et~al. (1992)]{gautier} 
Gautier, T. N., Boulanger, F., Perault, M., \& Puget, J. L. 1992, \aj, 
	103, 1313

\bibitem[Geballe et~al.  (2002)]{geballe02}
Geballe, T. R., et~al. 2002, \apj, 564, 466

\bibitem[Giacconi et~al. (2001)]{giacconi01} 
Giacconi, R., et~al. 2001, \apj, 551, 624

\bibitem[Gilli (2002)]{gilli02} 
Gilli, R. 2002 in {\it New X-ray Results 
from Clusters of Galaxies \& Black Holes}, ed. C. Done, E.M. Puchnarewicz, 
	\& M.J. Ward, Adv. Sp. Res., in press (astro-ph/0303115)

\bibitem[Gilli (2001)]{gilli} 
Gilli, R., Salvati, M., \& Hasinger, G. 2001, \aap, 366, 407

\bibitem[Gladders \& Yee (2003)]{gladders}
Gladders, M. \& Yee, H. 2003, in preparation

\bibitem[Granato et~al. (2001)]{granato} 
Granato, G., et~al. 2001, \mnras, 324, 757

\bibitem[Gregg et~al. (2002)]{gregg02} 
Gregg, M. D., et~al. 2002, \apj, 564, 133

\bibitem[Gruppioni et~al. (1999)]{gruppioni99} 
Gruppioni, C., et~al. 1999, \mnras, 305, 297

\bibitem[Guiderdoni et~al. (1998)]{guiderdoni} 
Guiderdoni, B., Hivon, E., Bouchet, F. R., \& Maffei, B. 1998, \mnras, 295,877

\bibitem[Haas et~al. (2000)]{haas00} 
Haas,  M., et~al. 2000, \aap, 354, 453

\bibitem[Hasinger et~al. (1998)]{hasinger98} 
Hasinger, G., Burg, R., Giacconi, R., Schmidt, M., Trumper, J., \&
	Zamorani, G. 1998, \aap, 329, 482

\bibitem[Hasinger et~al. (2001)]{hasinger01} 
Hasinger, G., et~al. 2001, \aap, 365, 45

\bibitem[Hauser \& Dwek (2001)]{hauser} 
Hauser, M. G. \& Dwek E. 2001, \araa, 39, 249

\bibitem[Helou \& Beichman (1990)]{helou} 
Helou, G. \& Beichman, C. 1990, in {\it From Ground-Based to Space-Borne 
	Sub-mm Astronomy}, ESA SP-314, (Noordwijk: ESA), p 117

\bibitem[Ivezic et~al. (2002)]{ivezic} 
Ivezic, Z., et~al. 2002, \aj, 124, 2364

\bibitem[Ivison et~al. (2002)]{ivison02} 
Ivison, R. J., et~al. 2002, \mnras, 337, 1

\bibitem[Jannuzi et~al. (2002)]{jannuzi} 
Jannuzi, B. T., Dey, A., Brown, M. J. I., Tiede, G. P., \& NDWFS Team 2002, 
	\baas, 34, 104.01
 
\bibitem[Jarrett, Dickman \& Herbst (1994)]{jarrett94} 
Jarrett, T. H., Dickman, R. L. \& Herbst, W. 1994, \apj, 424, 852

\bibitem[Jarrett et~al. (2003)]{jarrett} 
Jarrett, T. H., Chester, T., Cutri, R., Schneider, S., \& Huchra, J. P. 
	2003, \aj, in press

\bibitem[John (1988)]{john88} 
John, T. L. 1988, \aap, 193, 189

\bibitem[Kaiser (1984)]{kaiser} 
Kaiser, N. 1984, \apjl, 284, L9

\bibitem[Kajisawa \& Yamada (2001)]{kajisawa}
Kajisawa, M. \& Yamada T. 2001, PASJ, 53, 833

\bibitem[Kaviani et~al. (2003)]{kaviani} 
Kaviani, A., Haehnelt, M. \& Kauffmann, G. 2003, \mnras, 340, 739

\bibitem[Kawara et~al. (1998)]{kawara98} 
Kawara, K., et~al. 1998, \aap, 336, 9

\bibitem[Kennicutt et~al. (2003)]{kennicutt} 
Kennicutt, R. J., et~al. 2003, \pasp, this volume

\bibitem[Kenter et~al. (2002)]{kenter02} 
Kenter, A., Murray, S. \& Meehan, G. 2002, APRB, 17, 108

\bibitem[Kepner et~al. (1999)]{1999ApJ...517...78}
Kepner, J., Fan, X., Bahcall, N., Gunn, J., Lupton, R., \& Xu, G. 1999,
	\apj, 517, 78

\bibitem[Kessler et~al. (1996)]{kessler96} 
Kessler, M. F., et~al. 1996, \aap, 315, L27

\bibitem[Klaas et~al. (2001)]{klaas01} 
Klaas, U., et~al. 2001, \aap, 379, 823

\bibitem[Kodama et~al. (1999)]{1999MNRAS.302..152}
Kodama, T., Bell, E. \& Bower, R. 1999, \mnras, 302, 152

\bibitem[Koerner et~al. (1998)]{koerner} 
Koerner, D. W., Ressler, M. E., Werner, M. W., \& Backman, D. E. 1998, \apj, 
	503, L83

\bibitem[Kuraszkiewicz et~al. (2003)]{kuras03} 
Kuraszkiewicz, J. K., et~al. 2003, \apj, 590, in press (astro-ph/0302572)

\bibitem[Labbe et~al. (2003)]{labbe02} 
Labbe, I., et~al. 2003, \aj, 125, 1107

\bibitem[Lacy et~al. (2001)]{lacy} 
Lacy, M., Laurent-Muehleisen, S., Ridgway, S., Becker, R., \& White, R. L. 
	2001, \apjl, 551, L17

\bibitem[Lagache et~al. (2003)]{lagache03} 
Lagache, G., Dole, H., \& Puget, J-L. 2003, \mnras, 338, 555 

\bibitem[Lagrange et~al. (2000)]{lagrange} 
Lagrange, A-M., Backman, D., Artymowicz, P. 2000, in 
	{\it Protostars and Planets IV}, ed. V. Mannings, A. Boss, S., 
	Russell, (Tucson:U. Arizona Space Science Series), p639

\bibitem[Lahav et~al. (1990)]{lahav} Lahav O.,
Nemiroff, R. J. \& Piran T. 1990, \apj, 350, 11

\bibitem[Lahav et~al.(2002)]{2002MNRAS.333..961L} Lahav, O.~et al.\ 2002, 
\mnras, 333, 961 

\bibitem[Lari et~al. (2001)]{lari01} 
Lari, C., et~al. 2001, \mnras, 325, 1173

\bibitem[Leitherer et~al. (1999)]{starburst} 
Leitherer, C., et~al. 1999, \apjs, 123, 3

\bibitem[Magorrian et~al. (1998)]{magorrian} 
Magorrian, J., et~al. 1998, \aj, 115, 2285

\bibitem[Maiolino et~al. (2000)]{maiolino00} 
Maiolino, R., et~al. 2000, Adv. Sp. Res., 25, 809

\bibitem[Maiolino et~al. (1998)]{maiolino98} 
Maiolino, R., et~al. 1998, \aap, 338, 781

\bibitem[Manners et~al. (2003)]{manners03} 
Manners, J. C., et~al. 2003, \mnras, in press (astro-ph/0207622)

\bibitem[Mart\'in et~al. (2001)]{martin} 
Mart\'in, E. L., et~al. 2001, \pasp, 113, 529

\bibitem[Menanteau et~al. (1999)]{menanteau}
Menanteau, F., et~al. 1999, \mnras, 309, 208

\bibitem[Mirabel et~al. (1999)]{mirabel99} 
Mirabel, I. F., et~al. 1999, \aap, 341, 667

\bibitem[Mizumoto et~al. (2003)]{mizumoto} 
Mizumoto, Y., et~al. 2003, {\it http://www.naoj.org/staff/chris/SXDS/}

\bibitem[Moorwood (1999)]{moorwood99} 
Moorwood, A. F. M. 1999, in {\it The Universe as seen by ISO}, 
	ed. P. Cox \& M. Kessler, ESA SP-427 (Noordwijk: ESA), p825

\bibitem[Norberg et~al. (2002)]{norberg} 
Norberg, P., et~al. 2002, \mnras, 332, 827

\bibitem[Oliver et~al. (1996)]{oliver96} 
Oliver, S., et~al. 1996, \mnras, 280, 673

\bibitem[Oliver et~al. (2000)]{oliver00} 
Oliver, S., et~al. 2000, \mnras, 316, 749

\bibitem[Oliver et~al. (2003)]{oliver03} 
Oliver, S., et~al. 2003, in preparation

\bibitem[Owen et~al. (2003)]{owen03} 
Owen, F., et~al. 2003, in preparation

\bibitem[Oppenheimer et~al. (2000)]{oppenheimer} 
Oppenheimer, B. R., Kulkarni, S. R., \& Stauffer, J. R. 2000, in 
	{\it Protostars and Planets IV}, ed. V. Mannings, A. Boss, S., 
	Russell, (Tucson:U. Arizona Space Science Series), p639

\bibitem[Papovich \& Bell (2002)]{papovich02} 
Papovich, C. \& Bell E. F. 2002, \apjl, 579, 1

\bibitem[Pearce et~al. (2001)]{pearce} Pearce  et~al. 2001, \mnras, 326, 649

\bibitem[P\'{e}rez Garc\'{\i}a \& Rodr\'{\i}guez Espinosa (2001)]
{rodriguez01} P\'erez Garc\'{i}a, A. M. \& Rodr\'{i}guez Espinosa, J. M. 2001,
	\apj, 557, 39

\bibitem[Phillips et~al. (2001)]{phillips01} 
Phillips, J., et~al. 2001, \apj, 560, 15

\bibitem[Pierre et~al. (2002)]{pierre02} 
Pierre, M., Valtchanov, I. \&  Refregier, A. 2002, in {\it New Visions of the
	X-Ray Universe}, in press (astro-ph/0202117)

\bibitem[Pierre et~al. (2003)]{pierre03} 
Pierre M.,  et~al. 2003 \aap, in preparation

\bibitem[Polletta et~al. (2003)]{polletta03} 
Polletta, M., Lonsdale, C. J., Xu, C. K. \& Wilkes, B. J. 2003, \apj, 
	submitted  

\bibitem[Polletta et~al. (2000)]{polletta00} 
Polletta, M., Courvoisier, T. J.-L., Hooper, E. J. D., \& Wilkes, B. J. 
	2000, \aap, 362, 75

\bibitem[Poggianti et~al. (2001)]{poggianti01} 
Poggianti, B. M., Bressan, A., \& Franceschini, A. 2001, \apj, 550, 195

\bibitem[Priddey \& McMahon (2001)]{priddey} 
Priddey, R. \& McMahon, R. 2001, \mnras, 324, L17 

\bibitem[Reid et~al. (1999)]{reid} 
Reid, I. N., et~al. 1999, \apj, 521, 613

\bibitem[Rieke et~al. (2003)]{riekegto} 
Rieke, G., et~al. 2003, SIRTF Guaranteed Time Observation Program,\\ 
	{\it http://sirtf.caltech.edu/SSC/geninfo/gto/abs/pid81}

\bibitem[Rigopoulou et~al. (1999)]{rigopoulou99} 
Rigopoulou, D., Spoon, H., Genzel, R., Lutz, D., Moorwood, A., \& 
	Tran, Q.  1999, \aj, 118, 2625

\bibitem[Rodighiero et~al. (2003)]{rodighiero02} 
Rodighiero, G., Fadda, D., Alessandra, G., Lari, C., \& Franceschini, A. 2003,
	in {\it Exploiting the ISO Data Archive: Infrared Astronomy in the 
	Internet Age}, ESA SP-511 (Noordwijk: ESA), in press

\bibitem[Roelfsema et~al. (1996)]{roelfsema96} 
Roelfsema, P. R., et~al. 1996, \aap, 315, L289

\bibitem[Rosati et~al. (2002)]{rosati02} 
Rosati, P., et~al. 2002, \apj, 566, 667 

\bibitem[Rowan-Robinson et~al. (1997)]{mrr97} 
Rowan-Robinson, M., et~al. 1997, \mnras, 289, 490

\bibitem[Rowan-Robinson et~al. (2000)]{rr00} 
Rowan-Robinson, M., et~al. 2000 \mnras, 314, 375

\bibitem[Rowan-Robinson (2001)]{mrr01} 
Rowan-Robinson, M. 2001, \apj, 549, 745

\bibitem[Rowan-Robinson (2002)]{mrr02} 
Rowan-Robinson, M. 2003, \mnras, submitted

\bibitem[Rowan-Robinson et~al. (2003)]{mrr03} 
Rowan-Robinson, M., et~al. 2003, in preparation

\bibitem[Saunders et~al. (1991)]{s91} 
Saunders, W., Frenk, C., Rowan-Robinson, M., Lawrence, A., \& Efstathiou, G. 
	1991 \nat, 349, 32

\bibitem[Saunders et~al. (2000)]{pscz} 
Saunders, W., et~al. 2000 \mnras, 317, 55

\bibitem[Sawicki (2002)]{sawicki02} 
Sawicki, M. 2002, \aj, 124, 3050

\bibitem[Schlegel et~al. (1998)]{schlegel} Schlegel D. J.,  Finkbeiner D. P. 
	\& Davis M., 1998 \apj, 500 525

\bibitem[Schreier et~al. (2001)]{schreier01} 
Schreier, E. J., et~al. 2001, \apj, 560, 127

\bibitem[Scott et~al. (2002)]{scott} 
Scott, S., et~al. 2002, \mnras, 331, 817

\bibitem[Shupe et~al. (1998)]{shupe98} 
Shupe, D. L., Fang, F., Hacking, P.B., \& Huchra, J.P. 1998, \apj, 501, 597

\bibitem[Shupe et~al. (1996)]{shupe96} 
Shupe, D., Huber, A. K., \&  Hacking, P. 1996, SPIE Proceedings 2817, 
{\it Infrared Spaceborne Remote Sensing IV}, ed. M. Scholl \& B. Andresen, 
	p1057

\bibitem[Siana et~al. (2002)]{siana} 
Siana, B., Smith, H. E., Lonsdale, C. J., \& The SWIRE Team 2002, \baas, 
	34, 1195

\bibitem[Silk \& Rees (1998)]{silk} 
Silk J. \& Rees, M. 1998, \aap, 331, L1

\bibitem[Silva et~al. (1998)]{grasil} 
Silva L., Granato G. L., Bressan A., \& Danese L. 1998, \apj, 509, 103

\bibitem[Simcoe et~al. (2000)]{simcoe} 
Simcoe R. A., et~al. 2000, AAS, 196, 5209

\bibitem[Simpson \& Eisenhardt (1999)]{simpson} 
Simpson, C. \& Eisenhardt, P. 1999, \pasp, 111, 691

\bibitem[Soifer et~al.(1989)]{soifer89} 
Soifer, B. T., Boehmer, L., Neugebauer, G., \& Sanders, D. 1989, \aj, 98,
	766

\bibitem[Somerville et~al. (2001)]{somerville} 
Somerville R., Lemson G., Sigad Y., Dekel A., Kauffmann G., \&
	White S., 2001, \mnras,  320, 289

\bibitem[Somerville (2003)]{somervillepc} 
Somerville, R. S. 2003, private communication.

\bibitem[Strauss \& Willick (1995)]{strauss95} Strauss M.A., \&
         Willick, J. 1995, Physics Reports, 261, 271

\bibitem[Sturm et~al. (2000)]{sturm00} 
Sturm, E., et~al. 2000, \aap, 358, 481

\bibitem[Surace et~al. (2000)]{surace00}
Surace J. A., Sanders D. B. \& Evans A. S. 2000, \apj, 529, 170

\bibitem[Szalay, Connolly \& Szokoly (1999)]{szalay99} 
Szalay, A. S.,  Connolly, A. J., \& Szokoly, G. P. 1999, \aj, 117, 68

\bibitem[Tran (1998)]{tran98} 
Tran, D. 1998, PhD thesis, Universit\'e de Paris XI.

\bibitem[Vaccari et~al. (2003)]{vaccari03} 
Vaccari, M., et~al. 2003, in preparation

\bibitem[Valtchanov et~al. (2003)]{valtchanov} 
Valtchanov, I., et~al. 2003, \aap, in preparation

\bibitem[Verstraete et~al. (1996)]{verstraete96} 
Verstraete, L., Puget, J.-L., Falgarone, E., Drapatz, S., Wright, C., \& 
	Timmermann, R. 1996, \aap, 31

\bibitem[Williams et~al. (1996)]{hdf} 
Williams, R., et~al. 1996, \aj, 112, 1335

\bibitem[Willick \& Strauss(1998)]{1998ApJ...507...64W} Willick, J.~A.~\& 
Strauss, M.~A.\ 1998, \apj, 507, 64 

\bibitem[Willis et~al. (2003)]{willis} 
Willis, A. G., et~al. 2003, \aap, in preparation

\bibitem[Xu et~al. (2001)]{xu01} 
Xu, C. K., Lonsdale, C.J., Shupe, D.L., O'Linger, J., \& Masci, F. 2001, \apj, 
	562, 179

\bibitem[Xu et~al. (2003)]{xu03} 
Xu, C. K., Lonsdale, C. J., Shupe, D. L., Franceschini, A., Martin, D., 
	\& Schiminovich, D. 2003, \apj, 587, 90

\bibitem[Yang et~al. (2003)]{yang03} 
Yang, Y., Mushotsky, R., Barger, A., Cowie, L., Sanders, D. \& 
	Steffen, A. 2003, \apj, 585, L85

\bibitem[York et~al. (2000)]{sdss} 
York, D. G., et~al. 2000, \aj, 120 1579

\end{thebibliography}
\end{document}